\documentclass[published]{JHEP3}

\JHEP{ }

\usepackage{cite}
\usepackage{epsf}
\usepackage{latexsym}
\usepackage{epsfig,multicol}

\def\preal{{\rm Re\,}}
\def\pim{{\rm Im\,}}

\def\yzero{\smash{\hbox{$y\kern-4pt\raise1pt\hbox{${}^\circ$}$}}}
\def\p{\partial}
\def\a{\alpha}
\def\b{\beta}
\def\g{\gamma}
\def\d{\delta}
\def\th{\theta}
\def\beq{\begin{equation}}
\def\eeq{\end{equation}}
\def\beqa{\begin{eqnarray}}
\def\eeqa{\end{eqnarray}}
\def\Om{\Omega}
\def\om{\omega}
\def\th{\theta}
\def\vt{\vartheta}

\def\-{\hphantom{-}}
\def\ov{\overline}
\def\s2{\frac{1}{\sqrt2}}

\def\oh{\frac{1}{2}}
\def\beq{\begin{equation}}
\def\eeq{\end{equation}}
\def\beqa{\begin{eqnarray}}
\def\eeqa{\end{eqnarray}}
\def\tr{{\rm tr \,}}
\def\Tr{{\rm Tr \,}}

\def\IF{\relax{\rm I\kern-.18em F}}
\def\II{\relax{\rm I\kern-.18em I}}
\def\IP{\relax{\rm I\kern-.18em P}}
\def\IC{\relax\hbox{\kern.25em$\inbar\kern-.3em{\rm C}$}}
\def\IR{\relax{\rm I\kern-.18em R}}

\def\ci{{\cal I}}

\def\cn{{\cal N}}
\def\cam{{\cal M}}

\def\Dsl{\,\raise.15ex\hbox{/}\mkern-13.5mu D} 
\def\IZ{Z\kern-.4em  Z}

\def\inte{{\bf Z}}
\def\cpx{{\bf C}}
\def\real{{\bf R}}
\def\nat{{\bf N}}

\def\R{{\cal R}}
\def\ca{{\cal A}}

\def\cl{{\cal L}}
\def\cw{{\cal W}}
\def\lam{\lambda}
\def\Lam{\Lambda}
\def\D{\Delta}
\def\G{\Gamma}
\def\raw{\rightarrow}
\def\Raw{\Rightarrow}
\def\lraw{\leftrightarrow}
\def\eps{\epsilon}
\def\k{\kappa}
\def\z{\zeta}
\def\sig{\sigma}
\def\Sig{\Sigma}

\def\Oom{{\bf \Omega}}
\def\co{\cal O}
\def\Dbar{D{\hspace{-7.5pt}\slash}}

\newbox\pippobox

\title{Computing Yukawa Couplings from Magnetized Extra Dimensions}

\author{D.~Cremades$^*$, L.~E.~Ib\'a\~nez$^{*,**}$ and  F.~Marchesano$^{***}$ \vspace{2mm} \\
 	*\ Departamento de F\'{\i}sica Te\'orica C-XI
	and Instituto de F\'{\i}sica Te\'orica  C-XVI,\\ \vspace{2mm}
	Universidad Aut\'onoma de Madrid,
	Cantoblanco, 28049 Madrid, Spain.\\ \vspace{2mm}
	**\ Theory Division, CERN, 1211 Geneva 23, Switzerland.
	\\
	***\ Department of Physics, University of Wisconsin-Madison, WI 53706,
	USA.
}
\received{\today}               
\revised{}
\accepted{}               

\preprint{\hepth{0404229}}
\preprint{FTUAM-04/7 IFT-UAM/CSIC-04/15 MAD-TH-04-4}

\abstract{We compute Yukawa couplings involving chiral matter fields
in toroidal compactifications of
higher dimensional super-Yang-Mills theory
with magnetic fluxes. Specifically we focus
 on toroidal compactifications of D=10 super-Yang-Mills theory, 
 which may be obtained as the low-energy limit 
of Type I, Type II or  Heterotic strings.
Chirality is obtained  by turning on 
 constant magnetic  fluxes   in each of the 2-tori.
Our results are general and may as well be applied to 
lower $D=6,8$ dimensional field theories.
We solve Dirac and Laplace equations to find out the
explicit form of wavefunctions in extra dimensions.
The Yukawa couplings are computed as overlap integrals
of two Weyl  fermions and one complex scalar over the
compact dimensions.
In the case of Type  IIB (or Type I) 
 string theories,  the models are T-dual to (orientifolded) Type IIA
 with D6-branes intersecting at angles.
These theories may have phenomenological relevance
since particular models with SM group and three quark-lepton
generations have been recently  constructed. We find that
the Yukawa couplings so obtained are described by Riemann
$\vt$-functions, which depend on the
complex structure and Wilson line backgrounds.
Different patterns of Yukawa textures are possible depending
on the values of these backgrounds.
We discuss the matching of these results with the
analogous computation in models with intersecting D6-branes.
Whereas in the latter case a string computation is required,   
in our case only field theory is needed.
}


\begin{document}


\section{Introduction}

One of the most outstanding  puzzles of the standard model (SM) of
particle physics is the structure  of the Yukawa couplings
between the Higgs field and the SM fermions. A correct description of
the observed  masses and mixing of quarks and leptons seems
to require very different values for the Yukawa coupling constants
for the different generations. Although many approaches have been
attempted to describe the hierarchical structure of Yukawa
couplings it is fair to say that we do not have at the moment
a compelling theory for quark and lepton masses.
Thus the search for theoretical schemes which could explain
this Yukawa structure is of the utmost importance.

In the last 25 years the idea that there could be
more than four dimensions has been pursued intensively,
particularly due the study of string theory which is
naturally defined in 10 or 11 dimensions. Extra dimensions offer
in principle
the possibility of computing Yukawa couplings in terms of the
extra-dimensional geography. Indeed, starting from a (D+4)-dimensional
field theory  and compactifying D dimensions one may get  
massless modes with factorized wavefunctions 
$\psi(x)\times \phi (y)$, with $x$($y$) denoting Minkowski and
extra dimensions respectively. Gauge boson components $A^i$ in
extra dimensions give rise to scalars at low energies and
Yukawa couplings are thus expected to appear upon compactification from the
higher dimensional gauge vertex interaction
$A^M\Psi \Gamma _M \Psi$. Qualitatively the Yukawa coupling constants
may be computed from overlap integrals over extra dimensions of the form
\cite{SW,GSW2}
\beq
Y_{ij}\ =\ \int dy^D \phi_i (y)\phi_j (y)A(y) \ .
\label{overlap}
\eeq
Unfortunately it is not easy to find compactifications in which:

\begin{itemize}

\item
  One has chiral fermions,  as required in the SM

\item  One can
solve the Dirac and Laplace equations to obtain explicit expressions
for the wavefunctions in extra dimensions

\item One can explicitly work out the overlap integrals and
actually compute the Yukawa couplings
\end{itemize}

Our main goal in the present article is to perform such a type    
of computation in a class of theories of potential
phenomenological interest. We consider as our starting point
10-dimensional super-Yang-Mills (SYM) theory as the best motivated
extra dimensional field theory, since it appears in  the low-energy
limit of Type I, Type IIB and heterotic string theories. However
one can easily apply the results (i.e. the solutions of Dirac
and Laplace equations as well as the computation of overlap integrals)
to other lower dimensional $D=6,8$, as we discuss below.

We compactify D=10 SYM on a 6-torus $T^6$ and, in order to obtain 
chiral fermions,  we add constant  magnetic flux through the torus
\cite{bachas,bgkl,aads,afiru}.
The fact that  one may obtain chiral fermions in the presence of explicit
gauge field backgrounds is well known
\cite{chiral}.
However, in the present case we consider a simple toroidal geometry 
with constant gauge field backgrounds, so that one can explicitly
find the eigenfunctions of the Dirac and Laplace  equations in
extra dimensions. 
\footnote{Toroidal compactifications with 
constant field strenght are also known 
as `Torons' \cite{thooft,vanbaal1,vanbaal2,tony}. 
The connection with D-brane physics has 
been adressed in \cite{gr1,gr2,aki,troost,torons}. 
For other related computations see \cite{onofri}.}
We find that in the case of a factorized torus $T^{2n}$
the eigenfunctions are proportional
to products of $n$ Jacobi theta-functions with characteristics.
The profile of the corresponding  wavefunction densities in 
extra dimensions is Gaussian-like, with the location of 
the maximum controlled by the possible presence of Wilson-line
backgrounds (see figures 3, 4 and 5).

From these wavefunctions one can explicitly
perform the overlap integrals and obtain the Yukawa 
couplings.
The Yukawas so obtained are again products of $n$ $\vt$-functions
but this time they depend both on the complex structure moduli 
of the tori as well as on   the possible Wilson line
backgrounds present in the compactification.
In the case of a non-factorizable 6-torus one obtains 
analogous results but now one has instead generalized Riemann
theta functions.

One of the main results of the present paper is the final
expression for Yukawa couplings.
In the case of a factorized $2n$-torus,
the Yukawa couplings between three
fields labeled by integers $i,j,k$ have the following general form:
\beq
Y_{ijk} =
{g_{4+2n}} \prod_{r=1}^{n}
\left(\frac{\pim \tau^{(r)}}{2\pi \ca ^{(r)}} \right)^{1/4}
\left|\frac{\th_{I}^{(r)}\th_{J}^{(r)}}
{\th_{K}^{(r)}}\right|^{1/4}
\cdot
e^{H^{(r)}/2}
\cdot
 \vt
\left[
\begin{array}{c}
\d_{ijk}^{(r)} \\ 0
\end{array}
\right]
\left( \tilde{\z} ^{(r)}, \tau ^{(r)} |I_{I}^{(r)}
 I_{J}^{(r)} I_{K}^{(3)}| \right)
\label{yukseminal}
\eeq
Here $g_{4+2n}$ is the $(4+2n)$-dimensional gauge coupling constant
in the dilute flux approximation\footnote{This coincides with the
gauge couplings of the different gauge groups up to flux dependent
corrections which are suppressed in the large volume limit.},
$\tau ^{(r)},\ \ca^{(r)}$ are the complex structure
and the areas of the $r = 1,2,3$ compact 2-tori
whereas $\tilde{\z}^{(r)}$ describe the dependence on the Wilson line 
variables. $\prod_r I_{I}^{(r)}$ are the number of generations of
particles labeled by $i$. The quantity $\th_I^{(r)}$ is the magnetic flux
felt by such particle in the $r^{th}$ 2-torus, which also signals the scale
of new physics in this sector, more precisely the mass of the lightest 
massive replica of the chiral fermion $i$.
$H^{(r)}$ is a known function  (see eq. (\ref{Hna})) of
Wilson lines, complex structure and magnetic fluxes. It vanishes
in the absence of Wilson lines. Finally, $\vt $ are standard Jacobi
theta functions with characteristics,
with the  argument $\d_{ijk}^{(r)}=i/I_I^{(r)}+j/I_J^{(r)}+k/I_K^{(r)}$, 
being the only `flavour' ($i,j,k$) dependence of the Yukawa coupling.
In a SUSY compactification 
 the (holomorphic) superpotential
should be identified with this product of $\vt$-functions, whereas the rest
of the factors should come from D-term normalization of three-point functions.

As we said,
the natural setting for studying compactifications of D=10
 super Yang-Mills is string theory, since it appears naturally
in the effective field theory limit of Type I, Type IIB (with D9-branes)  and 
heterotic strings. In fact, although the mentioned compactification 
and computations 
may be performed  without any reference to string theory, in the 
present paper we will have always in mind the string theory point of view.
In particular it is well known that Type IIB theory with D9-branes
compactified on a (magnetized) $T^6$ is T-dual to Type IIA theory with 
D6-branes wrapping 3-cycles on $T^6$ and intersecting at angles.
\footnote{Analogously, Type I theory on $T^6$ is T-dual 
to a Type IIA orientifold with D6-branes at angles \cite{bgkl}.}  
In recent  years string models with intersecting
D-branes have been actively pursued in order to obtain
realistic models of particle physics
\cite{bgkl,afiru,afiru2,bkl,imr}. The main motivation for
this is that in these schemes chiral fermions naturally 
appear at the points in compact space in which Dp-branes 
intersect. Furthermore, since Dp-branes may intersect multiple times,
this gives a rationale for the family replication present in the SM.
A number of semirealistic models with the massless fermion spectrum 
of the SM or some $\cn=1$ extension with three generations have 
been obtained\cite{imr,bklo,cim1,cim2,more,susy}
\footnote{Notice that the two last refs. in \cite{more} do not correspond to geometrical intersecting D-brane models, but to Type II orientifold constructions on Gepner points. The model-building techniques in these particular cases are, nevertheless, similar to those in \cite{imr}. For previous literature on Gepner model orientifolds see \cite{Gepner}.}.

In fact one of the  motivations for the present investigation was to
check how T-duality maps the results obtained for intersecting 
D-brane models to the effective action of Type IIB with D9-branes
and magnetic fluxes. In particular,   
in ref.\cite{yukis} we computed the `classical' contribution to 
Yukawa couplings in Type IIA models with intersecting D6-branes.
In this case one has to perform a sum over string worldsheet 
instanton contributions to obtain the final expression of Yukawa couplings,
a pure stringy (non field-theoretical) computation. On the 
other hand T-duality tells us that equivalent Yukawa couplings 
should be obtained for the case of Type IIB theory compactified 
on $T^6$ with magnetic fluxes. In this second case the computation is
purely field theoretical since one just starts from 
$D=10$ SYM theory, compactifies and performs the overlap integral
to obtain explicitly the Yukawa couplings, as described above. 
Thus on this {\it flux side} of the T-duality the computation
is an exercise on Kaluza-Klein theory. Indeed we have found that 
the Yukawa couplings computed in both T-dual theories agree
after appropriate transformations of the moduli and Wilson line
variables. This is a non-trivial check given the completely 
different origin (stringy on the {\it intersection side},
field theoretical on the {\it flux side}) of both computations.
Furthermore, the computation obtained in the present paper 
goes beyond the results obtained in \cite{yukis} since we are
also able to obtain all normalization factors and coefficients
appearing in the Yukawa coupling. In the {\it intersection side}
in order to obtain the  normalization factors one has to 
compute the {\it quantum} piece of the appropriate
string correlators as in refs. \cite{cvetic,abel}. 
We find agreement with those normalization factors in the 
limit of small intersecting angles in which they should
coincide.\footnote{ Actually there is a small discrepancy 
in that we find that the $\Gamma$-functions appearing
in eq.(77) of ref.\cite{cvetic} should have a 1/4 power instead of 1/2.
This has also recently been independently pointed out in 
\cite{lmrs}.}

Due to the mentioned T-duality between the intersecting Dp-brane
models on one side and magnetized toroidal Type I and Type IIB 
compactifications on the other side, all intersecting
toroidal brane models constructed in 
ref.\cite{bgkl,afiru,afiru2,bkl,imr,bklo,cim1,cim2,more} 
admit a description in terms of the picture given in the present paper. 
In particular, the expressions here obtained 
may be used to compute the Yukawa couplings of the models 
in ref.\cite{imr} whose massless fermion spectrum is that of
the non-SUSY SM or to those of the (local) $N=1$ SUSY model
in ref.\cite{yukis}.
 With little changes they can 
also be applied to the computation of Yukawa couplings in the 
$\cn=1$ orbifold  models of refs.\cite{susy}. 

We would also like to emphasize that our results are also
relevant for the phenomenological extra dimension models 
considered in recent years \cite{extra}. In particular 
it has been suggested  \cite{split} that, if fermions of different
generations have Gaussian-like profile in one extra dimension
and  are localized at distant points, one may obtain hierarchical
Yukawa couplings from a one-dimensional overlap integral analogous to
(\ref{overlap}). Most of the  
 specific models constructed \cite{split,extra}
obtain chirality from solitonic scalar backgrounds. 
We find the approach  in the present paper more effective since it 
gives rise to chirality,
family replication and Gaussian-like profile in a natural and
explicit way.

The plan of the paper is as follows. In the next section we briefly
discuss the origin of Yukawa couplings in extra dimensional
theories, leaving the details of the required dimensional reduction 
to Appendix A, and the details of SUSY compactifications to the Appendix B. 
In Section 3 we address the computation
of the eigenstates of the Dirac and Laplace operators
in $T^{2n}$ toroidal compactifications. These provide us with the
wavefunctions of the light fields of the compactification, 
which are proportional 
to Riemann and Jacobi $\vt$-functions and have a Gaussian profile. 
In Section 4
we generalize our results to the case with non-Abelian Wilson lines
are present, in which the rank of the initial gauge group 
is in general lowered. Although quite interesting for rank reduction purposes, 
compactifications with non-Abelian Wilson line are technically much more involved and, up to some subtle points which allow to differentiate $\th_J$ from $I_J/\ca$, the final result for the Yukawa couplings is similar to the one obtained with the Abelian Wilson line case. The reader not interested in those details may safely skip this section.

Armed with the wavefunctions of
chiral fermions and light scalars we address in Section 5 
the computation of Yukawa couplings, by explicitly performing
the integrals which measures the overlap of three wavefunctions.
In the case of SUSY compactifications we rewrite our results in terms of
a superpotential and K\"ahler potential factors.
This analysis is extended in the following section to the case
of D-branes of lower dimension (e.g. D5-branes) which are quite
relevant for the construction of specific models.
A detailed comparison with the results obtained in the T-dual models (involving intersecting D-branes) is given in Section 7, where the Yukawa couplings in both sides of the T-duality are seen to match. We also comment on the action that T-duality has on chiral fields, as can be deduced from this matching.
As a simple application of the general results in this paper,
in chapter 8 we present a 3-generation MSSM-like
model and obtain the relevant Yukawa couplings using our formulas.
Some final comments and discussions are left for Section 9.

\section{Yukawa couplings in Kaluza-Klein theories}

In this section we motivate the study of magnetized compactifications 
in order to achieve $D=4$ chiral models from  
 extra dimensions. We describe as well the general strategy that 
we follow to compute three-point functions in such models. 
We refer the reader to the appendix \ref{reduction} or \cite{GSW2} 
for a more detailed discussion.

\subsection{$D=10$ {\cn=1} Super Yang-Mills compactifications with magnetic 
fluxes}

Let us consider $\cn=1$ $D=10$ supersymmetric Yang-Mills theory, whose Lagrangian density is simply given by
\beq
\cl_{SYM} = -\frac{1}{4g^2} \Tr \left\{F^{MN}F_{MN}\right\} + {i \over 2g^2} \Tr\left\{\bar{\lambda}\Gamma^M D_M \lambda\right\}
\label{SYM10}
\eeq
where the trace is performed in the adjoint representation of a gauge group $G$, and $M,N = 0, \dots, 9$. The gauge group field strength $F_{MN}$ and covariant derivative $D_M$ are given by
\beqa
F_{MN}& = & \partial_M A_N - \partial_N A_M - i[A_M, A_N] \\
D_M\lambda & = & \partial_M\lambda - i [A_M,\lambda]
\eeqa
both the ten-dimensional vector $A_M$ and the spinor $\lambda$ transforming in the adjoint of $G$. Notice that the Yang-Mills coupling constant $g$ has dimensions of (mass)$^{-3}$ in $D=10$.

In order to obtain a $D=4$ theory at low energies, 
we should consider the above theory compactified on a 
six-dimensional compact manifold $\cam_6$, so that 
we recover four-dimensional physics at energies below 
the compactification scale $M_c$. Indeed, in general 
the ten-dimensional fields $A_M$ and $\lam$ admit a decomposition of the form
\beqa
\lam (x^\mu,y^m) & = & \sum_n \chi_n (x^\mu) \otimes \psi_n (y^m)\\
A_M (x^\mu,y^m) & = & \sum_n \varphi_{n, M}(x^\mu) \otimes \phi_{n, M} (y^m)
\label{product10}
\eeqa
where $x^\mu$ $\mu = 0,\dots,3$ and $y^m$, $m = 4, \dots, 9$ 
stand for the non-compact and internal dimensions, respectively. 
The internal wavefunctions $\psi_n$, $\phi_{n\ M}$ can be chosen to be eigenstates of the corresponding internal wave operator
\beqa
i{\Dbar}_{6} \psi_n & = & m_n \psi_n \\
\D_{6} \phi_{n, M} & = & M_{n, M}^2 \phi_{n\ M}
\eeqa
By applying the equations of motion, we find that the Dirac mass of the four-dimensional spinor $\chi_n$ is given by $m_n$, and so on. The lightest Kaluza-Klein replica then sets the scale of energy below which we recover a $D=4$ physics governed by massless fermions and light scalars.
In practice, however, it turns out that plain 
compactifications on smooth $\cam_6$ leave the gauge 
group $G$ unbroken, the rank of $G$ being usually too 
high to accommodate semirealistic interactions
\footnote{Whereas at the classical level any  group $G$ 
(or a direct product of these) is acceptable, at the 
quantum level anomaly cancellation imposes restrictions 
on the allowed gauge groups.}. Moreover, 
the computation of the Dirac index shows that such 
compactifications lead to non-chiral spectra in four dimensions \cite{GSW2}.

Both unwanted features can be avoided by introducing 
non-trivial expectation values for the gauge field $A_M$
\cite{chiral}. Indeed, since 
we are only interested in preserving Poincar\'e invariance in the 
four non-compact dimensions, we are entitled to consider non-vanishing 
v.e.v.'s $\langle A_m (y) \rangle$, $m = 4, \dots, 9$. On the one hand, 
the gauge group $G$ will be reduced to a subgroup $H \subset G$ commuting 
with the subgroup $J$ which contains $\langle A_m (y) \rangle$. 
On the other hand, a non-trivial gauge field modifies the Dirac operator 
and hence the computation of the Dirac index, and may 
introduce a chiral asymmetry that allows for a chiral 
massless spectrum \cite{GSW2,chiral}. 
We hence find that compactifications 
with  non-trivial gauge fields $\langle A_m (y) \rangle$, or equivalently, 
magnetized $\cam_6$ compactifications with 
$\langle F_{mn}\rangle \neq 0$, provide a natural way 
of achieving $D=4$ chiral theories with reduced gauge 
group (See figure \ref{flux1}).

%
\EPSFIGURE{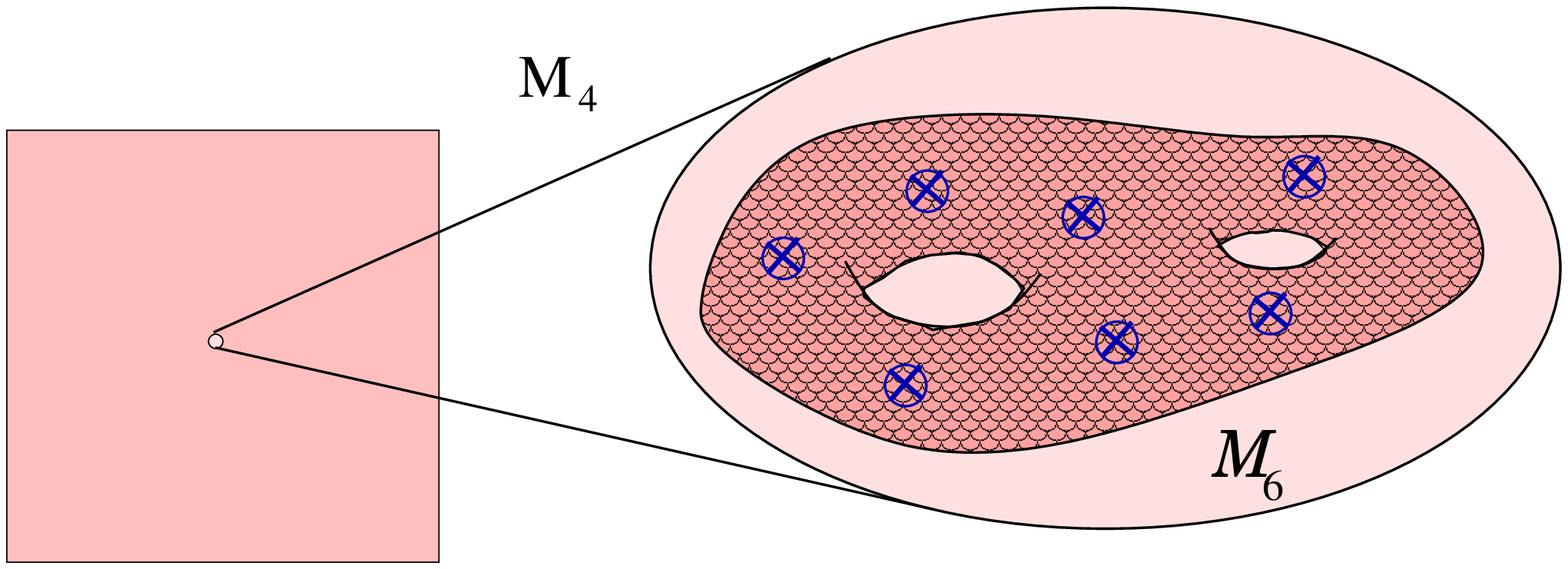, width=6in}
{\label{flux1}
Magnetized compactification. In the limit of large volume 
and diluted fluxes the field theory compactification 
with magnetic fluxes captures the underlying string theory physics.}
%

In addition, the introduction of a magnetic field in the 
compactification may not only lead to chiral matter but also 
to replication of chiral fermions, since the Dirac equation 
for the internal fermionic wavefunction $\Dbar_6 \psi = 0$ may 
yield several independent degenerate solutions, 
labeled by $\psi_j (y)$. In order to get canonical kinetic 
terms, these internal wavefunctions must satisfy 
\beq
\int_{\cam_6} d^{6}y\ {\psi_j (y)}^\dagger \psi_k (y)  =   \d_{jk}
\label{normKK10}
\eeq
the same condition applying to bosonic wavefunctions.

Finally, given the internal wavefunctions $\psi_j$, $\phi_k$ corresponding to the $D=4$ chiral fermions and lightest scalars, it is possible to compute the Yukawa couplings between them, as an overlap between three wavefunctions. Indeed, the fermionic part of the $D=10$ SYM action (\ref{SYM10}) contains a term of the form $A \cdot \lam \cdot \lam$, which upon dimensional reduction yields the Yukawa coupling\footnote{See Appendix \ref{reduction} for a more detailed discussion of the derivation of this formula.}
\beq
Y_{ijk} = \int_\cam\ \psi_{i}^{a \dag}\ \G^m\ \psi_j^b\ \phi_{k,m}^c\  f_{abc}
\label{yukfinal}
\eeq
where $f_{abc}$ are the structure constants of the initial gauge group $G$. 
Notice that formula (\ref{yukfinal}) provides us with the three-point 
function or normalized Yukawa coupling, i.e., not only contains 
the trilinear coupling of the superpotential $W_{ijk}$, but also 
all the normalization factors coming from the K\"ahler 
potential/kinetic terms. Moreover, this expression 
is completely general, in the sense that we are not making 
any assumption on the holonomy of the compact manifold $\cam_6$ 
or considering any particular embedding of the spin connection 
in the gauge group $G$.

\subsection{Models with fluxes on $D = 6,8$ dimensions}

The choice of $D=10$ SYM naturally arises from considering the 
low energy effective action arising from heterotic and Type I theories, 
which are the simplest superstring theories involving gauge interactions. 
From the field theoretical point of view, however, we could consider 
e.g. the Lagrangian (\ref{SYM10}) in $D$ dimensions, $D$ being even. 
The dimensional reduction scheme performed for $D=10$ can be generalized 
for arbitrary $D$, obtaining a formula for the three-point function similar 
to (\ref{yukfinal}), but now with ${\rm dim} (\cam) = D - 4$.

It is in fact easy to build string based models where 
such $D=6,8$ Lagrangians appear. 
Let us for instance 
consider type IIB string theory. Now, instead of compactifying it in 
a six-dimensional smooth manifold $\cam_6$, let us consider the 
possibility that $\cam_6$ presents a $\inte_N$  orbifold singularity, in such 
a way that in the vicinity of the singularity the metric can be 
written as $T^{2n} \times \cpx^{3-n}/\inte_N$. We can then place a 
stack of $N$ D$(3 + 2n)$-branes at such singular point, 
wrapping the $T^{2n}$ completely. In general, the worldvolume of 
$N$ D$(3 + 2n)$-branes in flat space will yield a massless sector 
containing a $\cn=1$ $D = 4 + 2n$ $U(N)$ gauge field theory, 
plus some extra matter transforming in the adjoint of this group. 
This extra matter is associated to the directions transverse to the D-branes, 
and signals the possibility of translating the D-brane in these directions. 
Now, since we are placing our D-branes at an orbifold point, these degrees of freedom are removed by the orbifold projection, and we are left with a much simpler massless spectrum, which still contains a $D$-dimensional gauge theory. 
We can now turn on non-trivial 
magnetic fluxes $F$ on the $T^{2n}$ which is wrapped by
the $D(3+2n)$-branes. The final chiral fields will be given by
the eigenstates of the Dirac equation on $T^{2n}$ which
survive the $\inte_M$ orbifold projection.
Notice that the chiral fields live on the worldvolume of the D-brane and are hence trapped in $4+2n$ dimensions. Thus the overall wavefunction of 
the massless  modes will have a non-trivial profile on $T^{2n}$ 
but will be a delta function in the rest of the dimensions.
We then can  apply the general formula 
(\ref{yukfinal}) in order to obtain the Yukawa couplings between 
chiral fermions and scalars. 

%
\EPSFIGURE{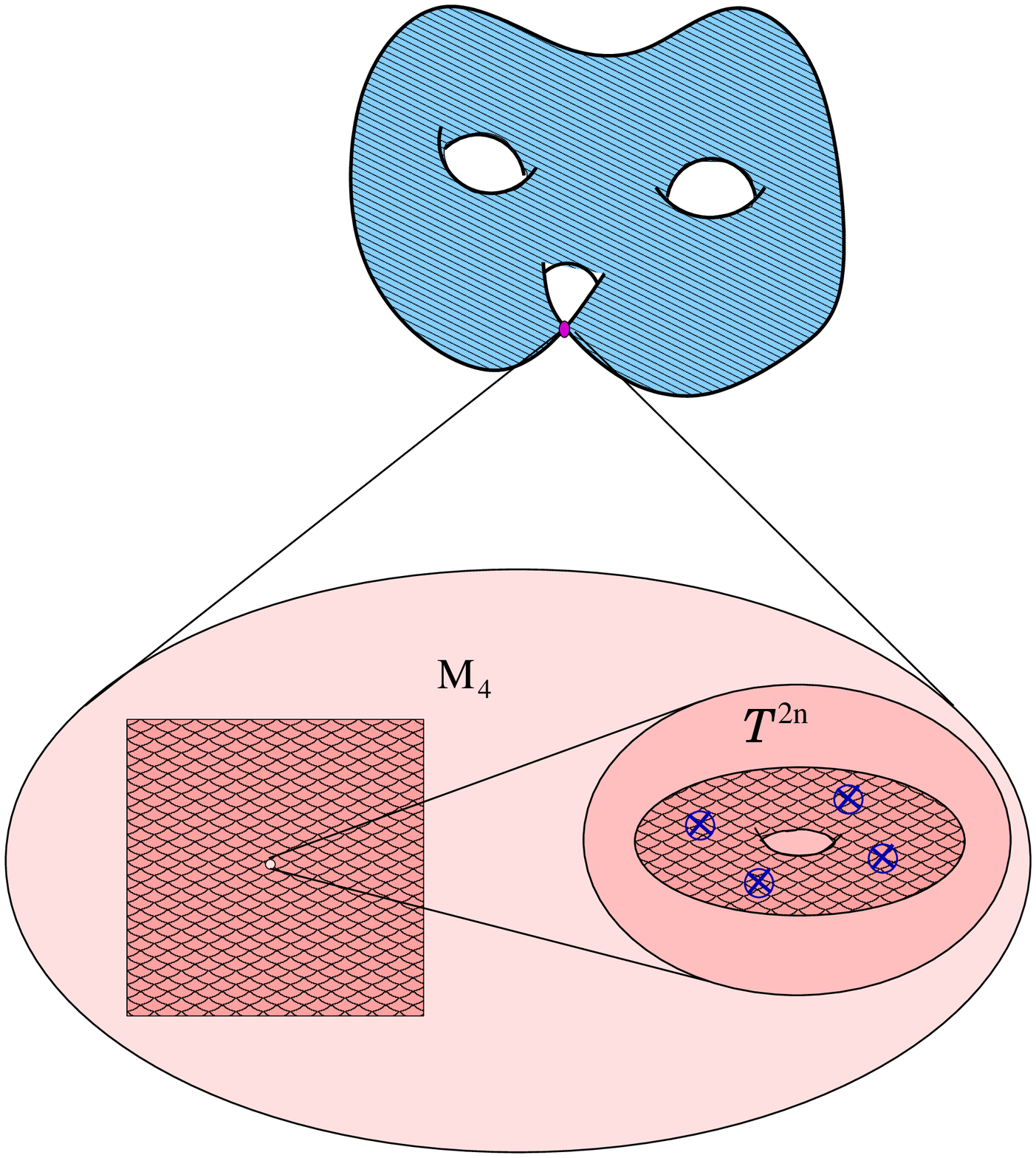, width=5in}
{\label{flux2}
General magnetized extra dimension scenario, 
involving compactifications with orbifold singularities. 
The gauge theory (D-brane) is just localized at the orbifold singularity.}
%

Actually, this kind of construction is related by T-duality 
to certain intersecting $D4$ and $D5$ brane models  
 proposed in \cite{afiru,afiru2}, 
where the orbifold singularity was essential in order to get 
a $D=4$ chiral spectrum. In fact, any of the semi-realistic models 
constructed in this intersecting D-brane picture can be translated 
to the magnetized extra dimension language, where the field theory 
techniques can be applied in order to compute quantities as, e.g., 
three-point functions.

Let us illustrate these facts with a simple example. Namely, 
let us consider a geometry which is locally of the form 
$T^2 \times \cpx/\inte_N$, and place $N_a$ D5-branes wrapping 
$T^2$ and expanding the four non-compact Minkowski dimensions. 
The action of the orbifold group on the open string spectrum is 
specified by a geometrical twist and the Chan-Paton action
\beq
\gamma_{\om,a} = {\rm diag} \left( {\bf 1}_{N_a^0}, 
e^{2 \pi i \frac{1}{N}}{\bf 1}_{N_a^1}, \ldots, 
e^{2 \pi i \frac{N-1}{N}}{\bf 1}_{N_a^{N-1}} \right),
\label{Chan}
\eeq
where we have to impose $\sum_{i=0}^{N-1} N_a^i = N_a$. 
If we choose a supersymmetric orbifold twist then 
we obtain a $D=6$ $\cn=1$ supersymmetric spectrum, given by \cite{afiru}

{\beq
\begin{array}{rl}
\vspace{0.1cm}
{\rm\bf Vector \ Multiplet} &\quad \prod_{i=1}^N U(N_a^i) \\
{\rm\bf Hypermultiplet} & \quad \sum_{i=1}^N (N_a^i,{\ov N}_a^{i+1})
\label{multiplets}
\end{array}
\eeq}

The addition of magnetic fluxes on $T^2$ will further break the 
gauge group to a product of unitary gauge groups.
 The techniques described in this paper 
can then be applied to this field theory. In particular,
the Yukawa couplings are obtained by considering
correlators of light fields respecting the $\inte_N$
projection and performing the overlap integral over 
the 2-torus. The final answer will then given by an expression
like (\ref{yukseminal}) for $n=1$.

Analogous $D=8$ flux models may be obtained starting from Type IIB 
$D7$-branes wrapping a 4-torus and located at a $\inte_N$ singularity
in the remaining two transverse dimensions. After addition of
magnetic flux on the $T^4$ the gauge group is broken and
chiral fields in bifundamental representations will appear.
The massless modes will have non-trivial profiles on $T^4$
and again the Yukawa coupling will be obtained by 
computing an overlap integral over $T^4$ and imposing
$\inte_N$ invariance. The final answer will again given by an expression
like (\ref{yukseminal}) but  for $n=2$.

Again, note that these two classes of models are T-dual to the 
$D4$ and $D5$ intersecting brane models of ref.\cite{afiru},
and hence our formulae below provide us with the Yukawa
couplings for these other classes of models.

\section{Toroidal wavefunctions I: Abelian Wilson lines}

In order to compute the Yukawa couplings from the 
formula (\ref{yukfinal}) we first need to have an 
explicit expression for the internal wavefunctions 
corresponding to massless modes in the low-dimensional theory. 
In the following we will compute such wavefunctions 
for toroidal magnetized compactifications. We will be 
particularly interested in how the wavefunctions depend 
on the Wilson lines involved in the 
compactification\footnote{This Wilson line 
dependence has been usually neglected in the 
previous literature. It is, however, of central 
interest when considering Yukawa couplings.}, 
since eventually the Yukawa couplings will have a rich dependence on those.

We have decided to divide the computation of wavefunctions 
in two different sections. In the present one, we first consider 
the simplest possible case of wavefunction. That is a 
particle in a $T^2$, charged under an Abelian gauge field, 
and in the presence of a magnetic flux. This example already 
illustrates the general form of the wavefunctions, 
which can be elegantly expressed in terms of 
Jacobi $\vt$-functions. We then proceed to consider 
a more interesting  example, which describes the wavefunctions 
in a magnetized compactification with gauge group 
breaking $U(N) \raw U(1)^N$.
In this case chiral fermions charged under any couple of $U(1)$'s 
appear. The results equally apply to a model with 
fluxes breaking $U(N)\raw \prod U(N_a)$ with $\sum N_a = N$,
in which case chiral fermions transforming as bifundamental
representations appear. This is of course the phenomenologically
interesting case in which  e.g., a  gauge group
$U(3)\times U(2)\times U(1)$ may be obtained, with chiral fermions 
transforming as bifundamentals.

To be complete in the next section we will consider 
the more general case where 
non-Abelian Wilson lines are present. In this case
the rank of the initial group will be reduced. 
Specifically, the final gauge group is of the 
general form $\prod_\a U(N_\a)$ and $\sum_\a N_\a \leq N$. 
This more general case is also of great interest when considering a initial gauge group whose rank is much larger than six. It involves, however, 
a  more technical computation and, for the specific purpose 
of computing Yukawa couplings enough insight can be gained 
by considering the `Abelian' case. The reader not interested 
in the technicalities of the general case 
may then safely skip Section 4.

\subsection{Eigenfunctions of the Dirac equation on $T^2$}

\subsubsection{Abelian gauge field}

Let us consider a flat two-dimensional torus $T^2 \simeq \cpx/\Lam$, where $\Lam$ is a two dimensional lattice generated by $e_1 = 2\pi R$ and $e_2 = 2\pi R \tau$, $\tau \in \cpx$.  The dual one-forms are defined as $dx^i(e_j) = \d^{i}_j$, $i, j = 1, 2$. The metric is then given by
\beq
ds^2 = g_{ij} dx^i dx^j = 2 h_{\mu\bar\nu} dz^{\mu} d\bar{z}^{\bar{\nu}}
\label{metric}
\eeq
\beq
\begin{array}{c}
g = (2\pi R)^2 \left(
\begin{array}{cc}
1 & \preal \tau \\ \preal \tau & |\tau|^2
\end{array}
\right)
\quad \quad
h = (2\pi R)^2 \left(
\begin{array}{cc}
0 & \oh \\ \oh & 0
\end{array}
\right)
\end{array}
\label{metric2}
\eeq
where $d z = dx_1 + \tau dx_2$. $T^2$ being a Riemann surface, any magnetic flux $F$ solving Yang-Mills equations must be constant. In particular, let us consider an Abelian magnetic flux such that $\int_{T^2} F = b$, that is
\beq
F = {b \over \pim \tau} \frac{i}{2} \left( dz \wedge d\bar{z} \right) = {b \over \sqrt{|g|}}\ \om,
\label{flux}
\eeq
where $\om$ is the K\"ahler form derived from (\ref{metric2}). Notice that, when expressed in terms of complex coordinates $(z, \bar{z})$, no dependence of the area $\ca = 4\pi^2 R^2 \pim \tau$ appears in the definition of $F$. This flux can be derived from the vector potential
\beq
A(z) = {b \over 2\pim \tau} \pim (\bar{z} dz),
\label{potential}
\eeq
whose transformations under lattice translations are
\vspace{.2cm}
\beq
\begin{array}{ccc} \vspace{.2cm}
A(z+1) = A(z) + {b \over 2\pim \tau} \pim dz & \Raw & \chi_1 = {b \over 2\pim \tau} \pim z \\
A(z+\tau) = A(z) + {b \over 2\pim \tau} \pim \bar{\tau}dz & \Raw & \chi_2 = {b \over 2\pim \tau} \pim \bar{\tau}z
\end{array}
\label{gauge}
\eeq
where we have deduced the corresponding gauge transformations $A \raw A + d\chi_i$. We now consider a complex field $\phi(z)$ with charge $q$ under the $U(1)$ gauge field given by $A$. Its transformation under torus translations are given by $\phi(z) \mapsto { exp\ } (iq \chi_i) \phi(z)$, that is
\beq
\begin{array}{c} \vspace{.2cm}
\phi(z+1) = e^{i q \chi_1(z)}\phi(z) =  exp\left\{i {q b \over 2\pim \tau} \pim z\right\} \phi(z) \\
\phi(z+\tau) =  e^{i q \chi_2(z)}\phi(z) =  exp\left\{i {q b \over 2\pim \tau} \pim \bar{\tau}z\right\} \phi(z)
\end{array}
\label{field}
\eeq
Consistency of such transformations under a contractible loop in $T^2$ implies Dirac's charge quantization
\beq
\frac{b}{2\pi} = M \in \inte.
\label{dirac}
\eeq

We can now implement Wilson lines in this language. The simplest way is introducing a constant complex number $\z = \z_1 + \tau \z_2$ such that
\beq
\begin{array}{c} \vspace{.2cm}
\chi_1 =  {\pi M \over \pim \tau} \pim (z + \z) \\
\chi_2 =  {\pi M \over \pim \tau} \pim \bar{\tau}(z + \z)
\end{array}
\label{wilson}
\eeq
Notice that we can identify these gauge transformations as Wilson loops
\beq
\begin{array}{cc} \vspace{.2cm}
\chi_1 = \oint_{\g_1} A(z), &  \g_1(s) = s \\
\chi_2 =  \oint_{\g_2} A(z), &  \g_2(s) = s \tau
\end{array}
\quad s \in [0,1]
\label{wilson2}
\eeq
which implies that now we have a vector potential of the form
\beq
A(z) = {\pi M \over \pim \tau} \pim \left((\bar{z} +\bar{\z})  dz\right).
\label{potential2}
\eeq

Any complex field $\phi$ with charge $q$ under such $U(1)$, no matter in which Lorentz representation, will be described by a wavefunction that transforms in terms of $\chi_1, \chi_2$ under lattice translations. We can express such wavefunction as
\beq
\phi(z) = e^{i \frac{q M \pi }{2\pim \tau} \pim [(z+\z)^2]} \cdot \th(z,\bar{z})
\label{phi}
\eeq
where the function $\th$ must have the transformation properties
\beq
\begin{array}{rcl}\vspace{.2cm}
\th(z+1) & = & \th(z) \\
\th(z+\tau) & = & e^{- \pi i q M \preal \tau} e^{- 2\pi i q M \preal (z+\z)}\ \th(z) 
\end{array}
\label{th}
\eeq

Given the transformation properties of the function $\th$, we can expand it in a Fourier series along one of the real coordinates of $z = x + \tau y$. More explicitly, we can write
\beq
\th(x,y) = \sum_{n \in \inte} c_n (y) e^{2\pi i n x}
\label{fourier}
\eeq
the second boundary condition in (\ref{th}) being translated into
\beq
c_n (y+1) = c_{n+qM} (y)\ e^{-\pi i q M \left[\preal \tau (1 + 2 y) + \preal \z \right]}
\label{recursion}
\eeq

\subsubsection{Dirac zero modes}

In order to compute the Dirac operator we need to specify a set of gamma matrices satisfying the Clifford algebra.
\beq
\{\G^a,\G^b\} = 2\d^{ab}
\label{gamma}
\eeq
for $T^2$ we can choose the hermitian matrices
\beq
\G^a = \left(
\begin{array}{cc}
0 & 1 \\ 1 & 0
\end{array}
\right), \quad \quad 
\G^b = \left(
\begin{array}{cc}
0 & -i \\ i & 0
\end{array}
\right), 
\label{gamma1}
\eeq
In order to translate this algebra to the holomorphic coordinate frame, we proceed in two steps. We first consider the vielbein $e$
\beq
g_{ij} = e^a_i e^b_j \d_{ab}, \quad {\rm i.e., } \quad g = e^{\rm t} \cdot {\bf 1}\cdot  e.
\label{vielbein}
\eeq
which in our case is
\beq
e = (2\pi R)
\left(
\begin{array}{cc}
1 & \preal \tau \\
0 & \pim \tau
\end{array}
\right)
\label{vielbein2}
\eeq
and which allows to express the Clifford algebra in the more geometrical coordinate frame by defining $\G^{i} = e_a^i \G^a$, $e_a^i$ being the inverse of the vielbein (\ref{vielbein}). These new matrices satisfy $\{\G^i,\G^j\} = 2g^{ij}$, $g^{ij}$ being the inverse of $g$ in (\ref{vielbein}). In order to express this algebra in holomorphic coordinates we perform a further transformation $f = (f^\mu_i, f^{\bar\mu}_i)$ such that 
\beq
h_{\mu\bar\nu} = f_\mu^i f_{\bar\nu}^j g_{ij} =  f_\mu^i e^a_i f_{\bar\nu}^j e^b_j \d_{ab}, \quad {\rm i.e., } \quad h = (e\cdot f)^{\rm t} (e \cdot  f).
\label{vielbein3}
\eeq
In our case
\beq
f^{-1} = 
\left(
\begin{array}{cc}
1 & \tau \\
1 & \bar\tau
\end{array}
\right)
\label{vielbein4}
\eeq
so that we finally obtain
\beq
\G^z = (2\pi R)^{-1}
\left(
\begin{array}{cc}
0 & 2 \\ 0 & 0
\end{array}
\right), \quad \quad 
\G^{\bar z} = (2\pi R)^{-1}
\left(
\begin{array}{cc}
0 & 0 \\ 2 & 0
\end{array}
\right).
\label{gamma3}
\eeq
We also need to specify the covariant derivative
\beqa\nonumber
\nabla_i & = & \p_i + \om_i + A_i \\
& = & \p_i + \frac{1}{2} \om_i^{kl}\Sig_{kl} + A_i^\a T_\a
\label{nabla}
\eeqa
where $\Sig^{kl} = \frac{1}{4} [\G^k,\G^l]$ are the anti-hermitian 
generators of Lorentz transformations and $T_\a$ those of gauge 
transformations. Since we are in flat space we will take the spin 
connection $\om_i^{kl}$ to vanish. In the case of an Abelian gauge field on 
$T^2$ with $T_\a =-i$ we hence have $\nabla_i = \p_i - i q A_i$. We are 
finally able to compute the Dirac operator %
\beq
i \Dbar = \bar{\Dbar}\ + \bar{\Dbar}^\dag  = i \sum_{\bar{a}} \G^{\bar{a}} \nabla_{\bar{a}} + i  \sum_{{a}} \G^{{a}} \nabla_{{a}} 
\label{diracgen}
\eeq
which in our case is given by
\beq
i\Dbar_2 = 
i \left(
\begin{array}{cc}
0 & -D^\dag \\  D & 0
\end{array}
\right) =
{i \over \pi R} 
\left(
\begin{array}{cc}
0 & \p - q {\pi M \over 2 \pim {\tau}} (\bar{z} + \bar{\z}) \\ 
\bar{\p} + q {\pi M \over 2\pim \tau} (z + \z) & 0
\end{array}
\right)
\label{diracop}
\eeq

Let us now consider a two-dimensional spinor in $T^2$ 
\beq
\Psi(z,\bar{z}) =
\left(
\begin{array}{c}
\psi_+ \\  \psi_-
\end{array}
\right),
\label{spinor}
\eeq
transforming under $U(1)$ gauge field (\ref{potential2}) with charge $q$. Hence both wavefunctions $\psi_{\pm}$ can be written in the form (\ref{phi}), (\ref{fourier}). Such spinor will contain a zero-mode of of the Dirac operator if it is annihilated by $\Dbar$, which implies
\beqa
D \psi_+ = 0 & \Raw & \left(\bar{\p} + {\pi i q M \over \pim \tau} \pim \left(z+\z\right)\right) \cdot \th(z,\bar{z}) \label{zeromode1} \\
D^\dag \psi_-  = 0 & \Raw & \left({\p} - {\pi i q M \over \pim \tau} \pim \left(z+\z\right)\right) \cdot \th(z,\bar{z}) \label{zeromode2}
\eeqa
Let us first consider eq.(\ref{zeromode1}). By using the Fourier decomposition (\ref{fourier}) the zero-mode condition can be translated into the first order differential equation
\beq
\begin{array}{c} \vspace*{.1cm}
{c_n'(y) \over c_n (y)} = - 2\pi q M (\pim \tau y + \pim \z) + 2\pi n\tau \\
\ \Raw \ c_n(y) = k_n\ e^{-\pi q M {(\pim (z + \z))^2 \over \pim \tau}} e^{2 \pi i n \tau y},
\end{array}
\label{ode}
\eeq
$k_n$ being a constant. Inserting this solution into (\ref{phi}), (\ref{fourier}) we find that our wavefunction is given by
\beq
\psi_+ = e^{i \pi q M (z + \z) {\pim (z + \z) \over \pim \tau}} \cdot \sum_{n \in \inte} k_n \ e^{2 \pi i n z}
\label{semifinal}
\eeq
thus being a holomorphic function up to a global function, as was to be expected from the general discussion in Appendix \ref{SUSYap}. Now, we can find the constant coefficients $k_n$ by imposing the recurrence relation (\ref{recursion}) into (\ref{ode}), obtaining
\beq
k_n = \cn_n\ e^{\pi i {n^2 \tau\over qM}} e^{2\pi i n \z}
\label{coef}
\eeq
where $\cn_n = \cn_{n + qM}$ are $|M|$ arbitrary constant coefficients. 

The existence of $|M|$ independent coefficients signals the fact that the Dirac equation has $|M|$ independent solutions, each of them having a different wavefunction $\psi_+^j$ $j = 0, \dots, |M|-1$. Indeed, we can obtain such wavefunctions by splitting the summation index in (\ref{semifinal}) as $n = r\cdot M + j$, and taking the overall factor $\cn_j$ out of the sum (\ref{semifinal}). We then learn that we can express the final solution in a rather elegant way. Namely, as
\beq
\psi_+^j(z) = \cn_j \cdot e^{i \pi q M (z + \z) {\pim (z + \z) \over \pim \tau}} \cdot
\vt
\left[
\begin{array}{c}
\frac{j}{qM} \\ 0
\end{array}
\right]
\left(q M (z + \z), q M \tau \right)
\label{psisoln}
\eeq
where $\vt$ is given by the Jacobi theta-function
\beq
\vt \left[
\begin{array}{c}
a \\ b
\end{array}
\right] (\nu,\tau) =  \sum_{l \in \inte} 
e^{\pi i (a + l)^2 \tau} \ e^{2\pi i (a + l)(\nu + b)}
\label{theta}
\eeq

The fact that the final set of solutions (\ref{psisoln}) satisfies the transformation properties defined by (\ref{wilson}) can be easily checked by using the modular transformation properties of $\vt$-functions:
\beqa
\vt
\left[
\begin{array}{c}
a \\ b
\end{array}
\right]
(\nu+n,\tau)
& = &
e^{2\pi i n a} \cdot 
\vt
\left[
\begin{array}{c}
a \\ b
\end{array}
\right] 
(\nu,\tau) 
\label{trans1}
\\
\vt   
\left[
\begin{array}{c}
a \\ b
\end{array}
\right]
(\nu+ n \tau,\tau)
& = &
e^{-\pi i n^2 \tau - 2\pi i n (\nu + b)} \cdot
\vt
\left[
\begin{array}{c}
a \\ b
\end{array}
\right] 
(\nu,\tau)
\label{trans2}
\eeqa

A parallel discussion can be done for (\ref{zeromode2}). In order to summarize our results, let us define the function 
\beq
\psi^{j, N} (\tau, \nu) = \cn_j\ e^{i\pi N \nu \pim \nu/\pim \tau} 
\cdot 
\vt
\left[
\begin{array}{c}
\frac{j}{N} \\ 0
\end{array}
\right]
(N \nu, N \tau)
\label{totalsoln}
\eeq
Where the constants $\cn_j$ will be soon specified. In terms of (\ref{totalsoln}) we can define the wavefunctions
\beqa
\psi^{j}_+ \equiv \psi^{j,qM}(\tau, z + \z) & & \quad  (\psi^{j}_+)^* = \psi^{-j, -qM}(\bar{\tau}, \bar{z} + \bar{\z}) \\
\psi^j_- \equiv \psi^{j,qM}(\bar{\tau}, \bar{z} + \bar{\z}) & & \quad (\psi^j_-)^* = \psi^{-j, -qM}(\tau, z + \z)
\label{four}
\eeqa
where the star denotes complex conjugation. Is easy to see that
\beqa
D \psi^j_+ = 0 \quad (q = +1) & & D^\dagger (\psi^j_+)^* = 0 \quad (q = -1) 
\label{chiralleft}\\
D^\dagger \psi^j_- = 0 \quad (q = +1) & & D (\psi^j_-)^* = 0 \quad (q = -1)
\label{chiralright}
\eeqa
We can interpret $\psi^j_+$ as the wavefunctions  corresponding to left-handed fermions in 4D, while $(\psi^j_+)^*$ represent their anti-particles. Right-handed fermions would then be given by $\psi^j_-$ and their anti-particles by $(\psi^j_-)^*$. 

Notice the important fact that the solutions (\ref{chiralleft}) and (\ref{chiralright}) are mutually exclusive, in the sense that the theta function defining $\psi_+^j$ will only converge if $M > 0$, whereas $\psi_-^j$ is only well-defined when $M < 0$. Hence, by introducing a non-trivial flux $M \neq 0$ we automatically select one of the two chiralities of the two-dimensional spinor (\ref{spinor}). Moreover, we obtain several replicas of such chiral fermions, by means of the $|M|$ independent solutions of the Dirac equation.

\subsubsection{Normalization}

Once that we have found a basis of linearly independent wavefunctions, we proceed to express everything in terms of a orthonormal basis. This will allow us to have canonically normalized kinetics terms in four-dimensional reduced action. In terms of the internal wavefunctions we just found, this amounts to impose the following normalization condition
\beq
\int_{T^2} dz d\bar{z} \psi_\pm^j (\psi_\pm^k)^* = \d_{jk},
\label{norm}
\eeq
which is nothing but condition (\ref{normKK10}) for the particular two-dimensional case at hand. For the sake of concreteness, let us impose the normalization condition for $\psi_+$. In terms of the wavefunctions (\ref{totalsoln}) we have
\beqa
\psi_+^j (\psi_+^k)^* 
 & = & \psi^{j,qM} (\tau,z+\z) \cdot \psi^{-k,-qM} (\bar{\tau},\bar{z}+\bar{\z}) \\ \nonumber
& \sim & \psi^{j,qM} (\tau,z) \cdot \psi^{-k,-qM} (\bar{\tau},\bar{z}) \\ \nonumber
& = & 
\cn_j \cn_k \cdot
e^{-2\pi qM \cdot (\pim z)^2/ \pim \tau} \cdot
\vt
\left[
\begin{array}{c}
\frac{j}{qM} \\0
\end{array}
\right]
(z qM, \tau qM) \cdot
\vt
\left[
\begin{array}{c}
\frac{-k}{qM} \\ 0
\end{array}
\right]
(- \bar{z} qM, - \bar{\tau} qM) \\ \nonumber
& = & 
\cn_j \cn_k \cdot
e^{-2\pi \cdot (\pim z')^2/ \pim \tau^\prime} \cdot
\vt
\left[
\begin{array}{c}
\frac{j}{qM} \\0
\end{array}
\right]
(z', \tau') \cdot
\left( \vt
\left[
\begin{array}{c}
\frac{k}{qM} \\ 0
\end{array}
\right]
(z', \tau') \right)^*,
\label{norm2}
\eeqa
where in the second line we have get rid of the constant $\z$, which will play no role when integrating over $z$. The last line is just the usual scalar product of theta functions, seen as holomorphic sections of a line bundle over $T^2$ \cite{gh}. Indeed, integration over $\preal z$ imposes the condition $j=k$, and equality on the summation indices of the theta functions
\beqa \nonumber
\int_0^1 d\left(\preal z\right) & \raw &
e^{-2\pi M \cdot [\pim z]^2/ \pim \tau} \cdot
\sum_n e^{2\pi qM \pim \tau \left(n + \frac{j}{qM}\right)^2}
e^{4\pi qM \left(n + \frac{j}{qM}\right) \pim z} \\
& = & \sum_n e^{-2\pi qM \pim \tau \left(n + \frac{j}{qM} + \frac{\pim z}{\pim \tau} \right)^2}
\label{intrez}
\eeqa
It is now easy to integrate over $\pim z/\pim \tau$, since
\beqa \nonumber
\int_0^1 d\left(\frac{\pim z}{\pim \tau}\right) \sum_n e^{-2\pi qM \pim \tau \left(n + \frac{j}{qM} + \frac{\pim z}{\pim \tau} \right)^2} & = &
\sum_n \int_0^1 d\left(\frac{\pim z}{\pim \tau}\right) e^{-2\pi qM \pim \tau \left(n + \frac{j}{qM} + \frac{\pim z}{\pim \tau} \right)^2} \\ & = &
\int_{-\infty}^\infty dx\ e^{-2\pi qM \pim \tau x^2}
\label{intimz}
\eeqa
We thus find that, in order to satisfy (\ref{norm}), that the wavefunctions (\ref{psisoln}) must be multiplied by the normalization factor
\beq
\cn_j = \left(\frac{2 \pim \tau |M|}{\ca^2} \right)^{1/4},\quad \forall j
\label{norm3}
\eeq
where we have used the fact that, for $\psi^j_+$, $qM = |M|$. The computation for $\psi_-$ give us the same result.

\subsection{Chiral matter eigenfunctions}

The previous section illustrates the general philosophy we will use in order to compute the internal fermionic wavefunctions in our magnetized compactifications. In the application to Magnetized Extra Dimensions, however, we need to consider a more general setup. Indeed, in general our effective field theory will be described by a $U(N)$ higher dimensional theory broken to $\prod_i U(N_i)$, when turning on non-trivial magnetic fluxes. The chiral spectrum will arise from fermions transforming in bifundamental representations $(N_a, \bar{N}_b)$. Let us first consider that we have a two-torus with a magnetic flux of the form
\beq
F_{z\bar{z}} = {\pi i\over \pim \tau}
\left(
\begin{array}{cccc}
{m_a}  \\
& {m_b} \\
& & {m_c} \\
& & & \ddots 
\end{array}
\right),
\label{fluxabif}
\eeq
where the $m_\a$ $\a = a,b,c,\dots$ are $N$ different numbers. Again, by Dirac's quantization condition the magnetic quanta $m_\a$ are given by integers, and it can be seen that turning on this kind of flux implies the gauge group breaking $U(N) \raw U(1)^N$.\footnote{See Section 4 for a more general, systematic discussion of these facts.} Let us then see how the previous computation of wavefunctions generalizes to this case.

\subsubsection{Fermions in bifundamentals}

In order to compute the wavefunctions involved in a $T^2$ compactification with the flux (\ref{fluxabif}), it is enough to consider the simpler case 
\beq
F_{z\bar{z}} = {\pi i\over \pim \tau}
\left(
\begin{array}{cc}
{m_a} & \\
& {m_b} 
\end{array}
\right).
\label{fluxabif2}
\eeq
That is, a gauge group breaking $U(2) \raw U(1)_a \times U(1)_b$. From the point of view of D-brane physics, this can be seen as two D-branes $a$ and $b$, each with a magnetic flux in its internal worldvolume proportional to $m_a$ and $m_b$, respectively.

\vspace{.2cm}

{\noindent \bf Dirac equation}

\vspace{.1cm}

In the case at hand the gauge connection that can be chosen to be
\beqa
A_{\bar{z}} = A_{\bar{z}}^\a T_\a & = & {\pi \over 2 \pim \tau}
\left(
\begin{array}{cc}
{m_a} (z + \z_a) & \\
& {m_b} (z + \z_b) 
\end{array}
\right),\\
A_{z} = A_z^\a T_\a & = & - {\pi \over 2 \pim \tau}
\left(
\begin{array}{cc}
{m_a} (\bar{z} + \bar{\z}_a) & \\
& {m_b} (\bar{z} + \bar{\z}_b) 
\end{array}
\right),
\label{potentialabif}
\eeqa
where $T_a = {\small \left( \begin{array}{cc} -i \\ & 0 \end{array} \right)}$ and $T_b = {\small \left( \begin{array}{cc} 0 \\ & -i \end{array} \right)}$ are the anti-hermitian generators of $U(1)_a \times U(1)_b \subset U(2)$. The Dirac operator is again given by
\beq
i\Dbar = 
i \left(
\begin{array}{cc}
0 & -D^\dag \\  D & 0
\end{array}
\right) =
{i \over \pi R} \left(
\begin{array}{cc}
0 & {\p + A_z} \\ 
{\bar{\p} + A_{\bar{z}}} & 0
\end{array}
\right)
\label{diracopbif}
\eeq
The zero modes of (\ref{diracopbif}) can be easily found. Let us consider a two-dimensional spinor in $T^2$ transforming in the adjoint of $U(2)$. 
\beq
\Psi(z,\bar{z}) =
\left(
\begin{array}{c}
\psi_+ \\  \psi_-
\end{array}
\right), 
\quad \quad
\psi_\pm = 
\left(
\begin{array}{cc}
A_\pm & B_\pm \\
C_\pm & D_\pm
\end{array}
\right)
\label{spinora}
\eeq
Again, $\psi_+$ will contain any zero-mode of the Dirac operator if it is annihilated by $D$, which implies
\beqa
D \psi_+ & = & (\pi R)^{-1} \left( \bar{\p} \psi_+ + [A_{\bar{z}},\psi_+] \right) \label{zeromode+a} \\ \nonumber
& = & (\pi R)^{-1} 
\left(
\begin{array}{cc}
\bar{\p} A_+ & \left(\bar{\p} + {\pi {I}_{ab} \over 2 \pim \tau} (z + \z_{ab}) \right) B_+\\
\left(\bar{\p} - {\pi {I}_{ab} \over 2 \pim \tau} (z + \z_{ab}) \right) C_+ & \bar{\p} D_+
\end{array}
\right) = 0
\eeqa
where we have defined
\beqa
I_{ab} & \equiv & m_a - m_b \neq 0 \label{defa1}\\
\z_{ab} & \equiv & ( m_a \z_a -  m_b \z_b)/I_{ab}
\label{defa2}
\eeqa
On the other hand, the condition for $\psi_-$ to contain zero modes is
\beqa
D^\dag \psi_- & = & (\pi R)^{-1} \left({\p} \psi_- + [A_{{z}},\psi_-] \right) \label{zeromode-a} \\ \nonumber
& = & (\pi R)^{-1} 
\left(
\begin{array}{cc}
{\p} A_- & \left({\p} - {\pi {I}_{ab} \over 2 \pim \tau} (\bar{z} + \bar{\z}_{ab}) \right) B_-\\
\left({\p} + {\pi {I}_{ab} \over 2 \pim \tau} (\bar{z} + \bar{\z}_{ab}) \right) C_- & {\p} D_-
\end{array}
\right) = 0
\eeqa
which is consistent with the fact that we must consider $\psi_- = \psi_+^\dag$, as $\Psi$ comes from a higher-dimensional gaugino. Notice that from (\ref{zeromode+a}) we deduce that $A_+, D_+$ have to be holomorphic functions, whereas $B_+, C_+$ have to be of the form
\beq
\cn \cdot e^{\pm i {\pi {I}_{ab} \over \pim \tau} (z + \z_{ab}) \cdot \pim (z + \z_{ab})} \cdot \xi(z),
\label{wave}
\eeq
respectively, where $\xi(z)$ is an arbitrary holomorphic function, and $\cn$ is a $z$-independent normalization factor. 

\vspace{.2cm}

{\noindent \bf Gauge transformations}

\vspace{.1cm}

In order to find the actual Dirac zero mode wavefunctions, however, we still have to impose them to be well defined in a gauge theory over $T^2$, that is, to have the appropriate $U(2)$ transformations under lattice translations. The transformations of an adjoint field $\Xi ={\small \left( \begin{array}{cc} A & B \\ C & D \end{array} \right)}$ are given by
\beq
\Xi(z) \quad \mapsto \quad \Om_i\cdot \Xi(z)\cdot \Om_i^\dag = 
\left(
\begin{array}{cc}
e^{i\chi_i^a(z)} & 0 \\ 0 & e^{i\chi_i^b(z)}
\end{array}
\right) \cdot
\Xi(z) \cdot
\left(
\begin{array}{cc}
e^{-i\chi_i^a(z)} & 0 \\ 0 & e^{-i\chi_i^b(z)}
\end{array}
\right)
\label{gaugeabif}
\eeq
where
\beq
\begin{array}{c} \vspace{.2cm}
\chi_1^\a (z) =  {\pi m_\a \over \pim \tau}\ \pim (z + \z_\a) \\
\chi_2^\a (z) =  {\pi m_\a \over \pim \tau}\ \pim \bar{\tau}(z + \z_\a)
\end{array}
\quad \a = a,b 
\label{transitionabif}
\eeq
We hence find that $A$ and $D$ are invariant under lattice translations, whereas $B$ and $C$ transform as
\beq
\begin{array}{ccc}
\begin{array}{lcr}\vspace*{.2cm}
B(z+1) & = & e^{i\chi_1^{ab}(z)} B(z), \\
C(z+1) & = & e^{-i\chi_1^{ab}(z)} C(z),
\end{array}
&  &
\begin{array}{lcr}\vspace*{.2cm}
B(z+\tau) & = & e^{i\chi_2^{ab}(z)} B(z) \\
C(z+\tau) & = & e^{-i\chi_2^{ab}(z)} C(z)
\end{array}
\end{array}
\label{boundT2abif}
\eeq
where now
\beq
\begin{array}{c} \vspace{.2cm}
\chi_1^{ab} (z) =  {\pi I_{ab} \over \pim \tau}\ \pim (z + \z_{ab}) \\
\chi_2^{ab} (z) =  {\pi I_{ab} \over \pim \tau}\ \pim \bar{\tau}(z + \z_{ab})
\end{array}
\label{transitionabif2}
\eeq

Notice that these boundary conditions are the same found in the previous subsection. Indeed, the transformation properties (\ref{boundT2abif}) can be understood in terms of the functions (\ref{wilson}), by making the substitution $M \mapsto I_{ab}$ and $\z \mapsto \z_{ab}$. Finally, we must take $q=1$ for the wavefunction $B$ and $q=-1$ for $C$. We then find that the wavefunctions of (\ref{spinora}) are given by
\beq
\begin{array}{ccc} 
B_+ = \psi^{j,I_{ab}} (\tau, z + \z_{ab}), &\quad  & C_+ = \psi^{j,-I_{ab}} (\tau, z + \z_{ab})
\end{array}
\label{totalsolnabif}
\eeq
and $B_\pm = (C_\mp)^*$. Notice that both wavefunctions in (\ref{totalsolnabif}) are again exclusive, and that $B_+$ will then vanish unless $I_{ab} > 0$, whereas $C_+$ will only be present for $I_{ab} < 0$. 

On the other hand, we have found that $A_\pm$, $D_\pm$ must be (anti)holomorphic and periodic under both lattice translations, for with the only possible solution is a constant function on $T^2$. These constant wavefunctions are to be identified with the gauginos of the unbroken gauge group $U(1)_a \times U(1)_b$, whereas the off diagonal entries of $\Psi$ are left and right-handed fermions transforming in the bifundamental representation of such gauge group. We then have
\beq
\begin{array}{cc}
B_+ < \infty \iff I_{ab} > 0 &\quad  {\rm\ Left-handed\ fermions\ in\ } (+1,-1)\ {\rm of\ } U(1)_a \times U(1)_b \\
C_+ < \infty \iff I_{ab} < 0 &\quad  {\rm\ Left-handed\ fermions\ in\ } (-1,+1)\ {\rm of\ } U(1)_a \times U(1)_b
\end{array}
\label{chiralitya}
\eeq
and thus we again obtain a chiral spectrum.

\subsection{Eigenfunctions of the Laplace equation}

The wavefunctions (\ref{totalsolnabif}) turn out to be not only solutions the Dirac equation, but also eigenfunctions of the Laplace operator. The eigenvalues of such eigenfunctions, and hence the mass of the corresponding scalars, will in general depend on the K\"ahler moduli of compactification. In order to see this, let us compute the square of the Dirac operator
\beq
\left(i\Dbar\right)^2 = 
\left(
\begin{array}{cc}
D^\dag D & 0 \\  0 & D D^\dag
\end{array}
\right) =
\oh \{ D^\dag, D\} + 
\left(
\begin{array}{cc}
\oh [ D^\dag, D ] & 0 \\ 
0  & - \oh [ D^\dag, D ]
\end{array}
\right) = 
\D + 
\left(
\begin{array}{cc}
{2 i F_{z\bar{z}} \over (2\pi R)^2} & 0 \\ 
0  & {2 i F_{\bar{z}z} \over (2\pi R)^2}
\end{array}
\right) 
\label{diracopsq}
\eeq
where the Laplace operator is given by
\beq
\D =  \sum_{\mu =1}^2 \nabla_\mu \nabla^\mu = \nabla_1^2 + \nabla_2^2,
\label{laplaceop}
\eeq
and $F_{z\bar{z}}$ by (\ref{fluxabif2}). We hence get that the action of the Laplacian on such $ab$ sector wavefunctions is
\beq
\D \psi^{j,\pm I_{ab}} = \pm {2\pi {I}_{ab} \over \ca} \psi^{j,\pm I_{ab}} = {2\pi |{I}_{ab}| \over \ca} \psi^{j,\pm I_{ab}},
\label{laplace}
\eeq
same for $\left(\psi^{j,\pm I_{ab}}\right)^*$.

These Laplace eigenfunctions turn out to be the ones with smallest eigenvalue. Actually, the whole tower of eigenfunctions and eigenvalues can be recovered from them and the harmonic oscillator algebra \cite{vanbaal2,troost}. Indeed, notice that for each $ab$ sector we have
\beqa\nonumber
\D & = & N + { 2\pi |{I}_{ab}| \over \ca},  \\
N & = & D^\dagger D, \\ \nonumber
[N,D^\dagger] & = &  { 4\pi |{I}_{ab}| \over \ca} D^\dagger
\label{defs}
\eeqa

This reminds of the harmonic oscillator quantum algebra, which can be recovered by defining
\beqa\nonumber
a & = & \sqrt{{\ca \over 4 \pi |{I}_{ab}|}} D \\
a^\dagger & = & \sqrt{{\ca \over 4 \pi |{I}_{ab}|}} D^\dagger
\label{osc}
\eeqa
The eigenfunctions and eigenvalues of the Laplacian are then
\beqa
\psi^{j,\pm I_{ab}}_r & = & (D^\dagger)^r \psi^{j,\pm I_{ab}}, \\
\lam_r & = & 2\pi  {|{I}_{ab}| \over \ca} (2r +1) > 0.
\label{eigen}
\eeqa

However, notice that this is not giving us the mass eigenvalues. From (\ref{4Dmass}) we know that the mass matrix is of the form
\beq
M^2 =
\left(
\begin{array}{cc}
\D & - i 4\pi  {{I}_{ab} \over \ca} \\
i  4\pi  {{I}_{ab} \over \ca} & \D
\end{array}
\right)
\label{mass}
\eeq
The eigenvalues of such matrix are
\beq
\tilde{\lam}_r = \lam_r \pm 4 \pi  {{I}_{ab} \over \ca}
\label{eigen2}
\eeq
which will give us one tachyonic scalar and the rest massive. The same tower of eigenstates will be present in the fermionic spectrum, the mass gap between the states being the same, but the lowest state corresponding to a massless fermion instead of a boson. This moduli-dependent spectrum describing a tower of massive fermions and scalars is the T-dual version of the 'gonions' described in \cite{afiru2}, for the case of D-branes intersecting at one angle on $T^2$. 
%
%

\subsection{Theta functions as wavefunctions}

Let us pause our derivation of toroidal wavefunctions 
for a while, and try to gain some intuition from our results 
so far. In the previous subsections we have found that, 
after introducing a constant magnetic flux in a 
pure $U(N)$ super Yang-Mills theory living in a two-torus, 
we obtain a broken gauge group $G \subset U(N)$ and a chiral 
spectrum. The wavefunctions of such spectrum are given by 
either constant wavefunctions, which are to be associated 
with gauge bosons and gauginos of $G$, or  by the non-trivial 
functions $\psi^j(z)$, which will represent both chiral 
fermions and scalars in the dimensionally reduced theory. 
The mass splitting between fermions and bosons will be 
proportional to the flux density \cite{bachas}.

The chiral matter eigenfunctions are given by (\ref{totalsoln}) which, 
up to some normalization and exponential factors, 
are holomorphic Jacobi theta-functions. The exponential 
factor presents (up to phases) a gaussian behaviour on the 
torus coordinate $y = \frac{\pim z}{\pim \tau} \in [0,1]$. 
On the other hand, the theta function dependence on $z$ seems 
much harder to visualize. One can develop some intuition 
by plotting the square modulus of the wavefunction 
$\psi^{j,N} (\tau, z)$, associated to the `probability density' 
$\rho(z) = |\psi(z)|^2$ of finding a quantum particle with such wavefunction.
%
\EPSFIGURE{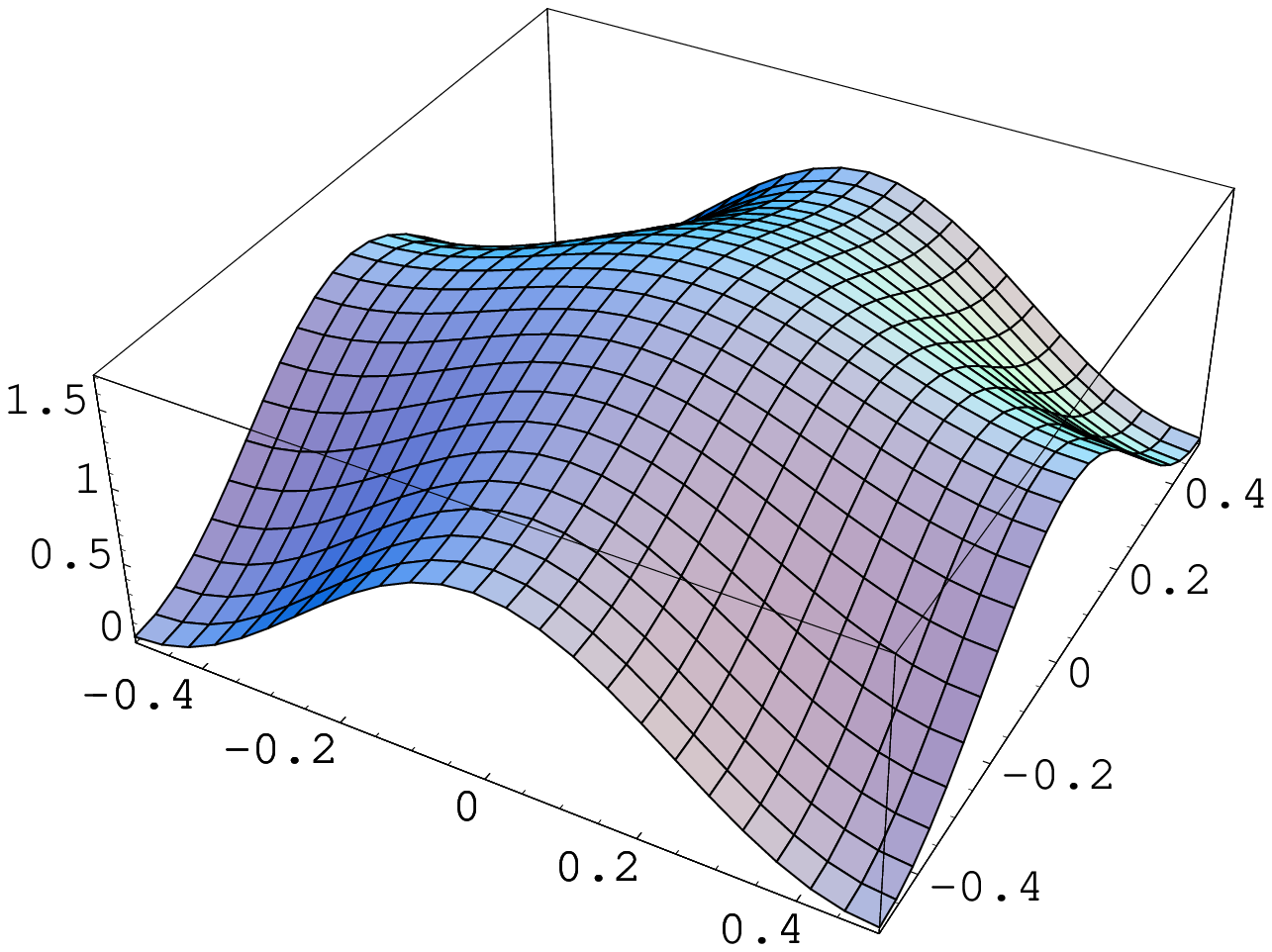, width=4.5in}
{\label{theta1}
Probability density $\rho(z) = |\psi(z)|^2$ for a square $T^2$ with one unit of magnetic flux and vanishing Wilson line. The density $\rho$ presents a symmetric, gaussian-like behaviour in the two axis of the $T^2$ $x$ and $y$, being peaked at $(0,0)$.}
%

We show such density in figure \ref{theta1}, for the particular case of a square torus ($\tau = i$) with one unit of magnetic flux $N = 1$ and no Wilson line $\z = 0$. We then find a square density which presents a gaussian-like behaviour in both axis of the $T^2$ $x$ and $y$ being, as well, a `periodic' function under lattice translations in both directions. The probability density of such function is peaked in $(x,y) = (0,0)$, while it vanishes at $(x,y) = (\oh,\oh)$. Of course, this is only in the particular case of $\z = 0$, and these two points will be conveniently shifted by varying the Wilson line. In general, the maximum and minimum of $\rho$ will be placed at $z = - \z$ and $z = \oh \tau - \z$. Notice that the minima and maxima of a wavefunction density may be crucial in the final computation of a Yukawa coupling, since a 3-point function is given by a overlap of three such functions, and the minima or maxima of such can lead to enhanced or suppressed Yukawa couplings.

\begin{figure}[ht]
\begin{center}
\begin{tabular}{cc}
\\
\epsfig{file=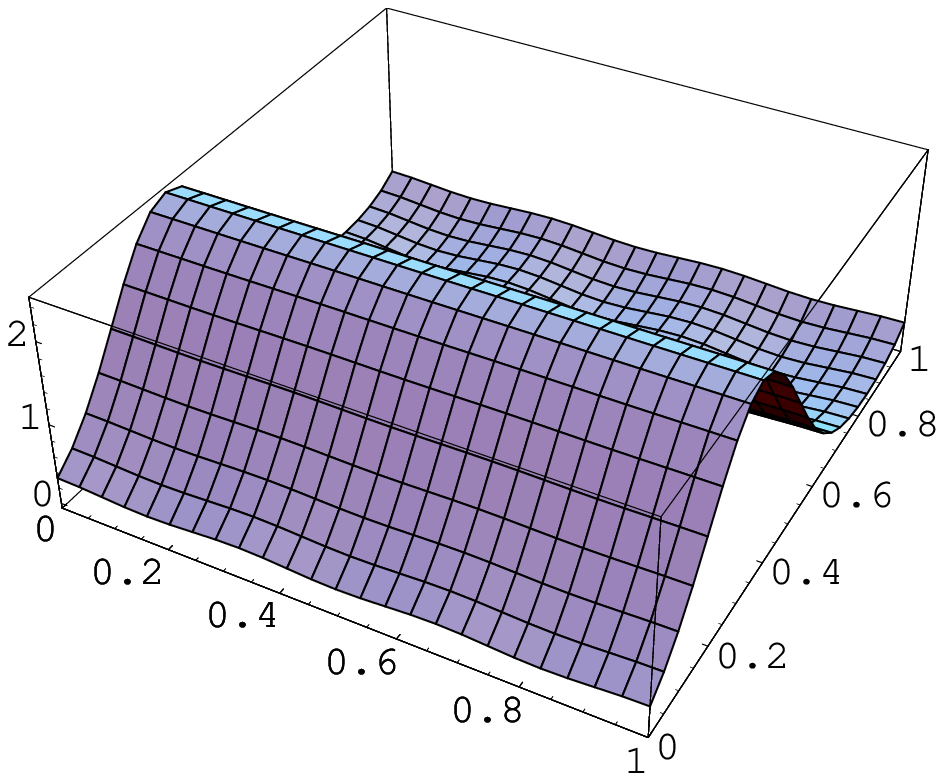, height=6cm} & 
\epsfig{file=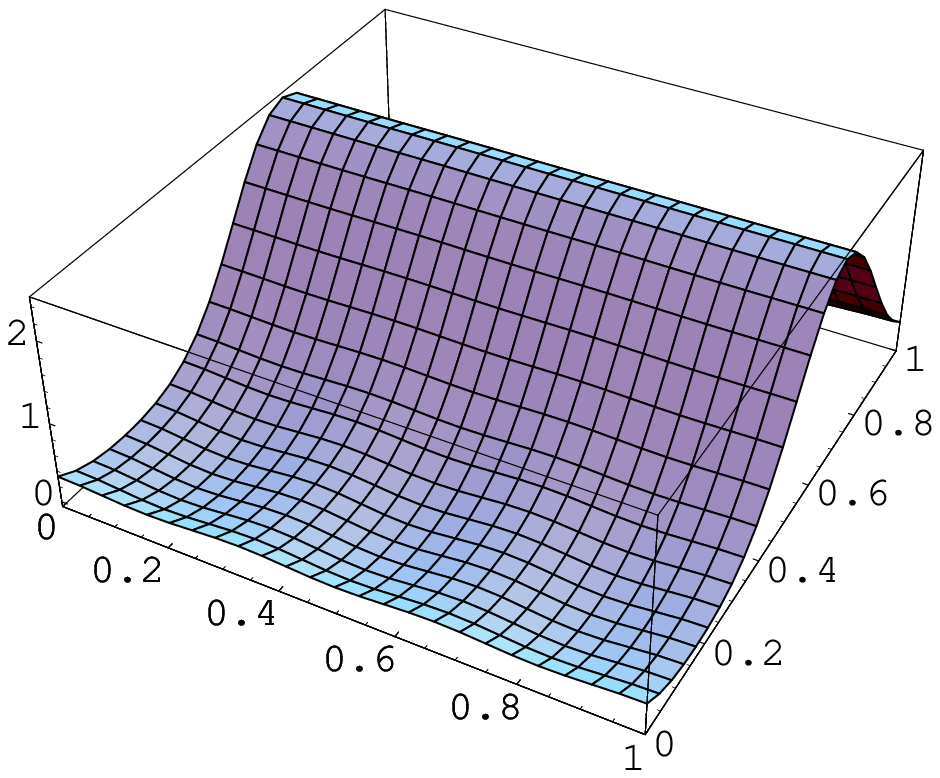, height=6cm} \\
{\Large $j = -1$} & 
{\Large $j = 1$} \\
\end{tabular}
\begin{tabular}{c}
\\
\epsfig{file=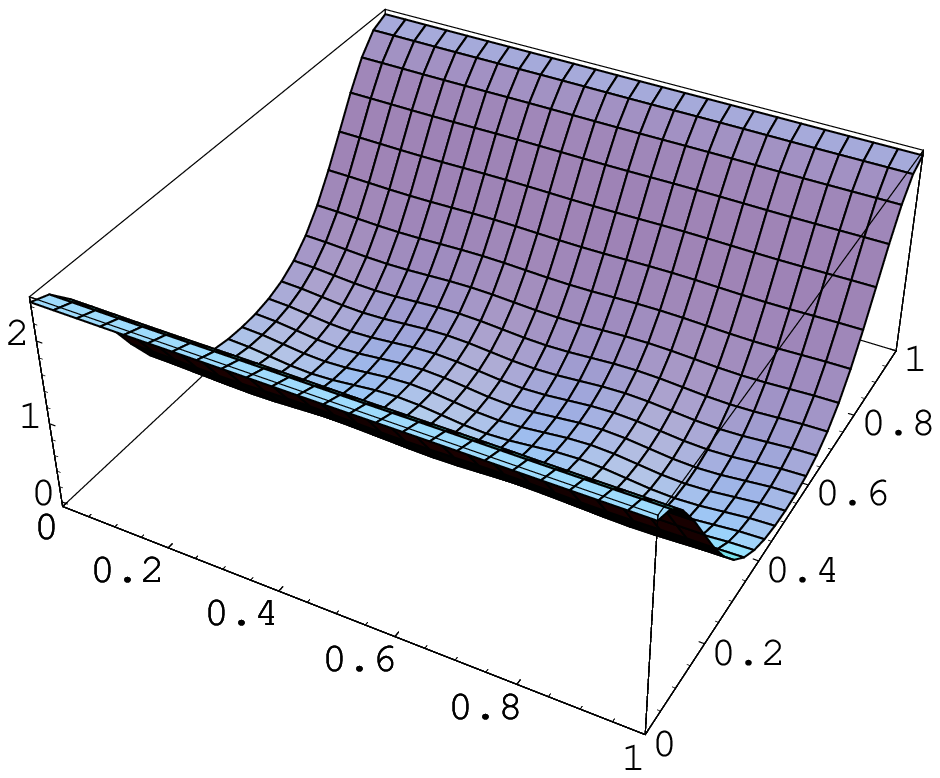, height=6cm} \\
{\Large $j = 0$} \\
\end{tabular}
\end{center}
\caption{\small{Probability densities $\rho^j(z) = |\psi(z)^{j,3}|^2$ for a square $T^2$ with triplication of the chiral spectrum. The gaussian-like behaviour is now present in only one axis of the torus, namely in the coordinate $y = \pim z / \pim \tau$. The gaussians are similar to each other, and centered at the points $y = - j/3$.}}
\label{3thetas}
\end{figure}

Let us now consider a more generic case, namely when 
$N \neq 1$ and the spectrum of wavefunctions is composed of several replicas 
of the same chiral fermion/scalar. Let us choose $N=3$
(corresponding to 3 generations of the given fermion), which 
is moreover a phenomenologically interesting case, 
and plot the probability density $\rho^j = |\psi^{j,3}|$ for $j = -1, 0, 1$.
\footnote{Recall that in general $j$ is an index defined 
mod $|N|$ so, for $N =3$, $j = -1 \sim j = 2$, etc.} 
We show our results in figure \ref{3thetas}.
%
%
What we observe is that the three wavefunctions are similar, but shifted with respect to each other in the $y = \pim z / \pim \tau$ direction, by units of $1/3$ times the length of this radius. We find, moreover, that we have lost the symmetric gaussian-like behaviour. Indeed, each of the wavefunctions' density has a gaussian profile in the $y$ axis of the torus, while it seems to be more or less constant in the $x$ direction. Since we are considering $\z = 0$, the gaussians are peaked at $y = - j/3$.

At first sight it might seem quite striking that we have lost the $x \lraw y$ symmetry of figure \ref{theta1}. After all, there is nothing special in our problem regarding the $y$ axis. A second thought reveals that this is just a matter of conventions. Namely, a matter of the choice of the particular basis which describes our space of wavefunctions. Indeed, a orthonormal basis for the vector space of wavefunctions with `weight' $N$ is giving by either \cite{tata}
\beq
\psi^{j, N} (\tau, \nu) = \cn \cdot e^{i\pi N \nu \pim \nu/\pim \tau} 
\cdot 
\vt
\left[
\begin{array}{c}
\frac{j}{N} \\ 0
\end{array}
\right]
(N \nu, N \tau), \quad \quad j=1,\dots,N
\label{basis1}
\eeq
or
\beq
\chi^{j, N} (\tau, \nu) = {\cn \over \sqrt{N}} \cdot
e^{i\pi N \nu \pim \nu/\pim \tau}
\cdot 
\vt
\left[
\begin{array}{c}
0 \\ \frac{j}{N}
\end{array}
\right]
(\nu, \tau/N), \quad \quad j=1,\dots,N
\label{basis2}
\eeq
where $\cn$ is given by (\ref{norm3}). These two bases are related by
a discrete Fourier transform 
\beqa
\chi^{j, N} & = & \frac{1}{\sqrt{N}} \sum_k e^{2\pi i \frac{jk}{N}} \psi^{k, N} \\
\psi^{k, N} & = & \frac{1}{\sqrt{N}} \sum_j e^{2\pi i \frac{kj}{N}} \chi^{j, N}
\eeqa

Each of these two bases is suitable for describing the chiral spectrum of our compactification, being just related by a global unitary rotation in the space of wavefunctions. By plotting the densities of the elements of the alternative basis (\ref{basis2}), we find a probability density where the roles of $x$ and $y$ have been interchanged, as figure \ref{rotatheta} shows.
%
\begin{center}
\EPSFIGURE{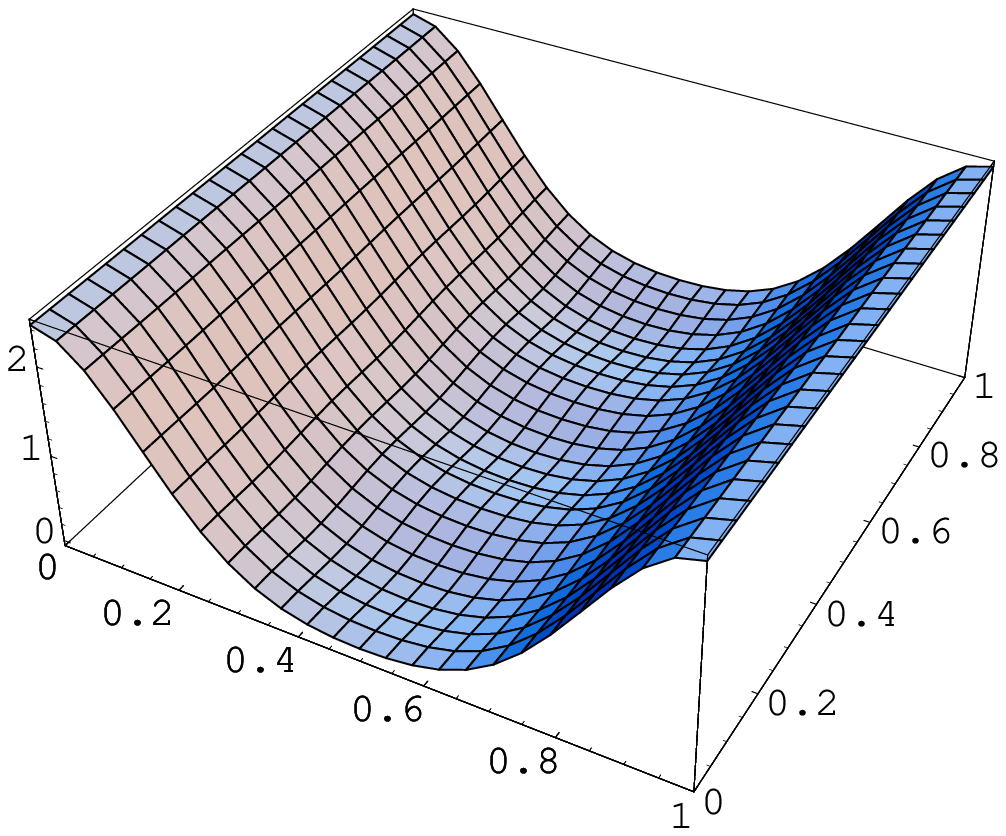, width=3.75in}
{\label{rotatheta}
Probability density of an element of the alternative basis (\ref{basis2}), $\tilde{\rho}(z) = |\chi(z)^{j,N}|^2$ for a square $T^2$. We have chosen $N =3$, $j=0$ and $\z =0$, so that we are describing the same spectrum as in the figure \ref{3thetas}. Notice that the squared density is a $\pi/2$ rotation of the wavefunction in figure \ref{3thetas}.}
\end{center}
%

It turns out that the choice of basis $\psi^{j,N}$ or $\chi^{j,N}$ has a nice physical meaning from the point of view of string theory. We will unveil it when comparing our results with those found in Intersecting D-brane scenarios by means of T-duality.

\subsection{Generalization to $T^{2n}$}

Let us now address how the previous computations generalize to magnetized compactifications in higher-dimensional tori. We will first address the case where $T^{2n}$ is a factorizable torus, and leave the more general case for the next section. Although the computations of wavefunctions becomes more technical, the final answer can be expressed as a product of $n$ wavefunctions in $T^2$.

A factorizable $2n$-dimensional torus can be decomposed as a product of two dimensional tori, that is
\beq
T^{2n} \simeq \ \overbrace{T^2 \times \dots \times T^2}^{n}.
\label{decompT2n}
\eeq
In that case, both the vielbein $e$ and holomorphic transformation $f$ matrices are given by a direct sum of $2 \times 2$ matrices
\beq
e  =  \bigoplus_{r=1}^n\ \left(2\pi R^{(r)}\right)
\left(
\begin{array}{cc}
1 & \preal \tau^{(r)} \\
0 & \pim \tau^{(r)}
\end{array}
\right), \quad \quad
f^{-1} = \bigoplus_{r=1}^n
\left(
\begin{array}{cc}
1 & \tau^{(r)} \\
1 & \bar\tau^{(r)}
\end{array}
\right).
\label{vielbeinT2n}
\eeq
where $r = 1, \dots, n$ labels each factor of $T^2$ in (\ref{decompT2n}). As a consequence, the Clifford algebra in holomorphic coordinates can be reproduced by the following set of gamma matrices
\beq
\begin{array}{rcl} 
\G^{r} & = & \left(2\pi R^{(r)} \right)^{-1} {\bf 1} \otimes \dots \otimes {\bf 1} \otimes \sig^z \otimes \sig^3 \otimes \dots \otimes \sig^3 \\
\G^{\bar{r}} & = & \left(2\pi R^{(r)} \right)^{-1} {\bf 1} \otimes \dots \otimes {\bf 1} \otimes \sig^{\bar{z}} \otimes \sig^3 \otimes \dots \otimes \sig^3 
\end{array}
\label{gammaT2n}
\eeq
where $\sig^{z}$, $\sig^{\bar{z}}$ are inserted in the $r^{\rm th}$ position, and we define
\beq
\sig^3 =
\left(
\begin{array}{cc}
1 & 0 \\
0 & -1
\end{array}
\right), \quad \quad
\sig^z =
\left(
\begin{array}{cc}
0 & 2 \\
0 & 0
\end{array}
\right),\quad \quad
\sig^{\bar{z}} =
\left(
\begin{array}{cc}
0 & 0 \\
2 & 0
\end{array}
\right).
\label{diracmat}
\eeq

Now, let us consider the addition of a constant magnetic flux $F$ in this background. We will consider $F$ to be a $(1,1)$-form\footnote{This condition is related to the hermitian Yang-Mills equations on gauge bundle compactifications. See Appendix \ref{SUSYap}.} in the complex structure defined by (\ref{decompT2n}) and (\ref{vielbeinT2n}). We will again deal with the case where the gauge group breaking is given by $U(N) \raw U(1)^N$. This implies having a magnetic flux of the form
\beq
F_{z^r\bar{z}^{\bar{r}}} = {\pi i\over \pim \tau^{(r)}}
\left(
\begin{array}{cccc}
{m_a^{(r)}}  \\
& {m_b^{(r)}} \\
& & {m_c^{(r)}} \\
& & & \ddots 
\end{array}
\right),
\label{fluxabifT2nf}
\eeq
where $m_\a^{(r)}$ are $n \times N$ integer numbers, different for fixed $r$. 

\vspace{.2cm}

{\noindent \bf Bifundamentals}

\vspace{.1cm}

In order to compute the wavefunctions of bifundamental fields, we again consider $F$ to be a diagonal $U(2)$ flux
\beq
F_{z^r\bar{z}^{\bar{r}}} = {\pi i\over \pim \tau^{(r)}}
\left(
\begin{array}{cc}
{m_a^{(r)}} & \\
& {m_b^{(r)}} 
\end{array}
\right),
\label{fluxabif2T2nf}
\eeq
Following similar steps that the ones for the $T^2$ case and using the gamma matrices (\ref{gammaT2n}), we find that the Dirac operator is now given by 
\beq
\begin{array}{rcl} \vspace*{.1cm}
i\Dbar & = & i\sum_{\bar{r}} \G^{\bar{r}} D_r - i\sum_r \G^{r} D_r^\dag \\
D_r & = & 
\bar{\p}_r  
+ {\pi \over 2 \pim \tau^{(r)}}
\left(
\begin{array}{cc}
{m_a^{(r)}} (z^r + \z_a^r) & \\
& {m_b^{(r)}} (z^r + \z_b^r) 
\end{array}
\right) 
\label{diracopabifT2nf}
\end{array}
\eeq
The Dirac operator will act in a $2n$-dimensional spinor, which can be expressed as a direct product of $n$ $2$-dimensional spinors. Each of the components of $\Psi$ is then given by $\psi_{\eps^1,\dots,\eps^n}$, where $\eps^r = \pm$. In this notation, the zero-mode equation is given by
\beq
\left.
\begin{array}{rcl} \vspace*{.1cm}
D_r \psi_{\eps^1,\dots,\eps^n} = 0 & \quad {\rm if}\quad & \eps^r = + \\
D_r^\dag \psi_{\eps^1,\dots,\eps^n} = 0 &\quad {\rm if}\quad & \eps^r = - 
\end{array}
\right\} \quad \forall r
\label{zeromodeT2nf}
\eeq

On the other hand, the the boundary conditions for a field $\Xi$ transforming in the adjoint of $U(2)$ is again given by (\ref{gaugeabif}), where now $i = 1, \dots, 2n$ and
\beq
\begin{array}{ccc}
\begin{array}{c} \vspace{.2cm}
\chi_{2r-1}^\a (z) =  {\pi m_\a^{(r)} \over  \pim \tau^{(r)}}\ \pim (z^r + \z_\a^r)  \\
\chi_{2r}^\a (z) =  {\pi m_\a^{(r)} \over \pim \tau^{(r)}}\ \pim \bar{\tau}^{(r)}(z^r + \z_\a^r) 
\end{array} 
& & \quad \a = a, b
\end{array}
\label{transitionabifT2n}
\eeq

The wavefunctions of the $2n$-dimensional fermions can be found by solving the differential equations (\ref{zeromodeT2nf}) and the boundary conditions imposed by (\ref{gaugeabif}) and (\ref{transitionabifT2n}). As in the $T^{2}$ case, many of these solutions will be exclusive, and the ones which are non-zero will depend on the signs of the numbers $I_{ab}^{(r)} = m_a^{(r)} - m_b^{(r)}$. The final answer is given by a tensor product of $T^2$ wavefunctions. In order to express it, let us first define the wavefunctions
\beq
\psi_{+}^{j^{(r)}, I_{ab}^{(r)}} =  \psi^{j^{(r)}, I_{ab}^{(r)}} (\tau^{(r)}, z^r + \z_{ab}^r),
\quad \quad
\psi_{-}^{j^{(r)}, I_{ab}^{(r)}} =  \left(\psi^{j^{(r)}, I_{ab}^{(r)}} (\tau^{(r)}, z^r + \z_{ab}^r) \right)^*.
\label{defsaT2n}
\eeq
where $j^{(i)} = 1, \dots, |I_{ab}^{(i)}|$, labels the `Landau level' coming from each complex dimension, and $\psi^{j^{(r)}, I_{ab}^{(r)}}$ is defined as in (\ref{totalsoln}). The wavefunctions will then be given by
\beq
\psi_{\eps_1, \dots, \eps_n} = 
\left(
\begin{array}{cc}
{\rm const.} & \prod_r \d_{\eps_r,s(I_{ab}^{(r)})}\ \psi_{\eps_r}^{j^{(r)}, |I_{ab}^{(r)}|} \\
 \prod_r \d_{-\eps_r,s(I_{ab}^{(r)})}\ \psi_{\eps_r}^{j^{(r)}, |I_{ab}^{(r)}|} & {\rm const.}
\end{array}
\right)
\label{psiaT2nf}
\eeq
where $s(I_{ab}^{(i)}) = {\rm sign} (I_{ab}^{(i)})$. We hence find $2^{n-1}$ $U(1)_a \times U(1)_b$ gauginos, associated to the constant wavefunctions, and $I_{ab} \equiv \prod_r |I_{ab}^{(r)}|$ chiral fermions, associated to chiral fermions in the bifundamental representation. The wavefunctions of the latter are given by a simple product of wavefunctions of the form (\ref{totalsolnabif}) and its complex conjugates.

\vspace{.2cm}

{\noindent \bf Laplace eigenvalues and masses}

\vspace{.1cm}

Just as in the $T^2$ case, the wavefunctions (\ref{psiaT2nf}) not only represent zero-modes of the Dirac operator, but also eigenfunctions of the Laplacian. Is easy to see that the eigenvalues are now given by
\beq
\D \left( \prod_r \psi_{\eps_r}^{j^{(r)}, |I_{ab}^{(r)}|} \right)
= \left( \sum_r  {2\pi |{I}_{ab}^{(r)}| \over \ca^{(r)}} \right) \cdot
\prod_r \psi_{\eps_r}^{j^{(r)}, |I_{ab}^{(r)}|}
\label{laplaceaT2nf}
\eeq

Again, one can recover the full spectrum of eigenfunctions 
and eigenvalues by considering the superposition of $n$ 
harmonic oscillator algebras, where the creation and 
annihilation operators $a_r$ and $a_r^\dag$ are defined 
in terms of $D_r$ and $D_r^\dag$ as in (\ref{osc}). 
The spectrum of scalar particles is much richer than in 
the $T^2$ case, and in particular {\it it does not have to be tachyonic}. 
We leave the derivation of the mass formulae issues for the general 
case treated in the next section.

\section{Toroidal wavefunctions II: non-Abelian Wilson lines}

In this section we consider the general case of 
introducing arbitrary fluxes, in particular those leading 
to a gauge reduction of the form 
$U(N) \raw \prod_i U(p_i)$, with $\sum_i p_i < N$. 
As we will see, this rank reduction occurs whenever we 
introduce non-Abelian Wilson lines in order to fulfill 
Dirac's quantization condition. In this case, there is a new 
technical issue when computing wavefunctions, which comes 
from the fact that  a field transforming in the bifundamental 
representation $(p_a, \bar{p}_b)$ is described by a 
$n_a \times n_b$ matrix instead of a $p_a \times p_b$ matrix, 
with $n_\a \geq p_\a$. On the other hand, many of the 
details of the computation of the wavefunctions are similar 
to those in the previous sections, so we will be more 
sketchy in their derivation.

\subsection{Non-Abelian gauge groups}

Let us first consider the case where we have a non-Abelian gauge group, say $U(N)$. Both the magnetic flux $F$ and the gauge potential $A$ that we introduce transform in the adjoint representation of this gauge group. The Yang-Mills equations applied to the particular case of $T^2$ imply that any irreducible component of $F$ must be proportional to the identity. Such irreducible magnetic flux is then given by
\beq
F_{z\bar{z}} = {\pi i \over \pim \tau} {M \over N}\cdot {\bf 1}_{N} 
\label{fluxna}
\eeq
Consider now a field $\Phi$ in the fundamental representation of $U(N)$. The main difference in this situation with respect to the Abelian case comes from the fact that the Wilson lines can be arbitrary elements of $U(N)$. That is, the most general gauge transformation is of the form
\beq
\Phi(z) \quad \mapsto \quad \Om_i(z) \Phi (z) = e^{i \chi_i(z)} \om_i \Phi(z) 
\label{gaugena}
\eeq
where
\beq
\begin{array}{c} \vspace{.2cm}
\chi_1 (z) =  {\pi M \over N \pim \tau}\ \pim (z + \z) \cdot {\bf 1}_{N}, \\
\chi_2 (z) =  {\pi M \over N \pim \tau}\ \pim \bar{\tau}(z + \z) \cdot {\bf 1}_{N},
\end{array}
\label{transitionna}
\eeq
and $\om_i$ are constant elements of $SU(N)$. 

Just as in the Abelian case, we can demand the gauge transformations to be consistent with the homology of $T^2$, and to reduce to the identity when they correspond to a closed and contractible loop. That is, we impose
\beq
\Om_2^{-1}(z + \tau) \cdot \Om_1^{-1}(z + 1 + \tau) \cdot \Om_2 (z + 1) \cdot\Om_1 (z) \Phi (z) = \Phi (z)
\label{diracna}
\eeq
Notice that this implies that the $SU(N)$ part of (\ref{diracna}) lies in the center of $SU(N)$, that is
\beq
\om_2^{-1} \cdot \om_1^{-1} \cdot \om_2 \cdot \om_1 = e^{2\pi i k/N} \cdot {\bf 1}_{N}, \quad k \in \inte.
\label{diracna2}
\eeq
After imposing this in (\ref{diracna}), and by a similar argument as in the Abelian case, we find that $M = k {\rm\ mod\ } N$ must be an integer, which is again Dirac's charge quantization condition. The $\om_i$ matrices can be chosen to be either \cite{vanbaal1,gr2}
\beq
\om_1 = Q, \quad \quad \om_2 = P^{M}
\label{choiceT2}
\eeq
or
\beq
\om_1 = Q^{M}, \quad \quad \om_2 = P
\label{choiceT2b}
\eeq
where we have defined
\beq
P = 
\left(
\begin{array}{cccc}
& 1 \\
& & 1 \\
& & & \ddots \\
1
\end{array}
\right), \quad \quad
 Q =
\left(
\begin{array}{cccc}
1 \\
& e^{2\pi i / N} \\
& & \ddots \\
& & & e^{2\pi i (N-1)/N}
\end{array}
\right).
\label{matrices}
\eeq
Both choices (\ref{choiceT2}) and (\ref{choiceT2b}) are, of course, equivalent and describe the same physics at low energies. Each of them may be more convenient, however, for showing different aspects of magnetized compactifications. For instance, notice that the presence of the non-Abelian Wilson lines $\om_1$, $\om_2$ imposes non-trivial constraints on the $U(N)$ gauge potentials $A_\mu$ (or gauginos) transforming in the adjoint
\beq
A_\mu = \om_1 \cdot A_\mu \cdot \om_1^{-1} = \om_2 \cdot A_\mu \cdot \om_2^{-1}
\label{adj}
\eeq
taking (\ref{choiceT2}) and following \cite{gr2}, is easy to see that the constraints (\ref{adj}) impose $A_\mu$ to be a diagonal $N \times N$ matrix, with only $P = {\rm g.c.d.}(N,M)$ independent elements. We thus see that introducing a magnetic flux reduces the rank of the gauge group as $U(N) \raw U(P)$.

In fact, notice that if we have the particular case $(N,M) = N\cdot(1,S)$, $S \in \inte$, then we can take  $\om_1=\om_2={\bf 1}$ in (\ref{diracna2}), and the gauge group is going to be $U(N)$, with no rank reduction. If we consider a $D=2$ spinor $\Psi$ in the fundamental of $U(N)$ there will appear $S$ replicas of a chiral spinor transforming in the fundamental representation. The wavefunction of such chiral fermions will again be given by $\psi^{j,S}$ in  (\ref{totalsoln}). We then see that in the case where there is no rank reduction the computation of wavefunctions boils down to the ones performed in the last section, even if non-Abelian gauge groups are involved.

Finally, in order to compute wavefunctions of chiral fermions and bosons transforming in bifundamental representations, as well as Yukawa couplings between them, the choice of non-Abelian Wilson lines (\ref{choiceT2b}) turns out to be more suitable. We will thus stick to that choice in the following.

\subsection{Fermions in bifundamentals}

Let us now consider $F$ to be a direct sum of two such `irreducible' representations. That is
\beq
F_{z\bar{z}} = {\pi i\over \pim \tau}
\left(
\begin{array}{cc}
\frac{m_a}{n_a} {\bf 1}_{n_a} & \\
& \frac{m_b}{n_b} {\bf 1}_{n_b} 
\end{array}
\right),
\label{fluxnabif}
\eeq
where $n_a + n_b = N$. This can be seen as two magnetic fluxes $F_a$ and $F_b$ over the same $T^2$, both corresponding to two different gauge groups $U(n_a)$ and $U(n_b)$, and with magnetic quanta $m_a$ and $m_b$ respectively. Actually, from the point of view of D-branes, each stack of D-branes corresponds to an 'irreducible' representation of $F$, so this system can be associated to two stacks of D-branes $a$ and $b$ wrapping $T^2$, with multiplicities $n_a$ and $n_b$, respectively, and with a magnetic flux turned on each stack, each flux being proportional to $m_a$ and $m_b$. 

As we have just seen, the low energy theory of such configuration will correspond to a gauge group $U(p_a) \times U(p_b)$, where $p_\a = {\rm g.c.d.} (n_\a, m_\a)$. As noted above, if $(n_\a, m_\a) = n_a(1, s_\a)$ then no non-Abelian Wilson lines appear and the gauge group will be given by $U(n_a) \times U(n_b)$. The computations of the previous section will readily apply to this case, and the wavefunction of the chiral fermions transforming in $(n_a, \bar{n}_b)$ will be given by (\ref{totalsolnabif}), with the substitution $I_{ab} \mapsto s_a - s_b$. We then proceed to consider the cases where $p_\a < n_\a$ and some gauge reduction is involved. In order to simplify the discussion, we will suppose $p_a = p_b = 1$, the generalization to arbitrary rank being straightforward.

\vspace{.2cm}

{\noindent \bf Dirac equation}

\vspace{.1cm}

Eq. (\ref{fluxnabif}) and condition (\ref{adj}) now imply a gauge connection of the form
\beqa
A_{\bar{z}} = A_{\bar{z}}^\a T_\a & = & {\pi \over 2 \pim \tau}
\left(
\begin{array}{cc}
\frac{m_a}{n_a} (z + \z_a) {\bf 1}_{n_a} & \\
& \frac{m_b}{n_b} (z + \z_b) {\bf 1}_{n_b} 
\end{array}
\right),\\
A_{z} = A_z^\a T_\a & = & - {\pi \over 2 \pim \tau}
\left(
\begin{array}{cc}
\frac{m_a}{n_a} (\bar{z} + \bar{\z}_a) {\bf 1}_{n_a} & \\
& \frac{m_b}{n_b} (\bar{z} + \bar{\z}_b) {\bf 1}_{n_b} 
\end{array}
\right),
\label{potentialap}
\eeqa
where $T_a = (-i){\bf 1}_{n_a}$ and $T_b = (-i){\bf 1}_{n_b}$ are anti-hermitian generators of $U(1)_a \times U(1)_b \subset U(N)$. The Dirac operator is again given by (\ref{diracopbif}). The zero modes of $\Dbar$ can be found by again considering a two-dimensional spinor of the form (\ref{spinora}) but now transforming in the adjoint of $U(N)$. The entry $A_\pm$ is no longer a number but a $n_a \times n_a$ matrix, etc. We now impose the Dirac equation on the spinorial component $\psi_+$ which implies
\beqa
D \psi_+ & = & (\pi R)^{-1} \left( \bar{\p} \psi_+ + [A_{\bar{z}},\psi_+] \right) \label{zeromode+na}\\ \nonumber
& = & (\pi R)^{-1} 
\left(
\begin{array}{cc}
\bar{\p} A_+ & \left(\bar{\p} + {\pi \tilde{I}_{ab} \over 2 \pim \tau} (z + \z_{ab}) \right) B_+\\
\left(\bar{\p} - {\pi \tilde{I}_{ab} \over 2 \pim \tau} (z + \z_{ab}) \right) C_+ & \bar{\p} D_+
\end{array}
\right) = 0
\eeqa
where we have now defined
\beqa
I_{ab} & \equiv &  - n_a m_b + n_b m_a \neq 0 
\label{defna1}\\
\tilde{I}_{ab} & \equiv & I_{ab}/n_an_b \label{defna2} \\ 
\z_{ab} & \equiv & (n_b m_a \z_a - n_a m_b \z_b)/I_{ab}
\label{defna3}
\eeqa
which is a generalization of the definitions (\ref{defa1}), (\ref{defa2}) for $n_\a > 1$, and allows to distinguish between the quantities $I_{ab}$ and $\tilde{I}_{ab}$. This difference will turn out to be quite important. The integer number $I_{ab}$ will again determine the multiplicity and chirality of the spectrum, and can be thought as a T-dual version of the intersection number from intersecting D-brane models\footnote{Notice, however, that we have defined $I_{ab}$ with a relative minus sign respect to \cite{afiru}.}. 

The conditions that the Dirac equation imposes on $\psi_-$ are also quite similar to the ones in the previous section, and can be obtained by making the substitution $I_{ab} \mapsto \tilde{I}_{ab}$ and taking the general definition of $\z_{ab}$ in (\ref{zeromode-a}). Again we deduce that $A_+, D_+$ have to be constant matrices associated to $U(1)_a \times U(1)_b$ gauginos, whereas $B_+, C_+$ have to be of the form
\beq
\cn \cdot e^{\pm i {\pi \tilde{I}_{ab} \over \pim \tau} (z + \z_{ab}) \cdot \pim (z + \z_{ab})} \cdot \xi(z),
\label{wavena}
\eeq
respectively, where $\xi(z)$ is now an arbitrary holomorphic matrix-valued function, and $\cn$ is a normalization factor. This time, instead of following the procedure of the previous section, we will consider the ansatze (\ref{wavena}) and impose the $U(N)$ gauge transformation properties that those fields must satisfy in order to be well-defined wavefunctions. It can be seen that both procedures lead us to the same final result.

\vspace{.2cm}

{\noindent \bf Gauge transformations}

\vspace{.1cm}

The gauge transformations for a field $\Phi(z)$ transforming in the $(n_a, \bar{n}_b)$ representation of $U(n_a) \times U(n_b) \subset U(N)$ are
\beq
\Phi(z) \quad \mapsto \quad \Om_i^a(z) \Phi (x) (\Om_i^b(z))^\dag = 
e^{i (\chi_i^a(z)- \chi_i^b(z))} \om_i^a \Phi(z) (\om_j^b)^\dag
\label{gaugebif}
\eeq
where
\beq
\begin{array}{c} \vspace{.2cm}
\chi_1^\a (z) =  {\pi m_\a \over n_\a \pim \tau}\ \pim (z + \z_\a) \cdot {\bf 1}_{n_\a} \\
\chi_2^\a (z) =  {\pi m_\a \over n_\a \pim \tau}\ \pim \bar{\tau}(z + \z_\a) \cdot {\bf 1}_{n_\a}
\end{array}
\quad \a = a,b 
\label{transitionbif}
\eeq
\beq
\begin{array}{rcl}
\om_1^\a & = & Q^{m_\a} \\
\om_2^\a & = & P
\end{array}
\quad \a = a,b 
\label{choiceT2bif2}
\eeq
Hence, the boundary conditions for the components $\phi_{k_a,k_b}$ of such bifundamental field are given by
\beqa 
\Phi(z+1) & = & e^{i(\chi_1^a-\chi_1^b)} Q^{m_a} \Phi(z) Q^{-m_b} \label{boundT2bif} \\ \nonumber
& \Raw & 
\phi_{k_a, k_b} (z+1) = e^{i \pi \tilde{I}_{ab} \frac{\pim (z + \z_{ab})}{\pim \tau}} e^{2\pi i \left(\frac{m_a}{n_a} k_a - \frac{m_b}{n_b} k_b\right)}\phi_{k_a, k_b} (z)
\\
\Phi(z+\tau) & = & e^{i (\chi_2^a-\chi_2^b)} P \Phi(z) P^{-1}
\label{boundT2bif2} \\ \nonumber
& \Raw &
\phi_{k_a,k_b} (z+\tau) = e^{i \pi \tilde{I}_{ab}
\frac{\pim \bar{\tau}(z + \z_{ab})}{\pim \tau}} \phi_{k_a+1,k_b+1}(z)
\eeqa
where $k_\a = 1, \dots, n_\a$. Notice that (\ref{boundT2bif}) and (\ref{boundT2bif2}) imply
\beqa
\phi_{k_a,k_b} (z+n_an_b) & = & e^{i \pi I_{ab} \frac{\pim (z + \z_{ab})}{\pim \tau}} \phi_{k_a,k_b}(z), \\
\phi_{k_a,k_b} (z+n_an_b\tau) & = & e^{i \pi I_{ab} \frac{\pim \bar{\tau}(z + \z_{ab})}{\pim \tau}} \phi_{k_a,k_b}(z).
\label{bound2T2bif}
\eeqa

Notice that these transformation properties will be satisfied by any field charged in the bifundamental representation $(n_a, \bar{n}_b)$. Let us, however, focus on the zero modes of the Dirac equation. In particular, let us look for solutions of $B_+$ in (\ref{spinora}), which must satisfy the ansatz (\ref{wavena}) with a $+$ sign. We find that the holomorphic matrix-valued function $\xi(z)$ is given by a theta function of the form
\beq
\left(\xi^{j,I_{ab}}(z)\right)_{l,l} = 
\vt
\left[
\begin{array}{c}
\frac{j}{I_{ab}} + \frac{l}{n_a n_b} \\ 0
\end{array}
\right]
((z + \z_{ab})I_{ab}, \tau I_{ab} n_a n_b),
\label{holosolnbif}
\eeq
where $l = 1,\dots, n_a n_b$ has to be understood ${\rm mod\ } n_a$, $n_b$ respectively, and $j=1,\dots, I_{ab}$. This solution is strictly valid and unambiguous only if ${\rm g.c.d.}(n_a,n_b) = 1$, which is the case that we will consider in the following. We are thus able to express the bifundamental $B_+$ in terms of a linear combination of the following wavefunctions
\beqa
\Phi^{j,I_{ab}} (z) & = & \cn \cdot e^{i {\pi \tilde{I}_{ab} \over \pim \tau} (z + \z_{ab}) \cdot \pim (z + \z_{ab})} \cdot \xi^{j, I_{ab}}(z),
\label{totalsolnbif}
\eeqa
and is easy to see that the hermitian conjugates $\left(\Phi^{j,I_{ab}}\right)^\dag$ expand a basis of wavefunctions for the bifundamental fields $C_-$.

Notice as well that (\ref{totalsolnbif}) will only converge if $I_{ab} > 0$. In case $I_{ab} < 0$ we will have fermions of opposite chirality, hence we should consider
\beqa
\Phi^{j,I_{ba}} & {\rm and} & \left(\Phi^{j,I_{ba}}\right)^\dag 
\label{fermbif}
\eeqa
as the wavefunctions coming from fermions transforming in $(\bar{n}_a,n_b)$ and $(n_a,\bar{n}_b)$, respectively.

\vspace{.2cm}

{\noindent \bf Normalization}

\vspace{.1cm}

The normalization condition in the case of bifundamental fields $\Phi^j$ is given by
\beq
\int_{T^2} dz d\bar{z}\  \Tr \left( \Phi^{i,I_{ab}} (\Phi^{j,I_{ab}})^\dag \right) = \d_{ij},
\label{normbif}
\eeq

On the other hand, notice that the boundary conditions (\ref{boundT2bif}) and (\ref{boundT2bif2}) imply that
\beqa
\phi_{k_a, k_b}^{i} (\phi_{k_a, k_b}^{j})^* (z+1) & = &
\phi_{k_a, k_b}^{i} (\phi_{k_a, k_b}^{j})^* (z) \\
\phi_{k_a, k_b}^{i} (\phi_{k_a, k_b}^{j})^* (z+\tau) & = &
\phi_{k_a+1, k_b+1}^{i} (\phi_{k_a+1, k_b+1}^{j})^* (z) \\
\label{normbif2}
\eeqa
As a result, if $I_{ab} \neq 0$, we can compute (\ref{normbif}) by integrating $\phi_{k_a, k_b}^{i,I_{ab}} (\phi_{l,l}^{j,I_{ab}})^*$ over a $T^2$ of complex structure $n_an_b\tau$.\footnote{We are again considering ${\rm g.c.d.}(n_a,n_b) = 1$.} That is,
\beqa \nonumber
\int_{T^2(\tau)} dz d\bar{z}\  \Tr \left( \Phi^i (\Phi^j)^\dag \right) & = &
\int_{T^2(\tau)} dz d\bar{z} \sum_{l} \phi^i_{l,l} (\phi^j_{l,l})^* \\ 
 = \int_{T^2(n_an_b\tau)} dz d\bar{z} \phi^i_{l,l} (\phi^j_{l,l})^*
& = & \left(2 \pim \tau |I_{ab}| n_an_b \right)^{-1/2} \cn^2 \d_{ij} 
\label{normbif4}
\eeqa
from where we can extract the normalization factor $\cn$.

\vspace{.2cm}

{\noindent \bf Summary}

\vspace{.1cm}

We have again found that the fermionic zero mode wavefunctions for chiral fields transforming in the bifundamental representation can be expressed in terms of theta functions. More precisely
\beq
B_+ = \Phi^{j,I_{ab}}, \quad \quad C_+ = \Phi^{j,-I_{ab}}
\label{wavebif}
\eeq
where $j = 1, \dots, |I_{ab}|$ and
\beqa
\left(\Phi^{j,I_{ab}}\right)_{k_a,k_b} & = & \left({2 \pim \tau |\tilde{I}_{ab}| \over \ca^2}\right)^{-1/4} e^{i {\pi \tilde{I}_{ab} \over \pim \tau} (z + \z_{ab}) \cdot \pim (z + \z_{ab})} \cdot \xi(z)_{k_a,k_b}^{j, I_{ab}} \label{totalsolnbif2}\\ \nonumber
\left(\xi^{j,I_{ab}}(z)\right)_{l,l}
& = & 
\vt
\left[
\begin{array}{c}
\frac{j}{I_{ab}} + \frac{l}{n_a n_b} \\ 0
\end{array}
\right]
((z + \z_{ab})I_{ab}, \tau I_{ab} n_a n_b).
\eeqa
These two solutions are exclusive in the sense that the theta-function series will not converge at the same time. Indeed\footnote{Recall that $U(n_a) \times U(n_b)$ is broken by the flux to $U(p_a) \times U(p_b)$, where $p_\a = {\rm g.c.d.}(n_\a, m_\a)$. The wavefunctions in (\ref{chiralityna}) will in general be bifundamentals of such gauge group. In the particular case at hand we are considering $p_a = p_b =1$.},
\beq
\begin{array}{cc}
B_+ < \infty \iff I_{ab} > 0 &\quad  {\rm\ Left-handed\ fermions\ in\ } (+1,-1)\ {\rm of\ } U(1)_a \times U(1)_b \\
C_+ < \infty \iff I_{ab} < 0 &\quad  {\rm\ Left-handed\ fermions\ in\ } (-1,+1)\ {\rm of\ } U(1)_a \times U(1)_b
\end{array}
\label{chiralityna}
\eeq
The anti-particles of such wavefunctions are given by the hermitian conjugates of (\ref{wavebif}).

\subsection{Eigenfunctions of the Laplace equation}

Just as previously pointed out, the wavefunctions (\ref{totalsolnbif2}) will be also eigenfunctions of the Laplace operator (\ref{laplaceop}). By a similar computation as the one in section 3.3 we obtain 
\beq
\D \Phi^{j,\pm I_{ab}} = \pm {2\pi \tilde{I}_{ab} \over \ca} \Phi^{j,\pm I_{ab}} = {2\pi |\tilde{I}_{ab}| \over \ca} \Phi^{j,\pm I_{ab}},
\label{laplacena}
\eeq
same for $\left(\Phi^{j,\pm I_{ab}}\right)^\dag$.

Notice that we recover the same eigenvalue than in the Abelian case of section 3, with the only replacement $I_{ab} \mapsto \tilde{I}_{ab}$. We can carry out the quantum harmonic oscillator algebra and compute the eigenvalues of the mass matrix by making such substitution in the expressions (\ref{defs}) through (\ref{eigen2}). We finally obtain that the lightest scalar particle is given by a tachyon of mass
\beq
m^2_{\rm tach} = - 2 \pi  {|\tilde{I}_{ab}| \over \ca}
\label{masstach}
\eeq
This spectrum should match with the one obtained in the T-dual picture, at least in the limit of large volume $\ca$ and diluted flux (small angles). By comparing both masses, the (approximate) analogue of the angle between two intersecting D-branes in the flux picture can be seen to be \cite{torons}
\beq
\th_{ab}^{app} = \frac{1}{\pi}\left(tan(\pi\th_a) - tan(\pi\th_b)\right) = 4\pi {\tilde{I}_{ab} \over (\ca/\a^\prime)} 
\label{twist}
\eeq
a quantity which only depends on the area of $T^2$ in string scale units and, in terms of the mathematical description of the magnetic flux as a bundle over $T^2$, is given by the $\mu$-slope of such bundle (see Appendix \ref{SUSYap} for the definition of $\mu$-slope).

\subsection{Generalization to $T^{2n}$}

Let us now address how the previous computations generalize to magnetized compactifications in higher-dimensional tori. The most general constant magnetic flux associated to a $U(N)$ gauge group is given by
\beq
F_{ij} = 2\pi {n_{ij} \over N a_i a_j}\cdot {\bf 1}_{N}
\label{fluxnaT2n}
\eeq
where $n_{ij} = - n_{ji}$. Here we represent $T^{2n}$ by the quotient $\real^{2n}/\Lam$, where $\Lam = \{ x \in \real^{2n} | x = n_i a^i; n \in \inte^{2n} \}$. We are then parametrizing the torus by the $2n$-dimensional hypercube $\{ x \in \real^{2n} | 0 < x_i \leq a_i \}$, where $a_i = || a^i ||$ are the lengths of the $2n$ lattice vectors. 

The flux (\ref{fluxnaT2n}) implies that the boundary conditions on a field $\Phi$ transforming on the fundamental of $U(N)$ are of the form
\beq
\Phi(x) \quad \mapsto \quad \Om_i(x) \Phi (x) = e^{i \chi_i(x)} \om_i \Phi(x) 
\label{gaugenaT2n}
\eeq
where
\beq
\chi_i(x) = exp \left\{\pi \sum_{j} {n_{ij} x^j \over N a_j} \right\} \cdot {\bf 1}_{N}
\label{transitionnaT2n}
\eeq
and $\om_i$ are constant elements of $SU(N)$. Consistency of the boundary conditions amounts to imposing
\beq
\Om_j^{-1}(x + a_i) \cdot \Om_i^{-1}(x + a_i + a_j) \cdot \Om_j (x + a_i) \cdot \Om_i (x) \Phi (x) = \Phi (x).
\label{diracnaT2n}
\eeq
This again implies that the $SU(N)$ part of (\ref{diracnaT2n}) lies in the center of $SU(N)$,
\beq
\om_j^{-1} \cdot \om_i^{-1} \cdot \om_j \cdot \om_i = e^{2\pi i c_{ij}/N} \cdot {\bf 1}_{N}, \quad c_{ij} \in \inte
\label{diracna2T2n}
\eeq
and that $n_{ij} = c_{ij} {\rm\ mod\ } N$. Following \cite{vanbaal1,gr2}, we consider $SU(N)$ constant matrices $P$ and $Q$ such that $P Q = Q  P e^{2\pi i/N}$. The $SU(N)$ part of the transition function can then be written as
\beq
\om_i = P^{s_i} Q^{t_i}, \quad \quad s_i, t_i \in \inte,
\label{SUWL}
\eeq
and the problem is reduced to finding $s_i$, $t_i$ such that
\beq
t_i s_j - t_j s_i = n_{ij} \ {\rm mod\ } N.
\label{SUWLcond}
\eeq
Such $P$ and $Q$ matrices can be taken to be (\ref{matrices}), the choices (\ref{choiceT2}) and (\ref{choiceT2b}) being solutions of (\ref{SUWLcond}) in the particular case of $T^2$. 

In a general $T^{2n}$ compactification, (\ref{SUWLcond}) will have a solution provided that all the higher Chern numbers are specified by the first Chern numbers of the flux \cite{torons}. In the following we will assume that this is the case. Indeed, when dealing with constant magnetic fluxes, lack of satisfaction of (\ref{SUWLcond}) means that the initial gauge group breaks as $U(N) \raw \prod_i U(P_i)$ (instead of $U(N) \raw U(P)$) after turning on $F$. The flux (\ref{fluxnaT2n}) can then be written as a direct sum of more `fundamental' fluxes, each one giving rise to a gauge group $U(P_i)$ \cite{gr1}. For our purposes, then, we can just consider fundamental fluxes satisfying (\ref{SUWLcond}) and direct sums of these.\footnote{In the mirror picture of intersecting D-branes, these fundamental constant fluxes correspond to D-branes wrapping $T^n$ submanifolds of $T^{2n}$.}

In the following, we will generalize our previous results in order to compute wavefunctions of chiral matter fields in higher dimensional tori. We will follow the same kind of strategy as used for $T^2$. We will first address the case where $T^{2n}$ is a factorizable torus. Although the computations of wavefunctions becomes more technical, the final answer can be expressed as a `tensor product' of $n$ wavefunctions in $T^2$. We will then address the case of a general $T^{2n}$, showing that the wavefunctions can then be expressed in terms of Riemann theta functions.

\subsubsection{Factorizable tori}

\vspace{.2cm}

{\noindent \bf Gauge group}

\vspace{.1cm}

Let us consider the factorizable background (\ref{decompT2n}) and the addition of a constant magnetic flux $F$ on it. Without loss of generality, we will consider $F$ to be fundamental in the sense described above and a $(1,1)$-form. This allows us to specify $F$ in terms of $2n$ integer numbers $(N^{(r)}, M^{(r)})$, $r =1, \dots, n$, such that \cite{torons}
\beq
\begin{array}{rcl} \vspace*{.1cm}
N & = & \prod_r N^{(r)} \\
n_{2r-1,2r} & = & N^{(1)}\dots N^{(r-1)} M^{(r)} N^{(r+1)}\dots N^{(n)}
\end{array}
\label{cherns}
\eeq
the rest of $n_{\mu\nu}$ vanishing. The components of the magnetic flux are then
\beq
F_{z^{r}\bar{z}^{\bar{r}}} = {\pi i \over \pim \tau^{(r)}} {M^{(r)} \over N^{(r)}}\cdot {\bf 1}_{N} 
\label{fluxnaT2nf}
\eeq
and the boundary conditions for a field $\Phi$ transforming in the fundamental are again given by $\Om_r ={\rm exp}(i\chi_r) \cdot \om_r$, where now
\beq
\begin{array}{c} \vspace{.2cm}
\chi_{2r-1} (z) =  {\pi M^{(r)} \over N^{(r)} \pim \tau^{(r)}}\ \pim (z^r + \z^r) \cdot {\bf 1}_{N}, \\
\chi_{2r} (z) =  {\pi M^{(r)} \over N^{(r)} \pim \tau^{(r)}}\ \pim \bar{\tau}^{(r)}(z^r + \z^r) \cdot {\bf 1}_{N},
\end{array}
\label{transitionnaT2nf}
\eeq
The action of the non-Abelian Wilson lines $\om_r$ on $\Phi$ can be more easily described by again using a tensor product representation, now acting on the gauge group indices. Indeed, since $N$ admits the decomposition (\ref{cherns}), we can express the elements $\Phi$ as
\beqa
\Phi_k, \quad k = 1,\dots, N & \quad \raw \quad & \Phi_{k^{(1)},\dots,k^{(n)}}, \quad k^{(r)} = 1,\dots, N^{(r)} 
\label{gaugedecomp}
\eeqa
we then have
\beq
\begin{array}{rcl}\vspace*{.1cm}
\om_{2r-1} & = & {\bf 1}_{N^{(1)}} \otimes \dots \otimes {\bf 1}_{N^{(r-1)}} \otimes Q^{M^{(r)}} \otimes {\bf 1}_{N^{(r+1)}} \otimes \dots \otimes {\bf 1}_{N^{(n)}} \\
\om_{2r} & = & {\bf 1}_{N^{(1)}} \otimes \dots \otimes {\bf 1}_{N^{(r-1)}} \otimes P \otimes {\bf 1}_{N^{(r+1)}} \otimes \dots \otimes {\bf 1}_{N^{(n)}}
\end{array}
\label{choiceT2nf}
\eeq
where $Q$ and $P$ are the obvious generalization of (\ref{matrices}) to $N^{(r)} \times N^{(r)}$ matrices. After introducing such flux, the gauge group will be broken from $U(N)$ to $U(\prod_r P^{(r)})$, where $P^{(r)} = {\rm g.c.d.}(N^{(r)}, M^{(r)})$ \cite{gr2}.

\vspace{.2cm}

{\noindent \bf Bifundamentals}

\vspace{.1cm}

In order to compute the wavefunctions of bifundamental fields, we need again to consider $F$ to be a direct sum of two fundamental fluxes:
\beq
F_{z^r\bar{z}^{\bar{r}}} = {\pi i\over \pim \tau^{(r)}}
\left(
\begin{array}{cc}
\frac{m_a^{(r)}}{n_a^{(r)}} {\bf 1}_{N_a} & \\
& \frac{m_b^{(r)}}{n_b^{(r)}} {\bf 1}_{N_b} 
\end{array}
\right),
\label{fluxnabifT2nf}
\eeq
where $N_\a = \prod_r n_\a^{(r)}$. The Dirac operator is now given by 
\beq
\begin{array}{rcl} \vspace*{.1cm}
i\Dbar & = & i\sum_{\bar{r}} \G^{\bar{r}} D_r - i\sum_r \G^{r} D_r^\dag \\
D_r & = & 
\bar{\p}_r  
+ {\pi \over 2 \pim \tau^{(r)}}
\left(
\begin{array}{cc}
\frac{m_a^{(r)}}{n_a^{(r)}} (z^r + \z_a^r) {\bf 1}_{N_a} & \\
& \frac{m_b^{(r)}}{n_b^{(r)}} (z^r + \z_b^r) {\bf 1}_{N_b} 
\end{array}
\right) 
\label{diracopbifT2nf}
\end{array}
\eeq
and will act in the $2n$-dimensional spinor, which can again be decomposed as the tensor product $\psi_{\eps^1,\dots,\eps^n}$. The zero more equation given by (\ref{zeromodeT2nf}), we find that the solution can be expressed in terms of the wavefunctions
\beq
\Phi_{+}^{j^{(r)}, I_{ab}^{(r)}} =  \Phi^{j^{(r)}, I_{ab}^{(r)}},
\quad \quad
\Phi_{-}^{j^{(r)}, I_{ab}^{(r)}} =  \left(\Phi^{j^{(r)}, I_{ab}^{(r)}}\right)^\dag.
\label{defsT2n}
\eeq
where $j^{(i)} = 1, \dots, I_{ab}^{(i)}$, and $\Phi^{j^{(r)}, I_{ab}^{(r)}}$ is defined as in (\ref{totalsolnbif2}). The wavefunctions will then be given by
\beq
\psi_{\eps_1, \dots, \eps_n} = 
\left(
\begin{array}{cc}
{\rm const.} & \prod_i \d_{\eps_r,s(I_{ab}^{(r)})} \bigotimes_r \Phi_{\eps_r}^{j^{(r)}, |I_{ab}^{(r)}|} \\
 \prod_r \d_{-\eps_r,s(I_{ab}^{(r)})} \bigotimes_r \Phi_{\eps_r}^{j^{(r)}, |I_{ab}^{(r)}|} & {\rm const.}
\end{array}
\right)
\label{psiT2nf}
\eeq
where $s(I_{ab}^{(i)}) = {\rm sign} (I_{ab}^{(i)})$. The tensor product in (\ref{psiT2nf}) is to be understood as
\beq
\left( \bigotimes_r \Phi_{\eps_r}^{j^{(r)}, |I_{ab}^{(r)}|} \right)_{\left(k_a^{(r)};   k_b^{(r)}\right)} = 
\prod_r \left( \Phi_{\eps_r}^{j^{(r)}, |I_{ab}^{(r)}|}\right)_{k_a^{(r)};   k_b^{(r)}} 
\label{tensorT2nf}
\eeq

\vspace{.2cm}

{\noindent \bf Laplace eigenvalues and masses}

\vspace{.1cm}

The eigenvalues of the Laplacian are in this case given by
\beq
\D \left( \bigotimes_r \Phi_{\eps_r}^{j^{(r)}, |I_{ab}^{(r)}|} \right)
= \left( \sum_r  {2\pi |\tilde{I}_{ab}^{(r)}| \over \ca^{(r)}} \right) \cdot
\bigotimes_r \Phi_{\eps_r}^{j^{(r)}, |I_{ab}^{(r)}|},
\label{laplaceT2nf}
\eeq
and again, one can recover the full spectrum of eigenfunctions and eigenvalues by considering the superposition of $n$ harmonic oscillator algebras. We then recover the eigenvalues found in \cite{troost}
\beq
\lam_{\left\{s^{(r)}\right\}_r} = 2\pi  \sum_r {|\tilde{I}_{ab}^{(r)}| \over \ca^{(r)}} (2s^{(r)} + 1), \quad \quad s^{(r)} \in \nat
\label{eigenT2n}
\eeq
whereas the eigenvalues of the mass matrix are given by
\beqa \nonumber
\tilde{\lam}_{\left\{s^{(r)}\right\}_r}^i & = & 
2\pi \left( \sum_r {|\tilde{I}_{ab}^{(r)}| \over \ca^{(r)}} (2s^{(r)} + 1) \pm 2 {\tilde{I}_{ab}^{(i)} \over \ca^{(i)}}  \right) \\
& = & (2 \a')^{-1} \left( \sum_r |\th_{app}^{(r)}| (2s^{(r)} + 1) \pm 2 \th_{app}^{(i)} \right)
\label{eigen2T2n}
\eeqa

Note that the lightest scalar excitations are obtained for $s^{(r)}=0$.
In the $T^2$ case it is always tachyonic, reflecting the fact that 
$N=1$ SUSY configurations are not possible in this case. In the $T^4$ 
case the lightest scalar is either massless or tachyonic. Finally,
in the $T^2\times T^2\times T^2$ case the lightest scalar may be 
massive, massless or tachyonic, depending on the values
of the slopes $\th_{app}^{(r)}$. In this latter case one
recovers a SUSY spectrum if the lightest scalar is massless
\footnote{See refs.\cite{afiru,imr,torons} for a detailed 
discussion of the different possibilities in the T-dual
language of intersecting D-branes.}.


\subsubsection{General tori}

Let us now consider the more general case where the $2n$-dimensional torus is not necessarily factorizable. For simplicity, we will restrict ourselves to fields charged under Abelian gauge groups. That is, we set $N=1$ in (\ref{fluxnaT2n}) for the rest of this section.

A generic flat $2n$-dimensional torus, $T^{2n} \simeq \cpx^n/\Lam$, inherits a complex structure from the covering space $\cpx^n$. Its geometry can hence be described in terms of a K\"ahler metric and complex structure as
\beq \begin{array}{rcl} \vspace*{.1cm}
ds^2 & = & h_{\mu\bar{\nu}} dz^{\mu} d\bar{z}^{\bar{\nu}} \\
dz^\mu & = & dx^\mu + \tau^{\mu}_{\nu} dy^\nu
\end{array}
\label{complexn}
\eeq
where $x^{\mu}, y^{\mu} \in (0,1)$, $\mu=1,\dots,n$,  parametrize the $2n$ vectors of the lattice $\Lam$. The natural generalization of the Jacobi theta function (\ref{theta}) to this higher-dimensional tori is known as Riemann $\vt$-function
\beq
\vt \left[
\begin{array}{c}
\vec{a} \\ \vec{b}
\end{array}
\right] (\vec{\nu} | \Oom)
=
\sum_{\vec{m} \in \inte^n} e^{\pi (\vec{m} + \vec{a}) \cdot \Oom \cdot (\vec{m} + \vec{a})}
e^{2\pi i (\vec{m} + \vec{a}) \cdot (\vec{\nu} + \vec{b})}
\label{tetaza}
\eeq
where $\vec{a}, \vec{b} \in \real^n$, $\vec{z} \in \cpx^n$ and $\Oom$ is an $n \times n$ complex matrix. The transformation properties of such $\vt$-function under lattice shifts are given by
\beqa
\vt
\left[
\begin{array}{c}
\vec{a} \\ \vec{b}
\end{array}
\right]
(\vec{\nu}+\vec{n} | \Oom)
& = &
e^{2\pi i \vec{a} \cdot \vec{n}} \cdot 
\vt
\left[
\begin{array}{c}
\vec{a} \\ \vec{b}
\end{array}
\right]
(\vec{\nu} | \Oom)
\label{trans1n}
\\
\vt
\left[
\begin{array}{c}
\vec{a} \\ \vec{b}
\end{array}
\right]
(\vec{\nu}+\Oom \cdot \vec{n} | \Oom)
& = &
e^{-\pi i \vec{n} \cdot \Oom \cdot \vec{n} - 2\pi i \vec{n} \cdot (\vec{\nu} + \vec{b})} \cdot
\vt
\left[
\begin{array}{c}
\vec{a} \\ \vec{b}
\end{array}
\right]
(\vec{\nu} | \Oom)
\label{trans2n}
\eeqa
where $\vec{n} \in \inte^n$. These transformation properties are very suggestive. Indeed, inspired by (\ref{totalsoln}) we can construct the following wavefunction
\beqa \nonumber
\psi^{\vec{j}, \nat}(\vec{z}, \Oom) & = & \cn \cdot exp \left\{
i\pi [\nat \cdot \vec{z}] \cdot (\nat \cdot \pim \Oom)^{-1} \cdot \pim[\nat \cdot \vec{z}]
\right\}
\cdot 
\vt
\left[
\begin{array}{c}
\vec{j} \\ 0
\end{array}
\right]
\left(\nat \cdot \vec{z}\ | \nat \cdot \Oom \right) 
 \\ 
& = & 
\cn \cdot e^{i\pi [\nat \cdot \vec{z}] \cdot (\pim \Oom)^{-1} \cdot \pim \vec{z} }
\cdot 
\vt
\left[
\begin{array}{c}
\vec{j} \\ 0
\end{array}
\right]
\left(\nat \cdot \vec{z}\ | \nat \cdot \Oom \right)
\label{totalsolnn}
\eeqa
here $\cn$ is a normalization factor and $\Oom_{\mu\nu} = \tau_{\mu\nu}$. The transformation properties of this wavefunction are given by
\beq \begin{array}{rcl} \vspace{.2cm}
\psi^{\vec{j}, \nat}(\vec{z} + \vec{n}, \Oom)
& = &
e^{i\pi [\nat \cdot \vec{n}] \cdot (\pim \Oom)^{-1} \cdot \pim \vec{z}}
\cdot \psi^{\vec{j}, \nat}(\vec{z}, \Oom)
\label{trans1psin}
\\ \nonumber
\psi^{\vec{j}, \nat}(\vec{z} + \Oom \cdot \vec{n}, \Oom)
& = &
e^{-i\pi [\nat^t \cdot \vec{n}] \cdot \left[ \preal \vec{z}  - \preal \Oom \cdot (\pim \Oom)^{-1} \cdot \pim \vec{z} \right]}
\cdot \psi^{\vec{j}, \nat}(\vec{z}, \Oom)
\end{array}
\label{trans2psin}
\eeq
provided that
\begin{itemize}

\item $\nat_{\mu\nu} \in \inte$

\item $\vec{j} \cdot \nat \in \inte^n$

\item $\left( \nat \cdot \pim \Oom \right)^t = \nat \cdot \pim \Oom$

\end{itemize}
and, of course, the series (\ref{totalsolnn}) will only converge if $\nat \cdot \pim \Oom$ is positive definite. 

The natural candidate for the wavefunction of a field with charge $q=+1$ is then $\psi^{\vec{j}, \nat}(\vec{z} + \vec{\z}, \Oom)$,  $\vec{\z}$ now representing the Wilson lines. This wavefunction satisfies the differential equations
\beq
\begin{array}{c}
D_a \psi^{\vec{j}, \nat}(\vec{z} + \vec{\z}, \Oom)\  =\  0, \quad \quad \forall a \\
D_a\ =\ \bar{\p}_a + \frac{\pi}{2} \left( [\nat \cdot (\vec{z} + \vec{\z})] \cdot (\pim \Oom)^{-1} \right)_a
\end{array}
\label{diracn}
\eeq
and hence satisfies Dirac equation. If $\nat \cdot \pim \Oom$ is positive definite, then it can be seen that these eigenfunctions satisfy the orthonormality condition
\beq
\int_{T^{2n}} (\psi^{\vec{j}, \nat})^* \psi^{\vec{k}, \nat} = \d_{\vec{j},\vec{k}},
\label{normn}
\eeq
by just fixing the normalization constant to
\beq
\cn_n = \left( 2^n | {\rm det} \nat | \cdot {\rm det} (\pim \Oom) \right)^{1/4} \cdot {\rm Vol} (T^{2n})^{-1/2}
\label{normn2}
\eeq
where ${\rm Vol} (T^{2n}) = \left|{\rm det\ } {\bf h} \right| \cdot {\rm det} (\pim \Oom)$.

In general, the integer-valued matrix $\nat$ will encode the quanta of the magnetic flux. To see this more precisely, let us compare the transformation properties (\ref{trans1psin}) with the transition functions (\ref{transitionnaT2n}) in the simple case of $T^4$. By identifying
\beq
\left( \frac{x^1}{a_1},  \frac{x^2}{a_2}, \frac{x^3}{a_3}, \frac{x^4}{a_4} \right) 
\sim
(x^1, y^1, x^2, y^2)
\label{id}
\eeq
we recover the same transformations in both sides if we impose
\beq
\nat = 
\left(
\begin{array}{cc}
n_{12} & n_{32} \\ n_{14} & n_{34}
\end{array}
\right),
\quad \quad
n_{24} = n_{13} = 0
\label{id2}
\eeq

Now, as proven in \cite{vanbaal2}, the degeneracy of states (i.e., the number of chiral fermions) is given by the absolute value of
\beq
{\rm Pf} (n) := \frac{1}{8} \eps^{ijkl} n_{ij} n_{kl} = {\rm det\ } \nat
\label{pf}
\eeq
at least in this particular case. In general, ${\rm det\ } \nat$ will give us the chiral spectrum obtained after turning on the flux: $|{\rm det } \nat|$ give us the degeneracy, whereas sign$({\rm det } \nat)$ give us the chirality.

Finally, notice that in order to have a well-defined wavefunction, the matrix $\nat$ and $\Oom$ must satisfy the following constraints
\begin{itemize}

\item $n_{24} = n_{13} =  0$

\item $\left( \nat \cdot \pim \Oom \right)^t = \nat \cdot \pim \Oom$

\item $\nat \cdot \pim \Oom > 0$

\end{itemize}
The first of this constraints is not such, since it can be satisfied by using the $SL(4,\inte)$ symmetry of $T^4$. The other two can be understood in terms of supersymmetry, in particular from the requirement that $F$ is a $(1,1)$-form (see Appendix \ref{SUSYap}). Indeed, in the case of $T^{2n}$, the sufficient and necessary conditions for (\ref{fluxnaT2n}) to be a $(1,1)$-form are known as Riemann Conditions \cite{gh}, which are
\beq
-i \Pi^t \cdot Q \cdot \bar{\Pi} =
\left(
\begin{array}{cc}
H & 0 \\
0 & -H^t
\end{array}
\right) 
\label{riemann1}
\eeq
$H$ being a $n \times n$ positive definite matrix. In our case, the matrices $\Pi$ and $Q$ are given by
\beq
\Pi =
\left(
\begin{array}{cc}
{\bf 1}_n & \Oom \\
{\bf 1}_n & \bar{\Oom} 
\end{array}
\right)^{-1}
=
\frac{i}{2}
\left(
\begin{array}{cc}
\bar{\Oom} & - \Oom \\
- {\bf 1}_n & {\bf 1}_n
\end{array}
\right)
\cdot(\pim \Oom)^{-1}, \quad \quad
Q = 
\left(
\begin{array}{cc}
 &  {\bf N}^t \\
- {\bf N} & 
\end{array}
\right)
\label{defmat}
\eeq
and the Riemann Conditions amount to
\beq
\begin{array}{c}\vspace*{.1cm}
\left(\nat \cdot \Oom \right)^t =  \nat \cdot \Oom \\
H =  \oh\ \nat^t \cdot \pim \Oom^{-1} > 0
\end{array}
\label{riemann2}
\eeq
which clearly imply the constraints above. Finally, they also imply that, up to a phase, we can rewrite our wavefunction as
\beq
\psi^{\vec{j}, \nat}(\vec{z} + \vec{\z}, \Oom) =
\cn \cdot e^{- 2\pi \pim (\vec{z} + \vec{\z}) \cdot H \cdot \pim (\vec{z} + \vec{\z})}
\cdot 
\vt
\left[
\begin{array}{c}
\vec{j} \\ 0
\end{array}
\right]
\left(\nat \cdot ( \vec{z} + \vec{\z}) | \nat \cdot \Oom \right).
\label{totalsolnn2}
\eeq

\section{Computing Yukawa couplings}

Once that we have derived both the fermionic and 
bosonic internal wavefunctions, and expressed them as an 
orthonormal basis, we are in position for computing 
the 3-point functions between them by using the general 
formula (\ref{yukfinal}). In this section we perform 
such computation for the toroidal compactifications 
previously considered. We will first focus on the 
simple case of $T^2$ and then generalize our results 
for higher-dimensional tori.

\subsection{Computing Yukawas on a $T^2$}

In order to get non-trivial Yukawa couplings we need to start 
with three gauge factors, allowing for three different types
of bifundamental matter fields.
Let us compute the Yukawa couplings in the simplest case, namely magnetic flux compactifications in $T^2$. In order to have non-trivial Yukawa couplings, we need to consider a flux of the form
\beq
F_{z\bar{z}} = {\pi i\over \pim \tau}
\left(
\begin{array}{ccc}
\frac{m_a}{n_a} {\bf 1}_{n_a} & & \\
& \frac{m_b}{n_b} {\bf 1}_{n_b} & \\
& & \frac{m_c}{n_c} {\bf 1}_{n_c} 
\end{array}
\right),
\label{fluxnatrif}
\eeq
with $n_\a \in \nat^+, m_\a \in \inte$, $\a = a, b, c$. As explained in Section 4, the initial gauge group is broken to $U(p_a) \times U(p_b) \times U(p_c)$, where $p_\a = {\rm g.c.d.}(n_\a,m_\a)$. Notice that, with the definitions 
\beqa
I_{\a\b} & \equiv &  - n_\a m_\b + n_\b m_\a, \label{inter}\\
\tilde{I}_{\a\b} & \equiv  & I_{\a\b}/n_\a n_\b,
\eeqa
the `differences of fluxes' $\tilde{I}_{\a\b}$ satisfy $\tilde{I}_{ab} + \tilde{I}_{bc} + \tilde{I}_{ca} = 0$. This implies that one $|\tilde{I}_{\a\b}|$ is bigger than the other two. Let us suppose that this is the case for $\tilde{I}_{bc}$, hence $|\tilde{I}_{bc}| = |\tilde{I}_{ab}| + |\tilde{I}_{ca}|$. This asymmetry will show up in the general formula for Yukawa couplings. 

We now have two possibilities, depending on the sign of $\tilde{I}_{bc}$. By the results of the previous sections, the fermionic wavefunction $\Psi$ is given by
\beq
\Psi =
\left(
\begin{array}{c}
\psi_+ \\  \psi_-
\end{array}
\right), \quad \quad
\left\{
\begin{array}{c} \vspace*{.5cm}
\psi_+ =
\left(
\begin{array}{ccc}
{\rm const.} & \Phi^{i, I_{ab}} & 0 \\
0 & {\rm const.} & 0 \\
\Phi^{j, I_{ca}} & \Phi^{k, I_{cb}} & {\rm const.}
\end{array}
\right)
\quad \quad {\rm if\ } \tilde{I}_{bc} < 0 \\
\psi_+ =
\left(
\begin{array}{ccc}
{\rm const.} & 0 & \Phi^{j, I_{ac}} \\
\Phi^{i, I_{ba}} & {\rm const.} &   \Phi^{k, I_{bc}} \\
0 & 0 & {\rm const.}
\end{array}
\right)
\quad \quad {\rm if\ } \tilde{I}_{bc} > 0
\end{array}
\right.
\label{fermionwave}
\eeq
with $\psi_- = (\psi_+)^\dag$, and where ${\rm const.} = {\ca}^{-1/2}$
 are  the gaugino's wavefunctions. 
The chiral wavefunctions $\Phi^{j, I_{\a\b}}$ $j= 0, \dots, |I_{\a\b}| -1$ 
have been computed in Section 4 and, in the particular case 
of Abelian Wilson lines, reduce to $\psi^{j, \tilde{I}_{\a\b}} 
\cdot M_{n_\a, n_\b}$, where $\psi^{j, \tilde{I}_{\a\b}}$ are 
the wavefunctions computed in Section 3 and $M_{n_\a, n_\b}$ is a 
$n_\a \times n_\b$ matrix taking care of gauge quantum numbers.

The general formulae (\ref{yukfinal}), (\ref{yukfinalap}) imply that the Yukawa couplings involving chiral massless fermions are computed by evaluating the integrals
\beq
\int_{T^2} dzd\bar{z}\ \Tr \left\{\psi_+ \cdot [\phi_-,\psi_+]\right\} \quad {\rm and} \quad \int_{T^2} dzd\bar{z}\ \Tr \left\{\psi_- \cdot [\phi_+,\psi_-]\right\}, 
\label{yukfirst}
\eeq
which are CPT conjugates of each other. Here $\phi_\pm$ are the wavefunctions of the bosonic fluctuations of the higher-dimensional gauge field $A_M$, with helicity $\pm 1$ in the internal coordinates of $T^2$. Notice that these are the only terms involving massless fermions allowed by Lorentz invariance in the internal coordinates. By our previous results we find that $\phi_\pm \sim \psi_\pm$ and that it corresponds to the lightest (in fact, tachyonic) $D=4$ scalar. Evaluating these expressions, we find that the Yukawa coupling involving left-handed fermions is given by
\beq
Y_{ijk} = \left\{
\begin{array}{cc}\vspace*{.3cm}
\sig_{abc}\ g \int_{T^2} dzd\bar{z}\ \Tr \left\{ \Phi^{i, I_{ab}} \cdot \Phi^{j,I_{ca}} \cdot \left(\Phi^{k,I_{cb}}\right)^\dag\right\} & \quad {\rm if\ } \tilde{I}_{bc} < 0 \\
\sig_{abc}\ g \int_{T^2} dzd\bar{z}\ \Tr \left\{ \Phi^{i, I_{ba}} \cdot \Phi^{j,I_{ac}} \cdot \left(\Phi^{k,I_{bc}}\right)^\dag\right\} & \quad {\rm if\ } \tilde{I}_{bc} > 0
\end{array}
\right.
\label{intT2}
\eeq
where we have restored the dependence of the 3-point functions on the gauge coupling constant $g$, by considering both (\ref{yukfinalap}) and the normalization  of the fields (\ref{normKK}). Here $\sig_{abc} = {\rm sign}(\tilde{I}_{ab}\tilde{I}_{bc}\tilde{I}_{ca}) = {\rm sign}(I_{ab}I_{bc}I_{ca})$ is a sign coming from fermionic statistics. From the point of view of $D=4$ physics, this term will yield a coupling between two chiral fermions of opposite chiralities, transforming in the $(p_a,\bar{p}_b)$, $(p_a,\bar{p}_c)$ bifundamental representations, and a complex tachyon in the $(p_{b},\bar{p}_{c})$ representation.\footnote{No more couplings are allowed from the choices made above and the action (\ref{action}) in $D=6$. In compactifications of higher dimensional theories, or those with a richer spectrum such as (\ref{multiplets}), however, more Yukawa between chiral fermions and scalars in different representations will appear. The computation of the 3-point functions in those cases will nevertheless be similar to the one we perform here.}

Again, the computation of the integrals (\ref{intT2}) is technically simpler in the case where only Abelian Wilson lines are present in the compactification. The computation of the 3-point function is simplified by the use of $\vt$-function identities. The case with non-Abelian Wilson lines has nevertheless the virtue that it can distinguish between two physically relevant quantities, which are the multiplicity of the spectrum, given by $I_{\a\b}$, and the slope of the flux (see Appendix \ref{SUSYap}), proportional to $\tilde{I}_{\a\b}$. The differentiation of both quantities will turn out to be quite relevant when interpreting our results from the point of view of the effective action.

\subsubsection{Abelian Wilson lines}

Let us first consider the case which only involves Abelian Wilson lines. That is, we consider that $(n_\a, m_\a) = n_\a(1, s_\a)\ s_\a \in \inte$. As shown in Section 4, turning on the flux (\ref{fluxnatrif}) provokes the gauge group breaking $U(N) \raw U(n_a) \times U(n_b) \times U(n_c)$, $n_a + n_b + n_c = N$, and only Abelian Wilson line is needed for having a well-defined gauge connection. Moreover, the degeneracy of chiral fermions in bifundamentals $(n_\a, \bar{n}_\b)$ is not given by $I_{\a\b}$ in (\ref{inter}) but rather by\footnote{$I_{\a\b}$ gives the degeneracy of chiral fermions {\em before} arranging them in bifundamental representations. $\ci_{\a\b}$ is the topological invariant to be identified with the intersection number in the T-dual picture of intersecting D-branes. Both numbers agree when $n_\a = 1,\ \forall \a$.} $\ci_{\a\b} = s_\a - s_\b$. Finally, the matrix-valued wavefunctions $\Phi^{i,I_{\a\b}}$ reduce to a $n_\a \times n_\b$ matrix times the wavefunctions $\psi^{i,\ci_{\a\b}}$ defined in (\ref{totalsoln}). The Yukawas then read
\beq
Y_{ijk} = \sig_{abc}\ g \int_{T^2} dzd\bar{z}\ \psi^{i, \ci_{ab}}(z+\z_{ab}) \cdot \psi^{j,\ci_{ca}}(z+\z_{ca}) \cdot \left(\psi^{k,\ci_{cb}}(z+\z_{cb})\right)^* 
 \label{intT2a}
\eeq
where we have chosen $\ci_{bc} < 0$ for definiteness. The Abelian Wilson lines $\z_{\a\b}$ are defined by (\ref{defa2}), but with the substitutions $m_\a \mapsto s_\a$ and $I_{\a\b} \mapsto \ci_{\a\b}$.

Now, in order to compute the integral (\ref{intT2a}), we will make use of the `addition formula' for theta functions, taken from \cite{tata}, Proposition II.6.4. (p. 221), and which was crucial in the categorical mirror symmetry computations of \cite{pz}. This formula says
\beqa 
\vt
\left[
\begin{array}{c}
\frac{r}{N_1} \\ 0
\end{array}
\right]
(z_1, \tau N_1)
\cdot
\vt
\left[
\begin{array}{c}
\frac{s}{N_2} \\ 0
\end{array}
\right]
(z_2, \tau N_2)  & = &
\sum_{m \in \inte_{N_1 + N_2}}
\vt
\left[
\begin{array}{c}
\frac{r + s + N_1m}{N_1+ N_2} \\ 0
\end{array}
\right]
(z_1+ z_2, \tau (N_1 + N_2)) \label{addition}  \\ \nonumber
& \cdot & 
\vt
\left[
\begin{array}{c}
\frac{N_2 r - N_1 s + N_1N_2 m }{N_1 N_2(N_1+ N_2)} \\ 0
\end{array}
\right]
(z_1N_2 - z_2 N_1, \tau N_1 N_2 (N_1 + N_2))
\eeqa

This identity is particularly useful to our purposes, since the wavefunctions $\psi^{j,N}$ are proportional to $\vt$-functions of this form. Indeed,
\beqa
 \psi^{i, \ci_{ab}} (z + \z_{ab}) \cdot  
 \psi^{j, \ci_{ca}} (z + \z_{ca})  & = &
 \ca^{-1} \left(2 \pim \tau \right)^{1/2}  |\ci_{ab} \ci_{ca}|^{1/4} 
\\ \nonumber 
 & \cdot & 
e^{i{\pi \ci_{ab}\over \pim \tau} (z + \z_{ab}) \pim(z + \z_{ab})}
e^{i{\pi \ci_{ca}\over \pim \tau} (z + \z_{ca}) \pim(z + \z_{ca})/\ca} \\ \nonumber
& \cdot & 
\vt
\left[
\begin{array}{c}
\frac{i}{\ci_{ab}} \\ 0
\end{array}
\right]
((z + \z_{ab})\ci_{ab}, \tau \ci_{ab})
\cdot
\vt
\left[
\begin{array}{c}
\frac{j}{\ci_{ca}} \\ 0
\end{array}
\right]
((z + \z_{ca})\ci_{ca}, \tau \ci_{ca}),
\label{prod}
\eeqa
so we can compute the product on the third line by using the formula (\ref{addition}). We only need to identify
\beq
\begin{array}{ccc}
r = i &  & s = j \\
N_1 = \ci_{ab} & & N_2 = \ci_{ca} \\
z_1 = (z + \z_{ab}) \ci_{ab} & & z_2 = (z + \z_{ca}) \ci_{ca}
\end{array}
\label{ident}
\eeq

Let us first consider the case without any Wilson line, i.e., set $\z_{ab} = \z_{ca} = 0$. We obtain
\beqa
& &
\vt
\left[
\begin{array}{c}
\frac{i}{\ci_{ab}} \\ 0
\end{array}
\right]
(z \ci_{ab}, \tau \ci_{ab})
\cdot
\vt
 \left[
\begin{array}{c}
\frac{j}{\ci_{ca}} \\ 0
\end{array}
\right]
(z \ci_{ca}, \tau \ci_{ca}) = \label{prod2} \\ \nonumber
& &
\sum_{m \in \inte_{\ci_{bc}}} 
\vt
\left[
\begin{array}{c}
\frac{i + j + \ci_{ab}m}{-\ci_{bc}} \\ 0
\end{array}
\right]
(z (-\ci_{bc}), \tau (-\ci_{bc}))
\cdot
\vt
\left[
\begin{array}{c}
\frac{\ci_{ca} i - \ci_{ab} j + \ci_{ab} \ci_{ca} m }{-\ci_{ab} \ci_{ca} \ci_{bc}} \\ 0
\end{array}
\right]
(0, \tau (-\ci_{ab} \ci_{bc} \ci_{ca}))
\eeqa
where we have made use of the fact that $\ci_{ab} + \ci_{ca} = \ci_{cb} = - \ci_{bc}$. Formula (\ref{prod2}) implies that
\beqa
\psi^{i, \ci_{ab}} (z)  \cdot 
\psi^{j, \ci_{ca}} (z) \quad & = &
\ca^{-1/2} \left(2 \pim \tau \right)^{1/4}  \left|{\ci_{ab} \ci_{ca} \over \ci_{bc}}\right|^{1/4}
\label{prod3}
\\ \nonumber
& \cdot & 
\sum_{m \in \inte_{\ci_{bc}}} \psi^{i+j+\ci_{ab}m,\ci_{cb}}(z) 
\cdot
\vt
\left[
\begin{array}{c}
\frac{\ci_{ca} i - \ci_{ab} j + \ci_{ab} \ci_{ca} m }{-\ci_{ab} \ci_{bc} \ci_{ca}} \\ 0
\end{array}
\right]
(0, \tau |\ci_{ab} \ci_{bc} \ci_{ca}|)
\eeqa

Now, the usefulness of these identities for performing the integration
in eq.(\ref{intT2a}) is that now the second theta function in the above
expression no longer depends on the coordinate $z$ and  
may be factored out from the integration. We are thus left we an integration
over two theta-functions which may be easily computed by using orthonormality
of wavefunctions, as we show below.
Notice as well that (\ref{prod3}) is a product of two wavefunctions, each one with a `weight' $\ci_{ab}$ and $\ci_{ca}$, expanded in a basis of a third class of wavefunctions, now with a `weight' $\ci_{cb} = \ci_{ca} + \ci_{ab}$. Moreover this basis behaves under gauge transformations as the third wavefunction involved in the Yukawa coupling (\ref{intT2a}), more precisely as its hermitian conjugate. This kind of identity is totally general for magnetized compactifications, and comes from the simple fact that if we can understand a Yukawa coupling as an integral of three wavefunctions
\beq
Y_{ijk} = \int_{\cam_{2n}} \Tr \left(\phi_{\a\b}^i \circ \phi_{\b\g}^j \circ \phi_{\g\a}^k \right)
\label{3pointdim}
\eeq
with the trace performed over gauge and internal Lorentz indices, then the integrand must be invariant under both gauge and Lorentz transformations. In particular, $\phi_{\g\a}^k$ must compensate the gauge transformations of $\phi_{\a\b}^i \circ \phi_{\b\g}^j$.  This is indeed the case, as can be easily seen, e.g.,  from the gauge transformation properties of $(N_\a, \bar{N}_\b)\cdot (N_\b, \bar{N}_\g) \sim (N_\g, \bar{N}_\a)^\dag$. This allows us to write the product of wavefunctions as\footnote{As is stands the expression (\ref{OPE}) is of field theoretical nature. Recall, however, that there is an underlying string theory in the whole construction, where the chiral fields $\phi^i$ will be represented by vertex operators ${\co}^i$. The expansion (\ref{OPE}) is then understood as the field-theoretical version of the OPE 
\begin{center}
${\co}^i (\om_1) \cdot {\co}^j (\om_2) = \sum_k C^{ij}_k (\om_2 - \om_1) {\co}^k (\om_2)$
\end{center}
in the underlying CFT (here $\om_i$ is a world-sheet coordinate).}
\beq
\phi_{\a\b}^i \circ \phi_{\b\g}^j = \sum_k c^{ij}_k \left(\phi_{\g\a}^k\right)^\dag
\label{OPE}
\eeq
where $k$ runs over an orthonormal basis of wavefunctions transforming in the representation $(N_\g, \bar{N}_\a)$. The same facts apply to Lorentz indices. In the particular case of $T^2$ compactifications, we see from (\ref{prod3}) that the coefficients $c^{ij}_k$ are given by $\vt$-functions.

Now, by considering the orthonormality condition (\ref{norm}), we can compute (\ref{intT2a}) to be
\beqa \nonumber
Y_{ijk}^{(\z=0)} & = & \sig_{abc}\ g
\ca^{-1/2} \left(2 \pim \tau \right)^{1/4} \left|{\ci_{ab} \ci_{ca} \over \ci_{bc}}\right|^{1/4} \\ \nonumber
& & \cdot
\sum_{m \in \inte_{\ci_{bc}}}
\d_{k,i+j+\ci_{ab}m} \cdot
\vt
\left[
\begin{array}{c}
\frac{\ci_{ca} i - \ci_{ab} j + \ci_{ab} \ci_{ca} m }{-\ci_{ab} \ci_{bc} \ci_{ca}} \\ 0
\end{array}
\right]
(0, \tau |\ci_{ab} \ci_{bc} \ci_{ca}|) \\
& = & 
\left(\frac{2 \pim \tau}{\ca^2} \right)^{1/4}  \left|\frac{\ci_{ab} \ci_{ca}}{\ci_{bc}}\right|^{1/4}
\cdot
 \vt
\left[
\begin{array}{c}
-\left(\frac{j}{\ci_{ca}} + \frac{k}{\ci_{bc}} \right)/\ci_{ab} \\ 0
\end{array}
\right]
(0, \tau |\ci_{ab} \ci_{bc} \ci_{ca}|)
\label{int3}
\eeqa

So we find that Yukawa couplings are proportional to theta functions, as was already pointed out in the T-dual picture \cite{yukis}. In order to compare both results, let us express the the $\vt$-function characteristic in a more symmetric form. Notice that
\beqa \nonumber
\frac{i}{\ci_{ab}} - \frac{j}{\ci_{ca}} + \frac{k}{\ci_{bc}} & = & \frac{k - j}{\ci_{ab}} + \frac{j}{\ci_{ca}} + \frac{k}{\ci_{bc}} \\
= \frac{1}{\ci_{ab}} \left( - \frac{j \ci_{bc}}{\ci_{ca}} + \frac{k \ci_{ca}}{\ci_{bc}} \right) & = &
- \frac{1}{\ci_{ab}} \left( \frac{j'}{\ci_{ca}} + \frac{k'}{\ci_{bc}} \right)
\label{hidden}
\eeqa
where in the first equality we have made use of $i = k - j$ mod $\ci_{ab}$, implicit in (\ref{int3}), and in the last one we have made a redefinition of the indices $j$ and $k$. This redefinition is always possible if g.c.d.$(\ci_{ab},\ci_{bc},\ci_{ca}) = 1$. In Section 7 we will see how this theta function characteristic matches with the result obtained in \cite{yukis}.

The inclusion of (Abelian) Wilson lines modifies the previous result to
\beqa \nonumber
Y_{ijk} & = & \sig_{abc}\ g
\left(\frac{2 \pim \tau}{\ca^2} \right)^{1/4}  \left|\frac{\ci_{ab} \ci_{ca}}{\ci_{bc}}\right|^{1/4}
\cdot
e^{i\pi \left(\ci_{ab} \z_{ab} \pim \z_{ab} + \ci_{bc} \z_{bc} \pim \z_{bc} + \ci_{ca} \z_{ca} \pim \z_{ca} \right)/\pim \tau}
 \\
& \cdot &
 \vt
\left[
\begin{array}{c}
-\left(\frac{j}{\ci_{ca}} + \frac{k}{\ci_{bc}} \right)/\ci_{ab} \\ 0
\end{array}
\right]
(-\ci_{ab} \ci_{ca} (\z_{ca} - \z_{ab}), \tau |\ci_{ab} \ci_{bc} \ci_{ca}|)
\label{int4}
\eeqa
but notice that, since $\ci_{ab} + \ci_{bc} + \ci_{ca} = 0$,
\beq
-\ci_{ab} \ci_{ca} (\z_{ca} - \z_{ab}) = \z_a s_a \ci_{bc} + \z_c s_c \ci_{ab} + \z_b s_b \ci_{ca},
\label{wilsonred}
\eeq
Actually, it turns out that the whole Wilson line dependence of the Yukawa coupling is a function of the linear combination of Wilson lines (\ref{wilsonred}). In order to see this, let us first express everything in terms of the following redefinition of complex Wilson line
\beq
\tilde{\z}_\a = s_\a \z_\a, \quad \a = a,b,c.
\label{redef}
\eeq

The Yukawa couplings now read
\beqa \nonumber
Y_{ijk} & = &  \sig_{abc}\ g
\left(\frac{2 \pim \tau}{\ca^2} \right)^{1/4}  \left|\frac{\ci_{ab} \ci_{ca}}{\ci_{bc}}\right|^{1/4}
\cdot
e^{i\pi \left(\frac{\tilde{\z}_{ab} \pim \tilde{\z}_{ab}}{\ci_{ab}} + \frac{\tilde{\z}_{bc} \pim \tilde{\z}_{bc}}{\ci_{bc}}
+ \frac{\tilde{\z}_{ca} \pim \tilde{\z}_{ca}}{\ci_{ca}}\right)/\pim \tau}
\\
& \cdot &
\vt
\left[
\begin{array}{c}
\d_{ijk} \\ 0
\end{array}
\right]
\left( \tilde{\z}, \tau |\ci_{ab} \ci_{bc} \ci_{ca}| \right)
\label{int5}
\eeqa
where we have defined
\beqa
\d_{ijk} & = & \frac{i}{\ci_{ab}} + \frac{j}{\ci_{ca}} + \frac{k}{\ci_{bc}} \\
\tilde{\z}_{ab} & = & \tilde{\z}_a - \tilde{\z}_b, \quad {\rm etc.} \\
\tilde{\z} & = & \ci_{ab}  \tilde{\z}_c + \ci_{bc} \tilde{\z}_a + \ci_{ca} \tilde{\z}_b
\label{defin}
\eeqa
Now notice that the exponential factor in (\ref{int5}) can be rewritten as:
\beq
e^{i\pi \left( \tilde{\z}_\a g_{\a\b} \pim \tilde{\z}_\b \right) /\pim \tau}
\label{expo}
\eeq
where $\a, \b = a, b, c$, and $g$ is a symmetric matrix given by
\beq
g = |\ci_{ab}\ci_{bc}\ci_{ca}|^{-1}
\left(
\begin{array}{ccc}
\ci_{bc}^2 & \ci_{bc}\ci_{ca} & \ci_{ab}\ci_{bc} \\
\ci_{bc}\ci_{ca} & \ci_{ca}^2 & \ci_{ca}\ci_{ab} \\
\ci_{ab}\ci_{bc} & \ci_{ca}\ci_{ab} & \ci_{ab}^2
\end{array}
\right)
\label{matrix}
\eeq
where we have again used the fact that $\ci_{ab} + \ci_{bc} + \ci_{ca} = 0$ and that $\ci_{ab} \ci_{bc} \ci_{ca} < 0$. This matrix is singular, having only one non-zero eigenvalue:
\beq
g \cdot {1 \over \sqrt{\k}} 
\left(
\begin{array}{c}
\ci_{bc} \\ \ci_{ca} \\ \ci_{ab}
\end{array}
\right)
=
|\ci_{ab}\ci_{bc}\ci_{ca}|^{-1}
\cdot
{1 \over \sqrt{\k}} 
\left(
\begin{array}{c}
\ci_{bc} \\ \ci_{ca} \\ \ci_{ab}
\end{array}
\right)
\label{eigenwilson}
\eeq
where $\k = \ci_{ab}^2 + \ci_{bc}^2 + \ci_{ca}^2$. This implies that the whole quantity in (\ref{expo}) must depend on the combination $\tilde{\z}$ above and, in particular, that we can rewrite (\ref{int5}) as
\beq
Y_{ijk} = \sig_{abc}\ g
\left(\frac{2 \pim \tau}{\ca^2} \right)^{1/4}  \left|\frac{\ci_{ab} \ci_{ca}}{\ci_{bc}}\right|^{1/4}
\cdot
e^{H(\tilde{\z},\tau)/2}
\cdot
 \vt
\left[
\begin{array}{c}
\d_{ijk} \\ 0
\end{array}
\right]
\left( \tilde{\z}, \tau |\ci_{ab} \ci_{bc} \ci_{ca}| \right)
\label{int6}
\eeq
where
\beq
H(\tilde{\z},\tau) = 2\pi i |\ci_{ab} \ci_{bc} \ci_{ca}|^{-1} {\tilde{\z} \cdot \pim \tilde{\z} \over \pim \tau}
\label{H}
\eeq
We obtain similar results for $\tilde{I}_{bc} > 0$.
Note that the gauge coupling $g$ in $D=6$ has dimension of length,
so that the Yukawa coupling in $D=4$ is indeed dimensionless, as it should.

\subsubsection{Non-Abelian Wilson lines}

Let us now turn to the case where the flux in (\ref{fluxnatrif}) satisfies $p_\a = {\rm g.c.d.}(n_\a,m_\a) = 1 < n_\a$, so that non-Abelian Wilson lines have to be introduced and the total rank $N = n_a + n_b + n_c$ is reduced. The chiral fields are now expressed in terms of the matrix-valued wavefunctions $\Phi^{i,I_{\a\b}}$ defined in (\ref{totalsolnbif2}), and the integral (\ref{intT2}) is given by
\beqa
Y_{ijk} & = & \sig_{abc}\ g \int_{T^2} dzd\bar{z}\ \Tr \left\{ \Phi^{j, I_{ca}} \cdot \Phi^{i,I_{ab}} \cdot \left(\Phi^{k,I_{cb}}\right)^\dag\right\} \\ \nonumber
& = & \sig_{abc}\ g \int_{T^2} dzd\bar{z} \sum_{k_a,k_b,k_c} \phi_{k_a,k_b}^{i,I_{ab}} \phi_{k_c,k_a}^{k,I_{ca}} 
\left(\phi_{k_c,k_b}^{j,I_{cb}}\right)^* \\ \nonumber
& = & \sig_{abc}\ g \int_{T^2} dzd\bar{z} \sum_{l=1}^{n_an_bn_c} \phi_{l,l}^{i,I_{ab}} \phi_{l,l}^{j,I_{ca}} 
\left(\phi_{l,l}^{k,I_{cb}}\right)^* 
\label{intT2na}
\eeqa
where in the third line we have also assumed that ${\rm g.c.d.}(n_a,n_b,n_c) = 1$.

The computation of this integral is harder than (\ref{intT2}), since we cannot use the theta-function identity (\ref{addition}) and the integral must be made by brute-force computation. Of some help is the fact that
\beq
\phi_{l, l}^{i,I_{ab}} \phi_{l, l}^{j,I_{ca}} (\phi_{l, l}^{k,I_{cb}})^* (z+\tau) = 
\phi_{l+1, l+1}^{i,I_{ab}} \phi_{l+1, l+1}^{j,I_{ca}} (\phi_{l+1, l+1}^{k,I_{cb}})^* (z)
\label{trick}
\eeq
And hence we can evaluate (\ref{intT2na}) by fixing $l=0$ and integrating over a torus of complex structure $n_an_bn_c\tau$, instead of performing the summation over $l$. We obtain the result
\beq
Y_{ijk} = \sig_{abc}\ g 
\left(\frac{2 \pim \tau}{\ca^2} \right)^{1/4}  \left|\frac{\tilde{I}_{ab} \tilde{I}_{ca}}{\tilde{I}_{bc}}\right|^{1/4}
\cdot
e^{H/2}
\cdot
 \vt
\left[
\begin{array}{c}
\d_{ijk} \\ 0
\end{array}
\right]
\left( \tilde{\z}, \tau |I_{ab} I_{bc} I_{ca}| \right) 
\label{guess}
\eeq
where the Wilson lines have been redefined as $\tilde{\z}_\a = m_\a \z_\a$, instead of (\ref{redef}). The exponential prefactor is now given by
\beqa \nonumber
H & = & 2\pi i |I_{ab} I_{bc} I_{ca}|^{-1} {\tilde{\z} \cdot \pim \tilde{\z} \over \pim \tau} \\
& = & 2\pi i \left(\tilde{I}_{ab} \z_{ab} \pim \z_{ab} + \tilde{I}_{bc} \z_{bc} \pim \z_{bc} + \tilde{I}_{ca} \z_{ca} \pim \z_{ca} \right)/\pim \tau
\label{Hna}
\eeqa

The expression (\ref{guess}) is quite similar to the one obtained in the case of Abelian Wilson lines. It is more general and contains more information, though, in the sense that we can now distinguish between the degeneracy of chiral states $I_{\a\b}$ (intersection number) and the quantity $\tilde{I}_{\a\b}$ ($\mu$-slope of the flux), which were both identified with $\ci_{\a\b}$ in the expression (\ref{int6}).

\subsection{Higher dimensional tori}

The computation of the Yukawa couplings for $2n$-dimensional tori can be easily deduced from the results obtained for $T^2$. In the particular case that $T^{2n}$ has the factorizable geometry (\ref{decompT2n}) the chiral matter wavefunctions are given by (\ref{psiaT2nf}) or (\ref{psiT2nf}). That is, by a product of $n$ wavefunctions of the form $\Phi^{I_{ab}^{(r)}}$ or $(\Phi^{I_{ab}^{(r)}})^*$ obtained in $T^2$ compactifications, the choice of $\Phi$ or $\Phi^*$ depending on the sign of $I_{ab}^{(r)}$. This implies that we can decompose the integral in (\ref{yukfinal}) as a product of $n$ integrals of the form (\ref{intT2}). More precisely, each integral will be given by the analogue of (\ref{intT2}) for the $r^{th}$ two-torus if $\sig_{abc}^{(r)} = {\rm sign } (I_{ab}^{(r)} I_{bc}^{(r)} I_{ca}^{(r)}) > 0$ and by its complex congujate if $\sig_{abc}^{(r)} < 0$. The final Yukawa coupling will be given by a product of $n$ contributions, one for each $T^2$, which are either of the form (\ref{guess}), either by its complex conjugate.

We then find that the Yukawa couplings for factorizable $T^{2n}$ magnetized compactifications are given by
\beq
Y_{ijk} = \sig_{abc}\ {g} \prod_{r=1}^{n}
\left(\frac{2 \pim \tau^{(r)}}{(\ca^{(r)})^2} \right)^{1/4}
\left|
\frac{\tilde{I}_{1}^{(r)}
\tilde{I}_{2}^{(r)}}{\tilde{I}_{1}^{(r)} + \tilde{I}_{2}^{(r)}}
\right|^{1/4}
e^{H^{(r)}/2}\
 \vt
\left[
\begin{array}{c}
\d_{ijk}^{(r)} \\ 0
\end{array}
\right]
\left( \tilde{\z}^{(r)}, \tau^{(r)} (I_{ab}^{(r)} I_{bc}^{(r)} I_{ca}^{(r)}) \right)
\label{yukT2n}
\eeq
where we must perform the substitution $J^{(s)} \mapsto \bar{J}^{(s)}$ and $\tilde{\z}^{(s)} \mapsto \bar{\tilde{\z}}_a^{(s)}$, whenever $\sig_{abc}^{(s)} = -1$. Here $\ca^{(r)}$, $\tau^{(r)}$ are the area and complex structure of the $r^{\rm th}$ $T^2$ in (\ref{decompT2n}) and 
\beqa
I_{ab} & = & \prod_{r=1}^n I_{ab}^{(r)},\\
i & = & \left(i^{(1)}, i^{(2)}, \dots, i^{(n)}\right), \\
i^{(r)} & = & 0, \dots, |I_{ab}^{(r)}| -1,
\label{totalinter}
\eeqa
is the total multiplicity of chiral fermions and scalars in the $ab$ sector, etc., and the appropriate labeling of them, with a different index $i^{(r)}$ for each $T^2$. On the other hand, $|\tilde{I}_1^{(r)}|$ and $|\tilde{I}_2^{(r)}|$ are the two smallest numbers among $|\tilde{I}_{ab}^{(r)}|$, $|\tilde{I}_{bc}^{(r)}|$ and $|\tilde{I}_{ca}^{(r)}|$. Finally,
\beqa
\sig_{abc} & = & \prod_r \sig^{(r)}_{abc} = 
\prod_r {\rm sign } (I_{ab}^{(r)} I_{bc}^{(r)} I_{ca}^{(r)}) = 
{\rm sign } (I_{ab}I_{bc}I_{ca}) \\
\d_{ijk}^{(r)} & = & \frac{i^{(r)}}{I_{ab}^{(r)}} 
+ \frac{j^{(r)}}{I_{ca}^{(r)}} 
+ \frac{k^{(r)}}{I_{bc}^{(r)}} \\
H^{(r)} & = & 2\pi i |I_{ab}^{(r)} I_{bc}^{(r)} I_{ca}^{(r)}|^{-1} {\tilde{\z}^{(r)} \cdot \pim \tilde{\z}^{(r)} \over \pim \tau^{(r)}} \\
\tilde{\z}^{(r)} & = & I_{ab}^{(r)}  \tilde{\z}_c^{(r)} + I_{bc}^{(r)} \tilde{\z}_a^{(r)} + I_{ca}^{(r)} \tilde{\z}_b^{(r)}
\label{others}
\eeqa

\subsection{Yukawas in supersymmetric models}

Although the derivation of (\ref{yukT2n}) is quite general, 
and in principle is valid for toroidal compactifications 
where supersymmetry might be broken explicitly, it should be 
possible to understand it as a 3-point function 
in a $\cn=1$ $D=4$ supersymmetric theory, at least in the 
cases where such construction can be achieved. 
The normalized Yukawa couplings that we have obtained should 
then fit in the general supergravity formula (see e.g.ref.\cite{bmi})
\beq
Y_{ijk} = \left(K_{ab} K_{bc} K_{ca}\right)^{-1/2} e^{K/2} W_{ijk}
\label{yukireal}
\eeq
where $W_{ijk}$ is the corresponding trilinear coupling of the superpotential, $K$ is the K\"ahler potential and $K_{ab} = \p_{ab}\bar{\p}_{ab} K$ are the kinetic terms of the chiral fields in the $ab$ sector, etc.

There are indeed several examples in the literature of $\cn=1$ $D=4$ chiral compactifications realized as Type IIA intersecting D6-brane models \cite{susy}, which are T-dual to Type I compactifications on a factorizable $T^6$ and with magnetic fluxes turned on.\footnote{Notice that, although all of these supersymmetric models are based on orbifolds of $T^6$ which freeze some compactification moduli and impose some discrete symmetries to the open string sector, our computations and results are general and will equally well apply to these restricted geometries.} Let us then consider the particular case of magnetized compactifications in $T^6$, i.e., the case of $n=3$, which involves $D=10$ $\cn=1$ super Yang-Mills compactifications with magnetic fluxes. This particular choice is not only relevant for Type I strings, but also for magnetized compactifications of heterotic strings and Type IIB involving D9-branes.

In this case the coupling constant $g$ is given 
by $g = e^{\phi_{10}/2} \cdot \alpha'^{3/2}$, where $\phi_{10}$ 
is the ten-dimensional dilaton and $\alpha'$ the string scale. 
The 3-point function then reads
\beq
Y_{ijk} = \sig_{abc}\ e^{\phi_{10}/2} \prod_{r=1}^{3}
\left(\frac{2 \pim \tau^{(r)}}{(\ca^{(r)}/\alpha^\prime)^2} \right)^{1/4}
\left|
\frac{\tilde{I}_{1}^{(r)}
\tilde{I}_{2}^{(r)}}{\tilde{I}_{1}^{(r)} + \tilde{I}_{2}^{(r)}}
\right|^{1/4}
e^{H^{(r)}/2}\
\vt
\left[
\begin{array}{c}
\d_{ijk}^{(r)} \\ 0
\end{array}
\right]
\left( \tilde{\z}^{(r)}, \tau^{(r)} |I_{ab}^{(r)} I_{bc}^{(r)} I_{ca}^{(r)}| \right)
\label{yukT6}
\eeq
and comparing this expression with (\ref{yukireal}) we are led to the identifications
\beqa
W_{ijk} & = & \prod_{r=1}^3 
\vt
\left[
\begin{array}{c}
\d_{ijk}^{(r)} \\ 0
\end{array}
\right]
\left( \tilde{\z}^{(r)}, \tau^{(r)} |I_{ab}^{(r)} I_{bc}^{(r)} I_{ca}^{(r)}| \right) \label{super} \\
\left(K_{ab} K_{bc} K_{ca}\right)^{-1} e^{K} & = & 
e^{\phi_{10}} \prod_{r=1}^{3}
\frac{\left(2 \pim \tau^{(r)}\right)^{1/2}}{\ca^{(r)}/\alpha^\prime}
\left|
\frac{\tilde{I}_{1}^{(r)}
\tilde{I}_{2}^{(r)}}{\tilde{I}_{1}^{(r)} + \tilde{I}_{2}^{(r)}}
\right|^{1/2}
e^{\tilde{H}^{(r)}}
\label{kahler}
\eeqa
where in (\ref{kahler}) we have neglected global phases and defined
\beq
\tilde{H}^{(r)} = - 2\pi  |I_{ab}^{(r)} I_{bc}^{(r)} I_{ca}^{(r)}|^{-1} {\left(\pim \tilde{\z}^{(r)}\right)^2 \over \pim \tau^{(r)}}
\label{Htil}
\eeq

Let us try reexpress the K\"ahler factors involved 
in the 3-point function in a more physical basis. 
Indeed, in terms of $D=4$ supergravity fields, (\ref{kahler}) reads
\beq
\left(K_{ab} K_{bc} K_{ca}\right)^{-1} e^{K} = 
{2 \over (2\pi)^{3}} (S + \bar{S})^{-1} \prod_{r=1}^{3}
\left(U + \bar{U}\right)^{1/2}
\left|
\frac{\tilde{I}_{1}^{(r)}
\tilde{I}_{2}^{(r)}}{\tilde{I}_{1}^{(r)} + \tilde{I}_{2}^{(r)}}
\right|^{1/2}
e^{\tilde{H}^{(r)}}
\label{kahlersugra}
\eeq
where the supergravity fields are defined as
\beqa
\preal S & = & (2\pi)^{-1} e^{-\phi_{10}} \prod_{r=1}^3 {\ca^{(r)} \over 4\pi \alpha^\prime}, \\
\preal T^{(r)} & = & (2\pi)^{-1} e^{-\phi_{10}}\ {\ca^{(r)} \over 4\pi \alpha^\prime}, \\
\preal U^{(r)} & = & \pim \tau^{(r)}
\label{sugravar}
\eeqa

Notice that in (\ref{kahlersugra}) there is no 
dependence in $\preal T^{(r)}$ and the 
only explicit dependence on the Wilson lines $\z^{(r)}_\a$ comes 
from $\sum_{r=1}^3\tilde{H}^{(r)}$. 
These Wilson lines are the vev's of the scalar fields in 
the adjoint of each gauge group, which belong to $D=4$ $\cn=1$ 
chiral multiplets. From the point of view of D-brane physics, 
they are open string degrees of freedom, so their vev's 
compose part of the open string moduli space. 
Now, the K\"ahler potential for the closed string moduli has the 
well-known form
\beq
K\ =\ -log\left(S+ \bar{S}\right)\ -\ log \prod_{r=1}^{3}
\left(T^{(r)} + \bar{T}^{(r)}\right) \ -\
 log \prod_{r=1}^{3}
\left(U^{(r)} + \bar{U}^{(r)}\right)
\label{kahlerillo}
\eeq
so that we get an  identity that  the K\"ahler metric fields 
must obey in the flux side
\beq
\left(K_{ab} K_{bc} K_{ca}\right)  =
{ (2\pi)^{3}\over 2}\prod_{r=1}^{3}
\left(T^{(r)} + \bar{T}^{(r)}\right)^{-1}
  \prod_{r=1}^{3}
\left(U^{(r)} + \bar{U}^{(r)}\right)^{-3/2}
\left|
\frac{\tilde{I}_{1}^{(r)}
\tilde{I}_{2}^{(r)}}{\tilde{I}_{1}^{(r)} + \tilde{I}_{2}^{(r)}}
\right|^{-1/2}
e^{-\tilde{H}^{(r)}}
\label{kahleron}
\eeq
%

%
%

\section{D-branes of lower dimension}

As we have stressed in the introduction, $\cn=1$ $D=10$ SYM can 
be seen as the effective theory arising from type I and type 
IIB string theories at low energies. This fact has allowed 
to study D9-brane models with magnetic fluxes by using 
field theory techniques. From the point of view of type I or 
type IIB, however, D9-branes are not the only objects which 
yield gauge groups and/or chiral fermions in a general model. 
Indeed, both theories contain D5-branes and other objects of 
lower dimension than the D9-branes, which usually will carry 
additional gauge factors. Being of lower dimension than 
the D9-branes, these objects will appear as point-like in 
some directions of the internal compact manifold, while still 
expanding the four non-compact dimensions of the theory. 
These sectors of lower dimension are seen by the effective 
theory as small instantons, as is the case for heterotic strings. 
From the point of view of open string theories, however, 
these D-branes of lower dimension should be considered as natural 
as the D9-branes on which have based our previous 
analysis. In fact, it may turn out to be quite 
relevant for model building when trying to get a 
semi-realistic compactification, as the explicit model of Section 8 shows.

Yielding a gauge group an chiral matter, it is fair to wonder which kind of Yukawa couplings do we obtain when considering D-brane models involving D5-branes. One would expect the general formula (\ref{yukT6}) to hold true, since under T-duality/Mirror symmetry both D9 and D5-branes are mapped to type IIA intersecting D6-branes, being on equal footing on this dual picture. Moreover, the Yukawa couplings between chiral fields in intersecting D6-brane models was computed in \cite{yukis}, and a general formula was obtained for every kind of D6-brane.

In any case, it seems natural to wonder which wavefunction should be associated to an open string stretching between, e.g., a D9-brane and a D5-brane, specially in the dimensions where such D5-brane is pointlike. Intuitively, since the D5-brane looks like a localized source of RR charges/chiral anomalies in the D9-brane worldvolume, one would expect that the associated wavefunction is proportional to a delta function. In the present section we will try to further motivate this intuitive picture.

In order to do this notice that, in the dimension where it is pointlike, a D5-brane can be formally thought as a D9-brane with infinite flux. From this point of view, the wavefunction associated to massless chiral fermion between a D9 with flux $N_{D9}$ and a D5-brane should be approximated by the wavefunction (\ref{totalsolnabif}) in the limit $\tilde{I}_{ab} \raw \infty$, whereas that of an open string between a D9 and an anti-D5-brane should be recovered from the limit $\tilde{I}_{ab} \raw - \infty$ (here $a$ labels the D5-brane and $b$ the D9-brane).

For the sake of concreteness, let us consider a D9-D5 system. We can then use take the formal limit $N \raw \infty$ in (\ref{totalsoln}). From the normalization procedure of wavefunctions, we know that
\beqa \nonumber
\int_{T^2} dx_1 (\psi_+^j)^* (\psi_+^j) & = & {(2 \pim \tau N)^{1/2} \over \ca} \sum_n 
e^{- 2\pi N \pim \tau \left( n + \frac{j}{N} + \frac{\pim (z + \z)}{\pim \tau} \right)^2} \\
& = & \ca^{-1} \eps^{-1/2} \sum_n e^{- \frac{\pi}{\eps} \left(n + \frac{j}{N} + \frac{\pim (z + \z)}{\pim \tau} \right)^2}
\label{norminf}
\eeqa
where we have defined $\eps^{-1} = 2 \pim \tau N$. Now, this last expression reminds of the family of gaussian functions which define a delta function
\beq
\d(x) = {\rm Lim}_{\eps \raw 0+} \frac{1}{\sqrt{\eps}} e^{-\pi\frac{x^2}{\eps}},
\label{deltalim}
\eeq
but with multiple images added (the sum over $n$) which is necessary to have a well defined function on $T^2$ instead of $\cpx$. Indeed, if we take the limit $N \raw \infty$, we recover
\beqa \nonumber
\int_{T^2} dx_1 (\psi_+^j)^* (\psi_+^j) & = & \ca^{-1} \sum_n \d\left(\frac{\pim (z + \z)}{\pim \tau} + n\right) \\
& \sim & \d_{T^2}\left(\frac{\pim (z + \z)}{\pim \tau}\right)
\label{norminf2}
\eeqa
This seems to suggest that in the $N \raw \infty$ limit, the wavefunction (\ref{psisoln}) should be proportional to (the square root of) $\d_{T^2}(\pim (z + d)/\pim \tau)$. Notice, however, that for a D5-brane there is nothing special about the coordinate $\pim z$ on the whole problem\footnote{As we have seen in Section3, this asymmetry may only appear when the linear space of wavefunctions has dimension bigger than one, which is not the case.}. It thus seems sensible to consider the following wavefunction
\beq
\psi^{N \raw \infty} (\tau, z + \z) \sim \sqrt{\d_{T^2} (z + \z)}
\label{limit}
\eeq
where some possible phase may be multiplying the whole thing. The former Wilson line $\z$ should be associated with the position of the $D5$ on the corresponding $T^2$.

In order to check that this is a sensible proposal, let us see which kind of Yukawa coupling we obtain by considering a system of two D9-branes, of fluxes $m_a$ and $m_b$, $I_{ab} = m_a - m_b > 0$ and a D5-brane defined by the point $\z_c$. The wavefunction coming from the $ab$ sector will be the usual one, $\psi^{i, I_{ab}} (\tau, z + \z_{ab})$, whereas those coming from the $bc$, $ca$ sectors would yield a delta function, given by (\ref{limit}). When integrating all of them we would then have
\beqa
Y_i & = & \int_{T^2} \psi^{i, I_{ab}} (\tau, z + \z_{ab}) \cdot \d(z + \z_c) \\
\nonumber
& = & 
{\left(2 \pim \tau |I_{ab}| \right)^{1/4} \over \sqrt{\ca}}
e^{i\pi I_{ab} (\z_{ab} - \z_c) \pim(\z_{ab} - \z_c)/\pim \tau} 
\cdot
\vt
\left[
\begin{array}{c}
\frac{i}{I_{ab}} \\ 0
\end{array}
\right]
\left((\z_{ab} - \z_c) I_{ab}, \tau I_{ab} \right)
\label{intD5}
\eeqa
Notice that the whole expression can be put in the form (\ref{int6}), (\ref{H}), if we just define
\beq
\begin{array}{lcr}
\tilde{\z}_c & = & -\z_c \\
I_{bc} & = & -1 \\
I_{ca} & = & 1
\end{array}
\label{defin4}
\eeq
which are indeed the intersection numbers arising in the T-dual construction\footnote{Recall that conventions on intersection numbers differ by a sign. Notice as well that in the particular class of models where D5-branes are involved the usual identity $I_{ab}+I_{bc}+I_{ca}=0$ does not longer hold.}.

\section{Comparison with intersecting brane computations}

We have emphasized several times along this paper the fact that magnetized compactifications describe the same kind of physics as intersecting D-brane models do. In particular, $D=10$ $\cn = 1$ SYM compactifications can be seen to capture the low energy physics of Type IIB string models involving D9-branes with magnetic fluxes. The latter are related by T-duality symmetry to Type IIA compactifications with intersecting D6-branes \cite{bdl}. This pairing of theories can be extended to orbifolded/orientifolded versions of the above which involve, e.g., Type I compactifications with fluxes and orientifolded Type IIA models. 

Although describing the same physics, it should be clear that 
the techniques involved in the computation of $D=4$ physical 
quantities is quite different in both pictures. As a result, 
some elements of the $D=4$ effective action may be much easier 
to compute in one side of the T-dual map than in the other side. 
This whole paper is meant to be an example of such important fact. 
Indeed, in \cite{yukis} the Yukawa couplings of Type IIA intersecting 
D6-branes models were computed for toroidal compactifications, 
giving explicit formulae in terms of theta functions. 
The techniques used for such computation involved a sum over 
world-sheet instantons connecting intersection points. 
The Yukawas were obtained up to a global factor, $h_{\rm qu}$ 
which encoded the contribution from quantum fluctuation of the world-sheet. 
The computation of such kinds of factors has been addressed in 
\cite{cvetic,abel} where CFT vertex operator techniques were used.

In the present paper we have succeeded in computing the same quantity, 
the 3-point function, by using a different method. Indeed, 
the derivation of Yukawa couplings in magnetized compactifications 
involve {\em just field theory} techniques, and no string theory is 
needed, at least in the appropriate limit of large volume and 
diluted fluxes. Notice as well that in this setup there is 
no `classical' $h_{\rm cl}$ and `quantum' $h_{\rm qu}$ contributions 
to the 3-point function: everything comes from the overlap of 
three wavefunctions.

In the present section we will check that the Yukawas computed 
from magnetized toroidal compactifications and D6-branes at 
angles match. More precisely, we will compare our present 
results with the ones obtained in \cite{yukis} by using 
T-duality transformations, in order to extract the `quantum' 
prefactor $h_{\rm qu}$ and compare it with the one obtained 
in \cite{cvetic}. We will do such comparison in two ways, namely by 
performing a T-duality in one of the $T^2$ radii and then 
in the other one, obtaining the same result in both cases. 
As a byproduct of this computation we will gain some 
understanding of the action of T-duality on the chiral 
fields of the compactification.

\subsection{Intersecting brane Yukawas}

As was shown in \cite{yukis}, the computation of Yukawas in the context of Intersecting Brane Worlds in toroidal compactifications can be nicely expressed in terms of theta functions. More precisely, it was found that the formula for Yukawa couplings in a factorizable $T^{2n}$ was found to be
\beq
Y_{ijk} = h_{\rm qu}\cdot h_{\rm cl} = h_{\rm qu}\ \sig_{abc}\
\prod_{r=1}^n
\vt \left[
\begin{array}{c}
\d^{(r)} \\ \phi^{(r)}
\end{array}
\right] (\nu^{(r)},\k^{(r)})
\label{thetaint}
\eeq
where the $\vt$-function parameters are given by 
\beqa
\d^{(r)} & = & \frac{i^{(r)}}{I_{ab}^{(r)}} + \frac{j^{(r)}}{I_{ca}^{(r)}} + \frac{k^{(r)}}{I_{bc}^{(r)}}
+ \frac{I_{ab}^{(r)} \eps_c^{(r)} + I_{ca}^{(r)} \eps_{b}^{(r)} + I_{bc}^{(r)} \eps_{a}^{(r)}}
{I_{ab}^{(r)} I_{bc}^{(r)} I_{ca}^{(r)}}, \\
\phi^{(r)} & = & I_{ab}^{(r)} \th_c^{(r)} + I_{ca}^{(r)} \th_b^{(r)} + I_{bc}^{(r)} \th_a^{(r)},  \\
\nu^{(r)} & = & 0, \\
\k^{(r)} & = &  \frac{B^{(r)}}{\a^\prime} I_{ab}^{(r)} I_{bc}^{(r)} I_{ca}^{(r)}+ i \frac{\ca^{(r)}}{4\pi\a^\prime} |I_{ab}^{(r)} I_{bc}^{(r)} I_{ca}^{(r)}|
= \left\{
\begin{array}{c}
{J^{(r)}} (I_{ab}^{(r)} I_{bc}^{(r)} I_{ca}^{(r)}) \quad {\rm if\ } \sig_{abc}^{(r)} > 0\\
{\bar{J}^{(r)}} (I_{ab}^{(r)} I_{bc}^{(r)} I_{ca}^{(r)}) \quad {\rm if\ } \sig_{abc}^{(r)} < 0
\end{array}
\right.
\label{param}
\eeqa
where $I^{(r)}_{ab}$ are the intersection numbers between the D-branes $a$ and $b$ in the $r^{th}$ two-torus, and $i^{(r)} = 0, \dots, |I_{ab}^{(r)}|-1$ label such intersections. The continuous parameters $\eps^{(r)}_a$, $\th_a^{(r)} \in [0,1]$ indicate the position and the Wilson line of the D-brane $a$ in the $r^{th}$ $T^2$, etc. whereas $J^{(r)} = (B^{(r)} + i \ca^{(r)}/4\pi^2)/\alpha'$ is the complexified K\"ahler parameter of the $r^{th}$ two-torus. Finally, $\sig_{abc}^{(r)} = {\rm sign}(I_{ab}^{(r)}I_{bc}^{(r)}I_{ca}^{(r)})$ and $\sig_{abc} = {\rm sign}(I_{ab}I_{bc}I_{ca})$. We are also supposing that ${\rm g.c.d.} (I_{ab}^{(r)},I_{bc}^{(r)},I_{ca}^{(r)}) = 1$ (otherwise $\d^{(r)}_{ijk}$ changes, see \cite{yukis}).

In order to compare (\ref{thetaint}) with the results of Section 5, it is convenient to express this quantity in terms v.e.v.'s of complex fields. Indeed, notice that we can rewrite (\ref{thetaint}) as 
\beq
Y_{ijk}^{\rm int} = h_{\rm qu}\ \sig_{abc}\
\prod_{r=1}^n
e^{H_{\rm int}^{(r)}/2}
e^{\pi i {(\nu^{(r)})^2 / (I_{ab}^{(r)} I_{bc}^{(r)} I_{ca}^{(r)}) J^{(r)}}}
\vt \left[
\begin{array}{c}
\d_{ijk}^{(r)} \\ 0
\end{array}
\right] \left(\nu^{(r)}, (I_{ab}^{(r)} I_{bc}^{(r)} I_{ca}^{(r)}) J^{(r)}\right) 
\label{thetaint2}
\eeq
where now the parameters are
\beqa
\d_{ijk}^{(r)} & = & \frac{i^{(r)}}{I_{ab}^{(r)}} + \frac{j^{(r)}}{I_{ca}^{(r)}} + \frac{k^{(r)}}{I_{bc}^{(r)}} \\
\nu^{(r)} & = & I_{ab}^{(r)} \nu_c^{(r)} + I_{bc}^{(r)} \nu_a^{(r)} + I_{ca}^{(r)} \nu_b^{(r)}, \\
\nu_a^{(r)} & = & \theta_a^{(r)} + J^{(r)} \epsilon_a^{(r)}, \ {\rm etc.}
\label{param2}
\eeqa
with $J^{(r)}$ defined as before and
\beq
H_{\rm int}^{(r)} = - 2 \pi i \left(I_{ab}^{(r)} I_{bc}^{(r)} I_{ca}^{(r)}\right)^{-1} \left( \frac{\pim (\bar{J}^{(r)}\nu^{(r)})}{\pim J^{(r)}} \right)^2 / J^{(r)}
\label{Hint}
\eeq

In fact, expressions (\ref{thetaint2}), (\ref{Hint}) are valid for $\sig_{abc}^{(r)} = 1$, $\forall r$. In case we choose $\sig_{abc}^{(s)} = -1$ for any $s$ we must substitute $J^{(s)} \mapsto \bar{J}^{(s)}$ and $\nu_a^{(s)} \mapsto \bar{\nu}_a^{(s)}$, which amounts to taking the complex conjugate of the $r^{th}$ factor in (\ref{thetaint2}). This is required by hermiticity of the effective Lagrangian, since changing the sign of the intersection numbers changes the chirality of the fields at the intersection. Notice that we have found the same feature in the T-dual picture of magnetized D-branes. For simplicity, in the rest of this section we will consider $\sig_{abc}^{(r)} = 1$, $\forall r$.

%
\EPSFIGURE{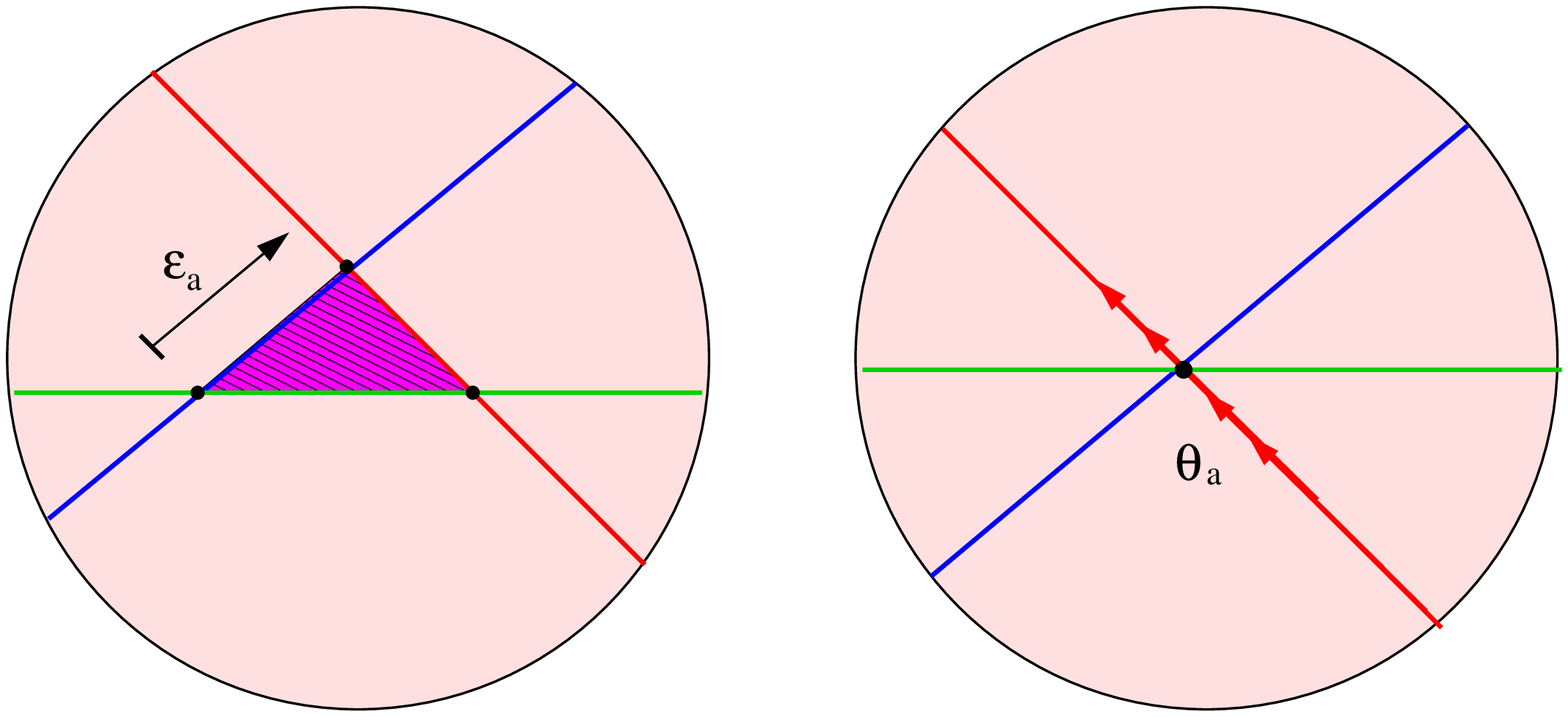, width=5in}
{\label{holomfig}
Evaluating worldsheet instantons involving several boundaries will in general not yield a holomorphic function of the open string moduli $\nu_a = \th_a + J \eps_a$, even in supersymmetric constructions, since their behaviour is asymmetric in $\th_a$ and $\eps_a$. In the figure we illustrate the evaluation of the smallest instanton between three different boundaries. Notice that the area of the instanton will change, and so will the action, if we vary the position $\eps_a$ of the D-brane $a$, and we move it away from the point where all three intersect. The action will not change, however, if we stay at this point and arbitrarily vary the Wilson lines. See \cite{yukis} for more details on these kind of compactifications.}
%

Finally, notice that (\ref{thetaint2}) is not holomorphic 
in the open string moduli $\nu_\a^{(r)}$. This may seem a 
little suspicious at first sight, since the `classical' 
contribution is nothing but a sum $\sum e^{-S}$ over holomorphic 
worlsheet instantons, and in heterotic models this provides a 
holomorphic function of the compactification moduli \cite{dsww}. 
This seems also to be the case for Type IIA compactifications 
on ${\bf CY_3}$ and with D6-branes wrapping Special 
Lagrangian cycles, at least if the open string instantons 
involve {\em only one boundary} \cite{vafa,kklm}. 
When several boundaries are involved, however, we don't expect 
this to be the case, since the position and the Wilson 
lines which make up the complex field $\nu_a$ do not 
have a symmetric role when evaluating the world-sheet 
action (see figure \ref{holomfig}). Actually, evaluating 
holomorphic instantons (that is, surfaces calibrated 
by the K\"ahler form $\om$) only guarantees a expression 
holomorphic in the K\"ahler parameters, in this case $J^{(r)}$, 
and not on the open string moduli $\nu_a^{(r)}$. 
Of course, in a supersymmetric compactification where we can understand 
Yukawa 
couplings as coming from a superpotential, the classical contribution $h_{\rm cl} = \sum e^{-S}$ (which encodes all the open string moduli dependence) must be holomorphic in all the fields up to a global term, that can be understood as a K\"ahler potential contribution to the 3-point function (\ref{yukireal}). The formula (\ref{thetaint2}) is a good example of all these facts.

\subsection{Matching by T-duality}

In order to relate (\ref{thetaint2}) to the results of Section 5 we need first to perform a T-duality transformation on the fields involved in the toroidal compactification. There are two different ways of doing this, corresponding two the two different radii of $T^2$. In each case the transformation properties are different, but we should obtain the same final result. We will perform both computations in the specific case of $T^2 \times T^2 \times T^2$ compactifications, but it is straightforward to generalize it to arbitrary number of tori. Before doing so, let us rewrite the 3-point function (\ref{yukT6}) in terms of more geometrical variables. First, instead of $\tilde{I}_{\a\b}$, let us consider the `slopes'
\beq
\th^{(r) app}_{\a\b} = 4\pi {\tilde{I}_{ab}^{(r)} \over (\ca^{(r)}/\a^\prime)} 
\label{slopes}
\eeq
which correspond to the spacing between particles in the tower of massive replicas (`gonions') of chiral fermions and bosons. Actually, these quantities specify the harmonic oscillator algebra (\ref{osc}) present at each chiral sector of the theory. We will encode such slope dependence in the function
\beq
\Theta^{(r)} = 
\frac{\th_{1}^{(r) app} \th_{2}^{(r) app}}
{\th_{1}^{(r) app} + \th_{2}^{(r) app}}
\label{mu}
\eeq
We can thus rewrite (\ref{yukT6}) as
\beq\nonumber
Y_{ijk} = \sig_{abc}\ 
(2\pi)^{-9/4} e^{\phi_{4}/2} \prod_{r=1}^{3}
\left(\pim \tau^{(r)}\right)^{1/4}
\left| \Theta^{(r)} \right|^{1/4}
e^{H_{\rm mag}^{(r)}/2}\
\vt
\left[
\begin{array}{c}
\d_{ijk}^{(r)} \\ 0
\end{array}
\right]
\left( \tilde{\z}^{(r)}, \tau^{(r)} |I_{ab}^{(r)} I_{bc}^{(r)} I_{ca}^{(r)}| \right)
\label{thetamag}
\eeq
where
\beq
H_{\rm mag}^{(r)}  =  2\pi i |I_{ab}^{(r)} I_{bc}^{(r)} I_{ca}^{(r)}|^{-1} {\tilde{\z}^{(r)} \cdot \pim \tilde{\z}^{(r)} \over \pim \tau}^{(r)}
\label{Htotal}
\eeq
and we have considered the four-dimensional dilaton $\phi_4$, defined as
\beqa
e^{\phi_4} & = & e^{\phi_{10}} \prod_{r=1}^3 \left( \pim J^{(r)} \right)^{-1/2}, \\
\pim J^{(r)} & = & {\ca^{(r)} \over 4\pi^2 \alpha^\prime},
\label{geomvar}
\eeqa
$J^{(r)}$ being the standard K\"ahler modulus of the $r^{th}$ two-torus\footnote{Despite the notation, $\pim J^{(r)}$ should not be seen as the imaginary part of the complex field $J^{(r)}$, whose real part is not even a dynamical field in Type I models. Indeed, as seen in \cite{cim1}, it pairs with RR-field to form the complex scalar of an $\cn =1$ chiral multiplet.}.

\subsubsection{Horizontal T-duality}

Let us first perform a simultaneous T-duality on the first radius of each $T^2$, that is along the direction given by $\pim (\tau\bar{z})$. We need to relate quantities by the T-duality transformation
\beqa\nonumber
\tau & \lraw & J \\
\tilde{\z} & \lraw & \nu
\label{transfs}
\eeqa
for each $T^2$ (for the sake of clarity, we will suppress the indices $(r)$ from now on). We now make the substitutions (\ref{transfs}) in (\ref{thetaint2}), obtaining that the theta functions match exactly, whereas the exponential factors become
\beqa \nonumber
2\pi i |I_{ab} I_{bc} I_{ca}|^{-1} \left( \nu^2 - \left(\frac{\pim (\bar{J}\nu)}{\pim J} \right)^2\right)/ J & = &
\pi i |I_{ab} I_{bc} I_{ca}|^{-1} \left({\nu \cdot \pim \nu \over \pim J} +  {\pim \nu \cdot \pim (\bar{J}\nu) \over (\pim J)^2} \right) \\
=
2\pi i |I_{ab} I_{bc} I_{ca}|^{-1} \left( {\tilde{\z} \cdot \pim \tilde{\z} \over \pim \tau} + {\pim \tilde{\zeta} \cdot \pim (\bar{\tau}\tilde{\zeta}) \over (\pim \tau)^2} \right)& \sim &
 2\pi i |I_{ab} I_{bc} I_{ca}|^{-1} {\tilde{\z} \cdot \pim \tilde{\z} \over \pim \tau}
\label{exps}
\eeqa
where $\sim$ stands for equality up to a global phase upon exponentiation. We thus see that the part of the Yukawa coupling (\ref{thetamag}) depending on the open string moduli $\tilde{\zeta}$ precisely matches the classical worldsheet contribution to the Yukawa couplings computed in \cite{yukis}. 

We are thus led to the identification
\beq
h_{\rm qu} \lraw
(2\pi)^{-9/4} e^{\phi_{4}/2} \prod_{r}
\left(\pim \tau\right)^{1/4}
\left| \Theta \right|^{1/4}
\label{quantum}
\eeq
The idea is now to relate this prefactor (\ref{quantum}) 
with the BCFT computation in \cite{cvetic}. 
Now, in \cite{cvetic} a  quite 
similar expression was obtained as a prefactor. Namely, a square root 
of products of gamma functions given by
\beq
{\G(1-\th_{ab}) \G(1-\th_{ca}) \G(\th_{ab}+\th_{ca})\over 
\G(\th_{ab}) \G(\th_{ca}) \G(1 - \th_{ab} - \th_{ca})}
= {\th_{ab} \th_{ca} \over \th_{ab}+\th_{ca}} \cdot
\left| \frac{B(-\th_{ab}, -\th_{ca})}{B(\th_{ab}, \th_{ca})}\right|
\label{gammas}
\eeq
The quotient of Beta functions quickly decreases to one for small angles, as can be seen by using the definition of the Beta function and the Weierstrass form for the gamma function
\beq
\G(x)^{-1} = xe^{\g x} \prod_{p=1}^{\infty} (1 +\frac{x}{p}) e^{-x/p}
\label{wei}
\eeq
which may also provide a way to measure the deviation of CFT computations from field theory results. We clearly recover the form of the prefactor (\ref{slopes}) for small angles. We do not, however, recover the square root of \cite{cvetic}, but rather a $1/4$ power instead.

Apart from the slope-dependent prefactor we are left with
\beq
(2\pi)^{-9/4} e^{\phi_{4}/2} \prod_{r}
\left(\pim \tau\right)^{1/4}
\label{lefts}
\eeq
The four-dimensional dilaton does not change under T-duality, so that we recover the prefactor
\beq
(2\pi)^{-9/4} e^{\phi_{4}/2} \prod_{r}
\left(\pim \tau\right)^{1/4} \quad \stackrel{(\ref{transfs})}{\longmapsto} \quad
(2\pi)^{-9/4} e^{\phi_{4}/2} \prod_{r}
\left(\pim J\right)^{1/4} = (2\pi)^{-9/4} e^{\phi_{10}/2}
\label{final}
\eeq
nicely matching the prefactors in \cite{cvetic}.

\subsubsection{Tilted T-duality}

We now consider the magnetized compactification again and perform a T-duality along the second radius of each $T^2$, that is along the tilted direction given by $\pim z$. After that, we get a intersecting D-brane model which has the same intersection numbers, and hence the same Yukawas, as the one we would obtain in the previous subsection. It will have, nevertheless, different wrapping numbers.

Before doing that, let us rewrite the Yukawa couplings (\ref{thetamag}) in terms of the elements of the `alternative' basis (\ref{basis2}).
\beq
Y_{lmn} = 
|I_{ab} I_{bc} I_{ca}|^{-1/2}
\sum_{i,j,k} e^{2\pi i\left(\frac{il}{I_{ab}} + \frac{jm}{I_{ca}} + \frac{nk}{I_{bc}} \right)}\ Y_{ijk}
\label{changeyuk}
\eeq
where $Y_{ijk}$ is given by (\ref{guess}). Notice that, if ${\rm g.c.d.}(I_{ab},I_{bc},I_{ca}) = 1$ then we can perform the following relabeling of indices
\beq
\begin{array}{lcr}
i' & = & i\cdot I_{bc}I_{ca}\ {\rm mod\ } I_{ab}\\
j' & = & j\cdot I_{ab}I_{bc}\ {\rm mod\ } I_{ca}\\
k' & = & k\cdot I_{ca}I_{ab}\ {\rm mod\ } I_{bc}
\end{array}
\label{relabel}
\eeq
so that we have
\beqa \nonumber
& & \sum_{i^\prime,j^\prime,k^\prime} e^{2\pi i\left(\frac{i^\prime l}{I_{ab}} + \frac{j^\prime m}{I_{ca}} + \frac{n^\prime k}{I_{bc}} \right)} 
\vt
\left[
\begin{array}{c}
\d_{i^\prime j^\prime k^\prime} \\ 0
\end{array}
\right]
\left( \tilde{\z}, \tau |I_{ab} I_{bc} I_{ca}| \right) \\ \nonumber
& = & \sum_{i,j,k} e^{2\pi i\left(il\frac{I_{bc}I_{ca}}{I_{ab}} + jm\frac{I_{ab}I_{bc}}{I_{ca}} + nk \frac{I_{ca}I_{ab}}{I_{bc}} \right)} 
\vt
\left[
\begin{array}{c}
\d_{ijk} \\ 0
\end{array}
\right]
\left( \tilde{\z}, \tau |I_{ab} I_{bc} I_{ca}| \right) \\
& = & 
\vt
\left[
\begin{array}{c}
0 \\ \d_{lmn}
\end{array}
\right]
\left( \tilde{\z}/|I_{ab} I_{bc} I_{ca}|, \tau/|I_{ab} I_{bc} I_{ca}| \right) 
\label{changetheta} 
\eeqa

The Yukawa couplings in the basis (\ref{basis2}) are given by
\beqa \nonumber
Y_{lmn} & = &
\sig_{abc}\ 
\frac{e^{\phi_{4}/2}}{(2\pi)^{9/4}} \prod_{r}
\left(\frac{\pim \tau}{|I_{ab} I_{bc} I_{ca}|^2} \right)^{1/4} 
\left| \Theta \right|^{1/4}
e^{H_{\rm mag}/2}\\
& & 
\cdot
\vt
\left[
\begin{array}{c}
0 \\ \d_{lmn}
\end{array}
\right]
\left( \tilde{\z}/|I_{ab} I_{bc} I_{ca}|, \tau/|I_{ab} I_{bc} I_{ca}| \right)
\label{guess2}
\eeqa

Now we identify fields under the T-duality transformations
\beqa\nonumber
\tau & \lraw & - {1 \over J} \\
\tilde{\z} & \lraw & {\nu \over J}
\label{transfl}
\eeqa
Making the substitutions (\ref{transfl}) in (\ref{guess2}), and using the transformation properties of the theta function
\beq
\vt
\left[
\begin{array}{c}
\a \\ \b
\end{array}
\right]
\left( \frac{\nu}{\k}, -\frac{1}{\k}\right)
=
(-i\k)^{1/2}
e^{2 i \pi\a\b + i\pi \nu^2/\k}\
\vt
\left[
\begin{array}{c}
\b \\ -\a
\end{array}
\right]
\left( \nu, \k \right)
\label{transfl2}
\eeq
we get that
\beqa \nonumber
\vt
\left[
\begin{array}{c}
0 \\ \d_{lmn}
\end{array}
\right]
\left( \tilde{\z}/|I_{ab} I_{bc} I_{ca}|, 
\tau/|I_{ab} I_{bc} I_{ca}| \right)
& = &
\vt
\left[
\begin{array}{c}
0 \\ \d_{lmn}
\end{array}
\right]
\left( {\nu\over \k}, 
-{1 \over \k} \right) \\
& = & 
(-i\k)^{1/2}
e^{i\pi \nu^2/\k}\
\vt
\left[
\begin{array}{c}
\d_{lmn} \\ 0
\end{array}
\right]
\left( \nu, \k \right)
\label{thetal}
\eeqa
where we have defined
\beq
\k = J \cdot |I_{ab} I_{bc} I_{ca}|.
\label{defk}
\eeq
We also get
\beq
2\pi i |I_{ab} I_{bc} I_{ca}|^{-1} {\tilde{\z} \cdot \pim \tilde{\z} \over \pim \tau} = 
- 2 \pi i |I_{ab} I_{bc} I_{ca}|^{-1} {\nu \over J} {\pim (\bar{J}\nu) \over \pim J}
\label{expl}
\eeq
Notice that the exponentials of (\ref{Hint}) and (\ref{expl}) match up to a phase. We are now left with
\beq
\frac{e^{\phi_{4}/2}}{(2\pi)^{9/4}} \prod_{r}
\left({\pim \tau} \right)^{1/4} 
\left| \Theta \right|^{1/4}
\left(-i \tau\right)^{-1/2}
\label{leftl}
\eeq
The identification of the slope-dependent prefactors works in the same way as in the previous T-duality transformation, so after identifying them we recover
\beq
\frac{e^{\phi_{4}/2}}{(2\pi)^{9/4}} \prod_{r}
\left(\frac{-\pim \tau}{\tau^2} \right)^{1/4} 
\label{leftl2}
\eeq
matching the previous result (\ref{final}) after the transformation (\ref{transfl}).

Notice that both T-dualities lead us to the same 'quantum' prefactor in the intersecting D-brane setup. Moreover, by substituting the field theory quantity $\Theta$ by the string theory analogue we recover a more symmetric expression in the angles.
\beqa \nonumber
h_{\rm qu} & = &
{e^{\phi_{10}/2} \over (2\pi)^{9/4}} \prod_{r}
\left(
\frac{|\th_{ab}^{(r) {\rm app}}| |\th_{ca}^{(r) {\rm app}}|}{|\th_{ab}^{(r) {\rm app}}| + |\th_{ca}^{(r) {\rm app}}|} 
\right)^{1/4} \\ 
& \simeq &
{e^{\phi_{10}/2}} \prod_{r}
\left({\G\left(1-\th_{ab}^{(r)}\right) \G\left(1-\th_{ca}^{(r)}\right) \G\left(1-\th_{bc}^{(r)}\right)
\over (2\pi)^3\ 
\G\left(\th_{ab}^{(r)}\right) \G\left(\th_{ca}^{(r)}\right) \G\left(\th_{bc}^{(r)}\right)} \right)^{1/4}
\label{quantumfinal}
\eeqa
where we have made the substitution $\th_{bc}^{(r)} = 1 - \th_{ab}^{(r)} - \th_{ca}^{(r)}$ implicit in (\ref{gammas}) \cite{cvetic}.

\subsection{Chiral fields and T-duality}

The previous matching of Yukawa couplings in magnetized D9-branes and intersecting D6-brane models suggest an intuitive picture of how T-duality acts on the chiral fields at intersection points. Indeed, notice that in order to match the Yukawas by means of a {\it horizontal} T-duality, we had to consider 3-point functions derived from the overlap of wavefunctions of the form (\ref{basis1}), whereas in order to compare results after a {\it tilted} T-duality we had to reexpress the Yukawas in terms of the wavefunctions (\ref{basis2}). Recall that (\ref{basis1}) and (\ref{basis2}) are the two canonical bases of wavefunctions on a $T^2$. The matching above then suggest that there is a one-to-one correspondence between them and the chiral fields localized at D6-brane intersections after performing T-dualities (\ref{transfs}) or (\ref{transfl}), respectively. 

Let us reverse the point of view, and consider an intersecting D6-brane model. We can relate it to a type IIB model with magnetized D9-branes by performing either three T-dualities of the form (\ref{transfs}) or those of the form (\ref{transfl}). After the T-duality transformation the chiral fields, which previously associated to a pointlike intersection of D6-branes, will be represented by either the wavefunctions (\ref{basis1}) (if we choose (\ref{transfs})) or those in (\ref{basis2}) (if we take (\ref{transfl})). Notice that in both cases the wavefunctions of those chiral fields have their profile {\em delocalized in the directions where the T-dualities have been performed}), whereas their profile strongly depends on the other directions (see figures \ref{3thetas} and \ref{rotatheta}). 
\vspace*{-2cm}
%
\begin{figure}[ht]
\begin{center}
\begin{tabular}{cc}
\\
\epsfig{file=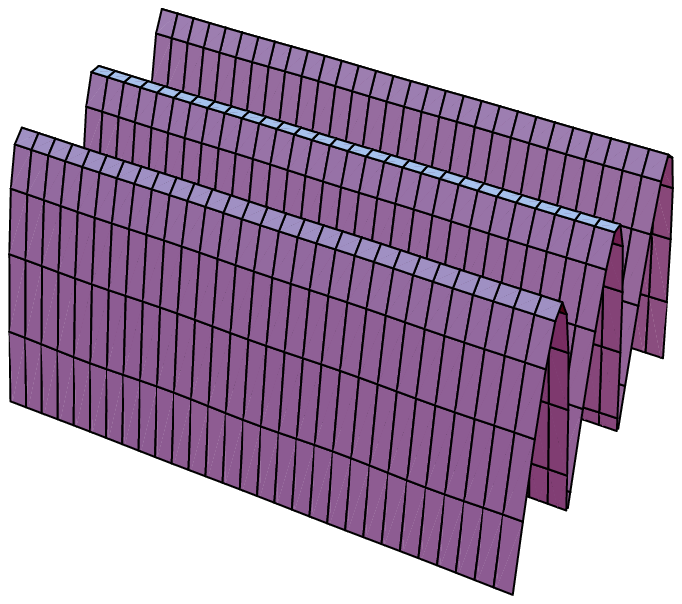, height=6cm} & 
\epsfig{file=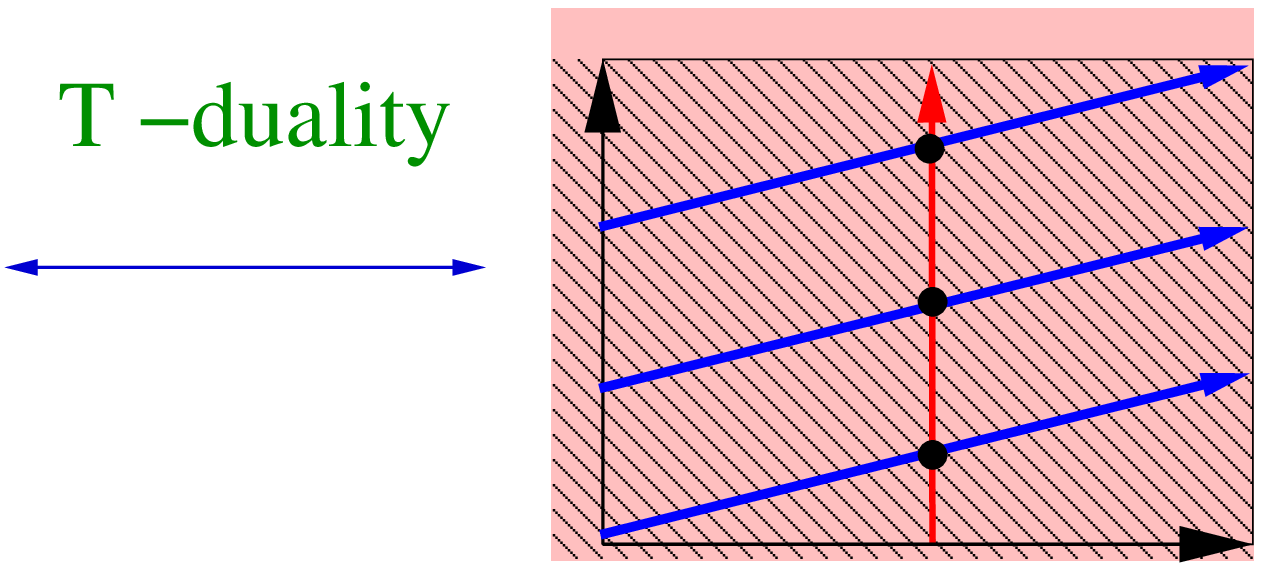, height=3.75cm} \\
{\Large Magnetized D-branes} & 
\hskip 4.25cm 
{\Large Intersecting D-branes} \\
\end{tabular}
\end{center}
\caption{\small{Action of T-duality on chiral fields. T-duality maps the chiral fields at intersection points to wavefunctions defined on the whole compactification space. However, their probability density is delocalized in the directions of the T-duality and peaked in the transverse directions. 
Here we have considered the case of a horizontal T-duality acting on the wavefunctions in figure 4.
}}
\label{chiral}
\end{figure}
%

This nicely matches with the intuitive picture that we have of T-duality of exchanging Dirichlet $\lraw$ Neumann boundary conditions, and hence changing the dimension of D-branes. Indeed, what in the type IIA picture was a D6-brane, localized in a 3-cycle of $T^6$, after a T-duality becomes a type IIB D9-brane wrapping the whole compactification space\footnote{Unless some dimension of the D6-brane is parallel to a T-duality direction, of course.}. Actually, it is amusing to notice that these well-known facts about D-brane physics can also be understood in terms of delocalization of wavefunctions, this time those of the gauginos, which are given by a constant function localized in the worldvolume of a D-brane.

\subsection{The K\"ahler metrics of chiral fields  revisited}

As we discussed in Section 5.3 in the magnetic flux picture,
agreement with the supergravity formulae 
in the case of SUSY compactifications gives a constraint that the 
K\"ahler metrics of the chiral fields should obey. In the case 
of the intersecting D6-brane picture the analogous to
eq.(\ref{kahleron}) is 
\beq
\left(K_{ab} K_{bc} K_{ca}\right)^{-1} e^{K} 
\  = \ 
e^{\phi_{4}} \prod_{r=1}^{3}
\left(\pim J^{(r)}\right)^{1/2}
\left| \Theta^{(r)}\right|^{1/2}
e^{\tilde{H}^{(r)}}
\label{ahiva}
\eeq
which yields
\beq
K_{ab} K_{bc} K_{ca} 
\  = \ 
e^{3\phi_{4}} \prod_{r=1}^{3}
\left(\pim J^{(r)}\right)^{-3/2}
\left(
{(2\pi)^3\ \G\left(\th_{ab}^{(r)}\right)
\G\left(\th_{bc}^{(r)}\right)\G\left(\th_{ca}^{(r)}\right)
\over
\G\left(1-\th_{ab}^{(r)}\right)
\G\left(1-\th_{bc}^{(r)}\right)\G\left(1-\th_{ca}^{(r)}\right)}
\right)^{1/2}
e^{-\tilde{H}^{(r)}}
\label{babilonio}
\eeq
This seems to suggest a dependence of the chiral fields K\"ahler metric on the Wilson lines and twist angles of the form
\beq
K_{ab}\ \propto\
\prod_{r=1}^3
\left(
{2\pi\ \G\left(\th_{ab}^{(r)}\right) 
\over 
\G\left(1-\th_{ab}^{(r)}\right)}
\right)^{1/2}
e^{2\pi\tilde{I}_{ab}^{(r)}{\left(\pim \z_{ab}^{(r)} \right)^2 \over \pim 
J^{(r)}}}
\label{kahlerguess}
\eeq
where we have used a decomposition of $\tilde{H}^{(r)}$ analogous to the second line of (\ref{Hna}). It would be interesting to compare these constraints with 
explicit expressions for the K\"ahler metrics of the chiral matter fields at intersections \cite{nath,lmrs}, as well as derivation of the Wilson line dependence of such metrics from direct BCFT computations, along the lines of \cite{lmrs}.

%

\section{A 3-generation  MSSM-like  Model}

As we mentioned above, 
the expressions here obtained
may be used to compute the Yukawa couplings of the 
T-dual of the intersecting brane models
in ref.\cite{imr} whose massless fermion spectrum is that of  
the non-SUSY SM  or, in general,
 all of the toroidal models in \cite{bgkl,afiru,afiru2,bkl,more,susy}. 
For definiteness let us discuss
here the T-dual of the model discussed in section 4 of ref.\cite{yukis}.
This model has a spectrum quite close to that of
a 3-generation MSSM and is simple enough so that we 
can display the explicit expressions for the Yukawa 
couplings without much complication. We leave a more 
systematic phenomenological analysis of the different 
models for future work.

%
\EPSFIGURE{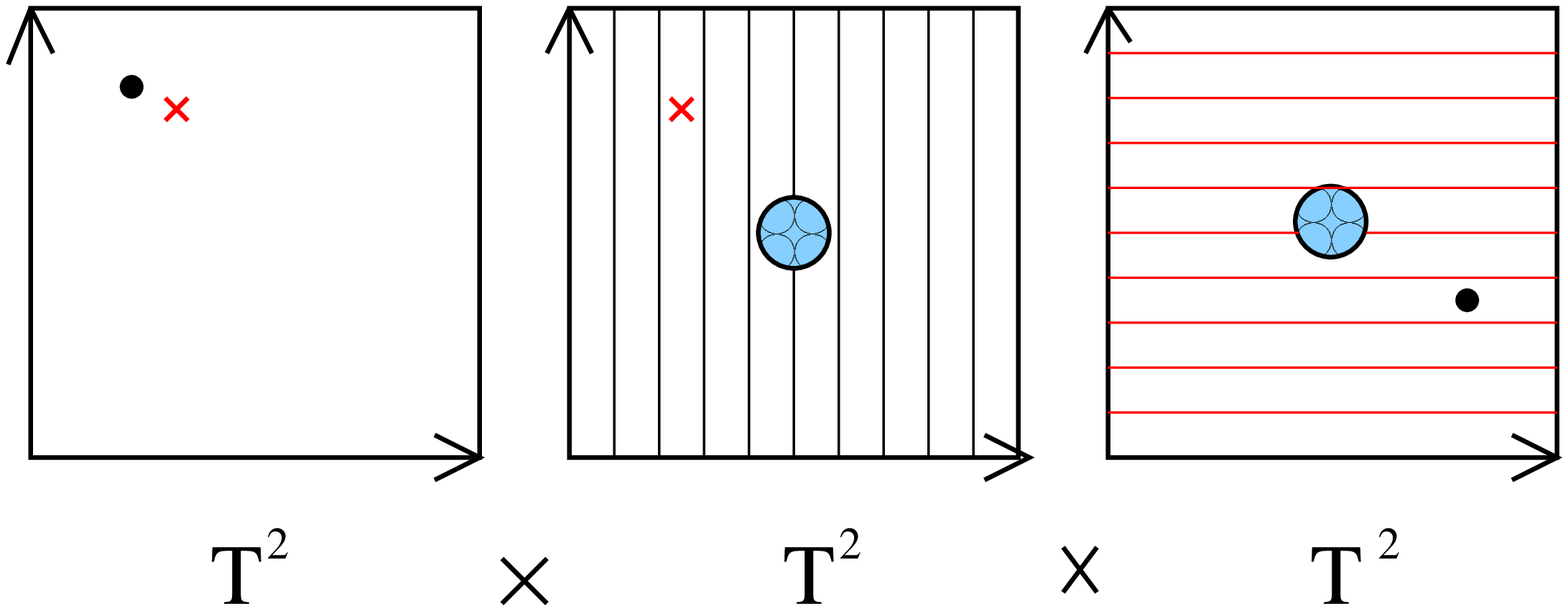, width=6in}
{\label{magnetica}
Structure of the 3-generation MSSM-like model described
in the text. D9-branes with $U(4)$ gauge group are
wrapping the 6-torus and are subject to 3 units of
magnetic flux in the  second and third tori. One 
$D5_b$($D5_c$)  brane is wrapping the second(third) 
torus and is pointlike (black dots and crosses respectively)
in the other two tori.}
%

The model may be constructed as follows 
(see fig.\ref{magnetica}). Consider as starting point 
a $U(4)$  $D=10$ Yang Mills theory and let us compactify it 
on $T^2\times T^2\times T^2$ \footnote{This may be considered 
as a subset  of a Type I string toroidal compactification. In fact, from Type I we should consider an $SO(8)$ theory, broken to $U(4)$ by the addition of a non-trivial flux/Wilson lines. We will not dwell in the details of the full Type I construction, though, since it would deviate us from the main point of this section.}.
To obtain chirality we add twelve units of magnetic flux 
along the second and third torus (circles in the figure), i.e., we have a flux of the form
\beq
\begin{array}{rcl}\vspace*{.2cm}
F_{z^2\bar{z}^2} & = & {12 \pi i\over \pim \tau^{(2)}} {\bf 1}_4 \quad \raw
(n_a^{(2)}, m_a^{(2)})\ =\ (4,12)\ =\ 4\cdot(1,3) \\
F_{z^3\bar{z}^3} & = & - {12 \pi i\over \pim \tau^{(3)}} {\bf 1}_4 \quad \raw
(n_a^{(3)}, m_a^{(3)})\ =\ (4,-12)\ =\ 4\cdot(1,-3)\\
\end{array}
\label{fluxguay}
\eeq
As it was seen in Section 4, this flux does not break the $U(4)$ gauge group nor needs of the addition of non-Abelian Wilson lines. Hence, we can think of this configuration as a stack of 4 $D9_a$-branes wrapping
the 6-torus and each of them subject to $3 = 12/4$ units of quantized flux
on the second and third torus. We now add a couple of D5-branes:
the first $D5_b$ is wrapping the second torus and is pointlike
in the first and third tori (black dots in the figure).
The second brane $D5_c$ is wrapping the third torus
and is localized in the other two (crosses in the figure). It is well
known that in Type I theory the gauge group in the worldvolume
of $n$ parallel 5-branes is $USp(2n)$. In our case with two sets 
of isolated $D5$-branes the overall
gauge group will be $U(4)_a\times SU(2)_b\times SU(2)_c$, since 
$USp(2)=SU(2)$. Now, solving Dirac equation on this 
background we will get bifundamental massless chiral fermions
transforming under this gauge group as $3(4,2,1)+3({\bar 4},1,2)$. 
We get three 
copies because of the three units of magnetic fluxes that
we added. Thus in the end we get a Pati-Salam type of
model with three generations of quarks and leptons.

This is just a T-dual version of the model described in Section 4 
of ref.\cite{yukis} and more details about its structure 
may be found in that reference. Let us just mention 
that the $U(1)$ in $U(4)$ is anomalous and becomes massive
through a generalized Green-Schwarz mechanism in the
standard way. Furthermore, one can break $SU(4)\rightarrow
SU(3)\times U(1)_{B-L}$ and $SU(2)_c\rightarrow U(1)$ 
by a judicious choice of Wilson lines on the worldvolume of the D5-branes, 
so that at the end of the day one is left with 3 generations 
of quarks and leptons and gauge group $SU(3)\times SU(2)_L
\times$$U(1)_Y\times U(1)_{B-L}$. The charged particle 
spectrum in this sector of the theory is supersymmetric 
if one chooses equal areas for the second and third tori
\footnote{This is the T-dual of the SUSY condition in the
case of intersecting branes, which requires equal 
complex structure in the second and third tori \cite{yukis}. 
See Appendix \ref{SUSYap}.}.
Let us finally mention that there is a minimal Higgs 
sector like that in the MSSM if both $D5$-branes sit on top
of each other in the first torus (couple of nearby points 
in the first torus in the figure)
\footnote{In other words, there is $\mu $ mass term for the
Higgs multiplets which is proportional to the distance 
between both $D5$-branes in the first torus.}.
All in all, the final spectrum in this subsector 
of the theory gets quite close to the form of the 
MSSM \footnote{It turns out that in the present model 
this subsector respects $\cn=1$ SUSY but the stringent
conditions of RR-tadpole cancellation requires the presence 
of other branes which do not respect SUSY, so that the complete
model is actually non-supersymmetric. Still its simplicity
makes it a good choice to show an explicit result for
Yukawa couplings.}.

Let us now discuss the structure of Yukawa couplings 
in the above model. To simplify the formulae let us
consider the initial Pati-Salam version with
gauge group $U(4)\times SU(2)\times SU(2)$,
in which there is a single set of Yukawa 
couplings $Y_{ij}(4.2,1)^i({\bar 4},1,2)^j(1,2,2)$. 
The only aspect which is a bit unfamiliar
is the presence of D5-branes but we already showed in
section 6  how to deal with them. Looking at fig.
(\ref{magnetica}) we see that the first torus does not give us
any flavour structure since both 5-branes are pointlike 
and on top of each other (so that there is a massless 
Higgs field transforming like $(1,2,2)$).
For the second and third tori we can apply the results in
previous sections without any modification. In the
present example we have (in the notation of Section 5.1) $s_a^{(r)} = m_a^{(r)}/4 = \pm 3$, so $\ci_{\a\b}^{(r)} = \tilde{I}_{\a\b}^{(r)} = I_{\a\b}^{(r)}/4$, $\a, \b = a, b, c$ and $\ci_{\a\b}$ are the relevant numbers to be introduced in the general formula (\ref{int6}). They are given by
\beq
\begin{array}{rcl}
\ci_{ab}^{(2)}\ & = &\ -\ci_{ac}^{(3)}\ =\ 3 \\
\ci_{ac}^{(2)}\ & = &\ \ci_{ab}^{(3)}\ =\ \ci_{bc}^{(2)}
\ =\ -\ci_{bc}^{(3)} \ =\ 1 \\
\tilde{\z}^{(2)}\ & = &\ 3(\z^{(2)}_a + \z^{(2)}_c); \quad \tilde{\z}_b^{(2)}\ =\ 0 \\
\tilde{\z}^{(3)}\ & = &\ 3(\z^{(3)}_a + \z^{(3)}_b); \quad \tilde{\z}_c^{(3)}\ =\ 0
\end{array}
\eeq
where we have also specified the relevant combinations of Wilson lines.
In this case we only have two such combinations, corresponding to the $U(1)$ inside $U(4)$ and the second and third tori, with some contribution of the D5-brane positions $\z_c^{(2)}$ and $\z_b^{(3)}$. The combinations are $\z^{(2)} = \z^{(2)}_a + \z^{(2)}_c$ and $\z^{(3)} = \z^{(3)}_a + \z^{(3)}_b$, and we can directly write down the Yukawa couplings in terms of them as
\beqa
Y_{ij} & = &
 e^{\phi/2} \sqrt{6} \prod _{r=1}^3 \left(\frac{\pim \tau^{(r)}}{(\ca 
^{(r)}/\a')^2} 
\right)^{1/4}  
\cdot  
 \prod _{r=2}^3
e^{i6\pi \frac {\bar{\z}^{(r)}\pim \bar{\z}^{(r)}}{\pim \tau^{(r)}}} 
\\ \nonumber & &
\cdot 
 \vt  
\left[
\begin{array}{c}
 i/3 \\ 0
\end{array}
\right]
\left(3\bar{\z}^{(2)}, 3\bar{\tau}^{(2)}\right)
\vt   
\left[ 
\begin{array}{c}
 j/3 \\ 0   
\end{array}     
\right]
\left(3\bar{\z}^{(3)}, 3\bar{\tau}^{(3)}  \right)
\label{ahuevo}
\eeqa
where $e^{\phi}$ is the $D=10$ dilaton.
One can easily show that these couplings are consistent with the 
ones obtained in \cite{yukis} for a T-dual version of this model.
\footnote{In order to relate both systems, one needs to perform a tilted T-duality described in Section 7.2.2 \cite{bkl}.} 
There it was shown that 
in this simple model there is a single quark-lepton generation 
which becomes massive if  the Higgs field gets a vev,
which is a good starting point to reproduce the observed structure of
quark and lepton masses.
The existence of a single massive family is 
 due to the factorization of the physics on the second 
and third tori of this particular model. Considering e.g.,
non-factorizable tori as discussed at the end of Section 4 
would modify this.

Note the following interesting point. 
Since the effective field theory from the heterotic 
$SO(32)$ string is also $D=10$, $N=1$ SYM theory,
a model similar to the above should be obtained 
starting from the heterotic string. The structure 
is identical, the only difference being that the
$SU(2)_b\times SU(2)_c$ gauge symmetries which came
from $D5$-branes will come now from small instantons
which are known to lead to simplectic groups.
Thus the above model admits a 
(non-perturbative) heterotic construction
involving small instantons.

\section{ Final comments and conclusions}

In this paper we have computed the explicit form of Yukawa   
coupling constants in toroidal compactifications of
$D=10$ SUSY Yang-Mills theories with constant magnetic fluxes. 
The results may also be applied to $D=6,8$ extra dimensional models.
The set-up studied is quite interesting, since it contains
chiral fermions yet is simple enough so that one can explicitly
obtain the wavefunctions of the light modes by explicitly  
solving Dirac and Laplace equations in the compact dimensions.
This allow us to compute the Yukawa couplings as
overlap integrals  over the compact toroidal dimensions.
Given the toroidal geometry is perhaps not surprising  that both
the wavefunctions and the Yukawa couplings obtained
are given by products of Jacobi theta-functions in the
case of a factorizable torus, and Riemann
theta-functions in the general case.   
One would expect  that in more complicated e.g. Calabi-Yau
compactifications the Yukawa couplings will
also be some type  of automorphic forms.

The class of models studied are T-dual to models
of intersecting D-branes recently studied in the literature.
Models with phenomenological interest have been constructed
using that approach in recent years.
We have found that, after appropriate redefinitions of the
moduli and Wilson lines, the results obtained in both
approaches agree in the dilute flux (small angle)
approximation. This is interesting since both calculations
are apparently quite different. In the flux case it is an exercise
in Kaluza-Klein compactification whereas in the intersecting
D-brane side is a stringy computation.

The Yukawa couplings obtained depend on the complex structure
and Wilson lines of the model. We have shown  that the wavefunctions
of chiral modes have a Gaussian profile in extra dimensions.
Since the Wilson lines
control the location of the maxima of these Gaussians,
 one can modify the results for Yukawa  couplings
by appropriately varying the Wilson line variables.
Large Yukawa couplings should appear when the
maxima of the three wavefunctions in the overlap integral
coincide. On the other hand small Yukawa couplings should
appear for wavefunctions with little overlap.   
This degree of freedom should be useful in order to
reproduce the hierarchical structure of Yukawa couplings in
a fully realistic model. Note that the Wilson line degree of
freedom may be also understood as contributions from
non-renormalizable Yukawa couplings. Indeed
in the present class
of models the Wilson lines correspond to the vacuum expectation
values of singlet (or adjoint) complex scalars so that,
expanding the Yukawa couplings on those fields one
gets effective non-renormalizable couplings involving
those scalars in addition to the fermions and the Higgs doublet.
This kind of structure has been abundantly used in phenomenological
analysis of Yukawa textures in the literature.

As an example of the ideas studied
in the paper we have briefly discussed a semirealistic model with
three quark-lepton generations. We leave for future research
a more systematic study of explicit models which could perhaps be
able to reproduce the observed quark/lepton masses and mixing.
Note as a general remark that, due to the complex nature of
both the tori complex structure as well as the Wilson lines, the
Yukawa couplings are complex quantities and thus CP violation   
should be a generic property in a general model of this type.

\medskip

\begin{center}

{\bf \Large Acknowledgments}

\end{center}

We thank  P.G. C\'amara, A. Font, R. Rabad\'an, G. Shiu and  A. Uranga,  
for useful   discussions.
L.E.I. thanks CERN's Theory Division for hospitality.
This work has been partially supported by the European Commission under 
the  RTN contract HPRN-CT-2000-00148 and the CICYT (Spain).
The work of D.C. is supported by  the Ministerio de Educaci\'on, Cultura 
y Deporte (Spain) through a FPU grant. 
The work of F.M. was supported in part by 
NSF CAREER Award No. PHY-0348093, and in part 
by funds from the University of Wisconsin.

\newpage

\appendix

\section{Dimensional Reduction of {\cn=1} Super Yang-Mills \label{reduction}}

Let us consider {\cn=1} Super Yang-Mills theory in $D$ dimensions, $D$ being even. Such theory is described by the action
\beq
S_D=\int d^Dw \ \cl_B + \cl_F
\label{action}
\eeq
where
\beq
\cl_B=-{1 \over 4g^2} \Tr\left\{F^{MN}F_{MN}\right\}, \quad \quad
\cl_F={i \over 2g^2} \Tr\left\{\bar{\lambda}\Gamma^M D_M \lambda\right\}
\label{actions}
\eeq
are the bosonic and fermionic part of the action, respectively, and $M,N = 0, \dots, D-1$. The gauge group field strength $F_{MN}$ and covariant derivative $D_M$ are given, as usual, by
\beqa
F_{MN}& = & \partial_M A_N - \partial_N A_M - i [A_M, A_N] \\
D_M\lambda & = & \partial_M\lambda - i [A_M,\lambda]
\eeqa
where both the $D$-dimensional vector $A_M$ and spinor $\lambda$ transform in the adjoint of the corresponding gauge group $G$. This action is invariant under supersymmetry as well as the gauge transformations
\beqa
A_M & \to & A_M + \partial_M \theta + i [\theta,A_M]\\
\lambda & \to & \lambda + i[\theta, \lambda]
\eeqa
$\th$ being an arbitrary function of $w$ taking values on adjoint of $G$.

For the sake of concreteness let us choose the gauge group $G = U(N)$. The Lie algebra basis of such group can be chosen to be $(U_a)^i_{j}=\d_{ai}\d_{aj}$, $(e_{ab})_{ij}=\d_{ai}\d_{bj}$. We can then expand the fields in the adjoint $A_M$ and $\lambda$ in terms of such basis of generators as
\beqa
A_M & = & B_M + W_M = B_{M}^a U_a + W_{M}^{ab} e_{ab}
\label{expansionB} \\
\lambda & = & \chi + \Psi = \chi^{a}U_a + \Psi^{ab}e_{ab}
\label{expansionF}
\eeqa
Hermiticity of the $U(N)$ generators imposes $B_M$ and $\chi$ to be real and $W_M^{ab} = (W_M^{ba})^*$, $\Psi^{ab}=(\Psi^{ba})^*$. Substituting (\ref{expansionF}) into the fermionic Lagrangian $\cl_F$ we find
\beqa
\cl_F=\cl_F^{(2)} + {\cl_Y}^\prime + {\cl_F}^\prime
\eeqa
where
\beqa \nonumber
\cl_F^{(2)} & = & {i \over 2g^2} \Tr\left\{\bar{\Psi}\Gamma^M\partial_M\Psi-i\bar{\Psi}\Gamma^M[B_M,\Psi]\right\},\\
{\cl_Y}^\prime & = & {1 \over 2g^2} \Tr\left\{\bar{\Psi}\Gamma^M[W_M,\Psi]\right\},\\ \nonumber
{\cl_F}^\prime & = & {i \over 2g^2} \Tr\left\{\bar{\chi}\Gamma^M \partial_M \chi - i\bar{\chi}\Gamma^M [W_M,\Psi] - i\bar{\Psi}\Gamma^M[W_M,\chi]\right\},
\eeqa
and performing an analogous computation with the expansion (\ref{expansionB}) and $\cl_B$ we get
\beqa
\cl_B={\cl_B^{(2)}}^\prime + {\cl_B^{(4)}}^\prime + {\cl_B}^\prime
\eeqa
with 
\beqa \nonumber
{\cl_B^{(2)}}^\prime & = & -{1\over 2g^2}\Tr\left\{D_MW_ND^MW^N-D_MW_ND^NW^M -iG_{MN}[W^M,W^N] \right\}\\
{\cl_B^{(4)}}^\prime & = & {1\over 4g^2}\Tr\left\{[W_M,W_N][W^M,W^N]\right\}\\ \nonumber
{\cl_B}^\prime & = & {i \over 2g^2}\Tr\left\{(D_MW_N-D_NW_M) [W^M,W^N]\right\}-{1\over 4g^2}\Tr\left\{G_{MN}G^{MN}\right\}
\eeqa
where we have defined
\beqa
G_{MN}&=&\partial_MB_N-\partial_NB_M\\
D_MW_N&=&\partial_MW_N-i[B_M,W_N].
\label{defD}
\eeqa

Now, let us expand all fields in terms of their Lie algebra components as in (\ref{expansionB}), (\ref{expansionF}). We find
\beqa
{\cl_B^{(2)}} & = &
{i \over 2g^2} \left(G_{MN}^a - G_{MN}^b\right) 
\left((W^{M ab})^* W^{N ab} - (W^{N ab})^* W^{M ab}\right)\nonumber\\
& & - {1\over 2g^2}\left[(D_M W_{N})^{ab *}(D^M W^N)^{ab}-
(D_M W_{N})^{ab *}(D^N W^{M})^{ab}\right] \nonumber\\
{\cl_B^{(4)}} & = &{1 \over 2g^2} \left[W_{M}^{ab}W_{N}^{bc}W^{M cd}W^{N da}-W_M^{ab}W_N^{bc}W^{N cd}W^{M da}\right] \label{lagrangians} \\
\cl_F^{(2)} & = &{i \over 2g^2}\bar{\Psi}_{ba}\Gamma^M (D_M \Psi)^{ab}\nonumber\\
{\cl_Y}&=&{1 \over 2g^2}\left(\bar{\Psi}^{ab}\Gamma^M W_{M}^{bd}\Psi^{da}-\bar{\Psi}^{ab}\Gamma^MW_{M}^{ca}\Psi^{bc}\right)
\nonumber
\eeqa
where $D_M\Psi$ is defined as in (\ref{defD}).
%
%

Notice that, up to now, we have done nothing but reexpressing equation (\ref{action}) in terms of the new fields $B_{M}^{a}(w)$, $W_{M}^{ab}(w)$, $\chi^a(w)$ and $\Psi^{ab}(w)$. Let us now compactify this theory in a $D-4$ dimensional manifold $\cam_{D-4}$ performing the dimensional reduction from $D$ to 4 dimensions in two steps. First, let us decompose the $D$-dimensional fields $B_{N}^{a}(w)$ and $W_{N}^{ab}(w)$ into the usual components $B_{\mu}^{a}(w)$, $B_{i}^{a}(w)$, $W_{\mu}^{ab}(w)$ and $W_{i}^{ab}(w)$ $\mu=0,...,3$, $i=4,...,D-1$. Since we are interested in maintaining Poincar\'e invariance in the $\mu$ coordinates, we are free to give non-vanishing vevs for $B_{i}^{ab}(w)$ and $W_{i}^{ab}(w)$:
\beqa
B_{i}^{a}(w)&=&\langle B_{i}^{a} \rangle (y) + C_{i}^{a}(w)\\
W_{i}^{ab}(w)&=& \langle W_{i}^{ab}\rangle (y) + \Phi_{i}^{ab}(w)
\eeqa
Notice that these vevs can only depend on the compact coordinates $y$ if we want to preserve four-dimensional Poincar\'e invariance. These non-vanishing vevs will generically break gauge invariance. These expectation values correspond to the turning of magnetic field in the compact dimensions. In the following we will consider $\langle W_i^{ab} \rangle (y)=0$, that is, we will restrict ourselves to Abelian fields in the compact space. Moreover, the case we will ultimately be interested of is $constant$ magnetic fields. 

Since our interest in this paper is to compute Yukawa couplings in the lower dimensional theory, we focus on the terms in (\ref{lagrangians}) involving $D=4$ fermions and scalars. Such scalars are given by the fields $C_{i}^{a}(w)$ and $\Phi_{i}^{ab}(w)$, so let us rewrite everything as
\beqa \nonumber 
{\cl_B^{(2)}}^\prime & = & {\cl}_B^{(2)}+\tilde{\cl}_B^{(2)} \\ 
& = & {i \over 2g^2} \left(G_{ij}^a - G_{ij}^b \right) \left((\Phi^{i ab})^*\Phi^{j ab}- (\Phi^{j ab})^*\Phi^{i ab} \right) \nonumber\\
 & & - {1\over 2g^2}\left[(D_\mu \Phi_{i}^{ab})^*(D^\mu \Phi^{i ab})+(\tilde{D}_i\Phi_{j}^{ab})^*(\tilde{D}^i\Phi^{j ab}) \right. \nonumber \\
 & & \quad - \left. (D_\mu \Phi_{i}^{ab})^*(\tilde{D}^i W^{\mu ab})-(\tilde{D}_i\Phi_{j}^{ab})^*(\tilde{D}^j\Phi^{i ab})\right] +\tilde{\cl}_B^{(2)}
\nonumber\\
{\cl_B^{(4)}}'& = & {\cl}_B^{(4)}+\tilde{\cl}_B^{(4)} = {1\over 2g^2}\left[\Phi_{i}^{ab}\Phi_{j}^{bc}\Phi^{i cd}\Phi^{j da}-\Phi_{i}^{ab}\Phi_{j}^{bc}\Phi^{j cd}\Phi^{i da}\right] +\tilde{\cl}_B^{(4)} \\
\label{lagrangians2}
\cl_F^{(2)} & = & {i \over 2g^2}\bar{\Psi}^{ba}\Gamma^\mu D_\mu \Psi^{ab}+ 
{i \over 2g^2}\bar{\Psi}^{ba}\Gamma^i \tilde{D}_i \Psi^{ab} \nonumber\\
{\cl_Y}' & = & {\cl}_Y+\tilde{\cl}_Y = {1 \over 2g^2}\left(\bar{\Psi}^{ab}\Gamma^i\Phi_{i}^{bd}\Psi^{da}-\bar{\Psi}^{ab}\Gamma^i\Phi_{i}^{ca}\Psi^{bc}\right) +\tilde{\cl}_Y
\nonumber
\eeqa
where the $\tilde{\cl}$'s contain terms irrelevant for our subsequent discussion and we have defined the 'average' covariant derivative by $\tilde{D}_i = \p_i - ig \langle B_i \rangle$. Notice that $\Phi^{ab}$, $\Psi^{ab}$ transform in the bifundamental representation of the ($D$-dimensional) gauge group $U(1)_a \times U(1)_b$, hence this derivative acts as
\beq
\tilde{D}_i \Phi_{j}^{ab} = \partial_i \Phi_{j}^{ab} - i\langle B_{i}^{a} \rangle W_{j}^{ab} + i \langle B_{i}^{b} \rangle W_{j}^{ab}
\eeq
same for $\Psi^{ab}$. $\cl_B^{(2)}$ and $\cl_F^{(2)}$ contain all the possible quadratic terms\footnote{In the case of non-zero vevs for non-Abelian internal gauge fields $<W>_{iab}(y)\neq 0$ more terms like e.g. $D_MW_N[W^M,W^N]$ should be included in $\cl_B^{(2)}$ and $\cl_F^{(2)}$.} in $\Phi_i$ and $\Psi$, that can give rise to effective mass operators for the dimensional reduction of these fields.

The second step is to expand the $D$ dimensional fields on a basis of eigenstates of the corresponding internal wave operator: 
\beqa
\Psi^{ab} (w)& = & \sum_n \chi_n^{ab}(x) \otimes \psi_n^{ab}(y)\\
\Phi_{i}^{ab}(w)& = & \sum_n \varphi_{n\ i}^{ab}(x) \otimes \phi_{n \ i}^{ab}(y)
\label{product}
\eeqa
and so on for the rest of the fields, satisfying
\beqa
i\tilde{\Dbar}_{D-4} \psi_n^{ab} & = & i\G^{(4)} \Gamma^i\tilde{D}_i\psi_n^{ab} = m_n \psi_n^{ab} \\
\D_{D-4} \phi_{n\ i}^{ab} & = & - \tilde{D}_j \tilde{D}^j \phi_{n\ i}^{ab} = M_{n\ i}^2 \phi_{n\ i}^{ab}
\eeqa
where $\G^{(4)} = i \G^1\G^2\G^3\G^4$. We are choosing the internal wavefunctions $\psi_n^{ab}, \phi_{n\ i}^{ab}$ to be dimensionless. The eigenvalues $m_n$, $M_{n\ i} ^2$ are directly related to the four-dimensional mass of the fields $\chi_n^{ab}$, $\varphi_{n\ i}^{ab}$. Indeed, by applying the equations of motion we find
\beqa
i\G^{(4)}\Dbar_{4} \chi_n^{ab} & = & -  m_n  \chi_n^{ab} \\
\D_4 \varphi_{n\ i}^{ab} & = &  M_{n\ i} ^2 \varphi_{n\ i}^{ab} - 2i  \int_{\cam} \langle G_{i}^{a\ j} - G_{i}^{b\ j} \rangle  \varphi_{n\ j}^{ab}
\label{4Dmass}
\eeqa

Notice, as well, that in order to get canonical kinetic terms the $(D-4)$-dimensional fields must satisfy \cite{SW}
\beqa
g^{-2} \int_\cam d^{D-4}y\ {\phi_{n\ i} ^{ab}(y)}^* \phi_{n\ i} ^{cd}(y) & = & \d_{ac}\d_{bd}\\
g^{-2} \int_\cam d^{D-4}y\ {\psi_n^{ab}(y)}^\dagger \psi_n^{ab}(y) & =  & \d_{ac}\d_{bd}
\label{normKK}
\eeqa

Functional integration of the massive fields would yield a four dimensional theory including just the lightest modes. The massless four dimensional fields would be given by $U(n_\a)$ gauge bosons $A^\a_\mu$, the gauginos $\lambda_\a$, the scalars in the adjoint $c_i^\a$, and bifundamental spinors $\chi^{ab}_0$. There will also be bifundamental scalars $\varphi^{ab}_i$, which may be massive, massless or tachyonic depending on the details of the compactification.
In a semi-realistic scenario, the $D=4$ gauge group obtained after dimensional reduction is identified with Standard Model gauge group, the bifundamental fermions with the SM quarks and leptons and one bifundamental scalar with the Higgs boson. Finally, the several replicas of each field that may exist due to multiplicity of zero modes are identified with the different particle generations.

One of the main motivations of this paper is to compute the generic form of the Yukawa couplings for the lightest $D=4$ fields in the class of compactifications described above. Using (\ref{lagrangians2}) and (\ref{product}) we find that such couplings are given by
\beqa
S_Y & = & {1 \over 2g^2} \sum_{IJK}\left\{ \int d^4x\ \bar{\chi}_I^{ab}\ \varphi_{J\ i}^{bd}\ \chi_K^{da} \cdot \int_\cam d^{D-4}y\ {\psi_I^{ab}}^\dagger\ \phi_{J\ i}^{bd}\ \Gamma^i\ \psi_K^{da} \right. \nonumber\\
& & - \left. \int d^4x\ \bar{\chi}_I^{ab}\ \varphi_{J\ i}^{ca}\ \chi_K^{bc} \cdot \int_\cam d^{D-4}y\ {\psi_I^{ab}}^\dagger\ \phi_{J\ i}^{ca}\ \Gamma^i\ \psi_K^{bc} \right\}
\label{yukfinalap}
\eeqa
where the $\{I,J,K\}$ index the replicas that may exist for each of these fields. 

\section{Fluxes and supersymmetry \label{SUSYap}}

The above discussion is also general in the sense that it does not depend whether the $D=4$ low energy-theory is a supersymmetric field theory or not. When coupling SYM theory to gravity, however, it may be useful to consider compactifications where $\cn = 1$ supersymmetry is preserved in the effective theory, at least at the perturbative level. The amount of supersymmetry preserved at low energies depends on the geometrical details of the compactification. In the following, we briefly review the conditions for supersymmetric magnetized compactifications. We refer the reader to Chapter 15 of \cite{GSW2} or ref. \cite{doug1,doug2} for more detailed discussions.

\subsection{Hermitian Yang-Mills Equations}

Consider $D=10$ SYM theory compactified in a $2n$-dimensional manifold $\cam_{2n}$, coupled to $\cn = 1$ supergravity. Under the assumptions of $H = d\phi = 0$, the conditions for a local unbroken supersymmetry amount to having a covariantly constant spinor $\eta$. This in turn implies that the compact manifold $\cam_{2n}$ is a Ricci-flat K\"ahler manifold, i.e., a Calabi-Yau $n$-fold ${\bf CY_n}$. On the other hand, a non-trivial Yang-Mills gauge field $A$ must have a field strength $F= dA$ which obeys
\beq
\d \lam = \G^{ij}F_{ij} \eta = 0
\label{SUGRA}
\eeq
where $\lam$ stands for a ten-dimensional gaugino. If $\cam_6$ is a complex manifold (as the existence of the covariantly constant $\eta$ would require) then we can rewrite (\ref{SUGRA}) as \cite{GSW2}
\beqa
F_{ab}\ =\ F_{\bar{a}\bar{b}} & = & 0
\label{SUGRA1} \\
g^{a\bar{b}}F_{a\bar{b}} & = & 0
\label{SUGRA2}
\eeqa
where $g_{a\bar{b}}$ is the hermitian metric on $\cam_{2n}$. These two conditions are quite strong. For our purposes we may consider a slightly less constraining system of equations, which comes from coupling $D=10$ SYM to $\cn = 2$ supergravity. Eq. (\ref{SUGRA}) then generalizes to 
\beq
\d \lam = \G^{ij}F_{ij} \eta + \eta' = 0
\label{SUGRAN=2}
\eeq
where $\eta'$ is the other covariantly constant spinor coming from extended supersymmetry. Eqs. (\ref{SUGRA1}), (\ref{SUGRA2}) then relax to  
\beqa
F_{ab} \ =\ F_{\bar{a}\bar{b}} & = & 0
\label{HYM1} \\
g^{a\bar{b}}F_{a\bar{b}} & = & c \cdot {\bf 1} 
\label{HYM2}
\eeqa
where $c$ is a constant which encodes which $\cn =1$ subalgebra of $\cn=2$ is preserved by the SYM theory. Eqs. (\ref{HYM1}) and (\ref{HYM2}) are known as hermitian Yang-Mills equations, and imply that the gauge field $A$ solves the super Yang-Mills equations, hence giving rise to a supersymmetric gauge theory upon dimensional reduction. If $\cam_6$ is a K\"ahler manifold (which is again implied by the existence of a covariantly constant spinor), we can rewrite (\ref{HYM2}) as
\beqa
F \wedge \om^{n-1} & = & c\ \om^n \cdot {\bf 1}
\label{HYM2b}
\eeqa
where $\om$ is the K\"ahler form of $\cam_6$. 

Equation (\ref{HYM1}), provides a generalization of a holomorphic function on $\cam_{2n}$. Indeed, notice that (\ref{HYM1}) implies $F_{\bar{a}\bar{b}} = i [D_{\bar{a}}, D_{\bar{b}}] = i [\p_{\bar{a}} - iA_{\bar{a}}, \p_{\bar{b}} - iA_{\bar{b}}] = 0$, so if we assume that the gauge field $A$ is hermitian, we can write it as
\beq
A_{\bar{a}} = - i \p_{\bar{a}} V \cdot V^{-1}, \quad \quad A_{a} = - i (\p_{a} V^{\dag -1})  \cdot V^{\dag},
\label{connection}
\eeq
where $V$ is a matrix-valued function on the coordinates $(z^a, \bar{z}^a)$. Then, if we define a field $f$ to be "holomorphic" by satisfying
\beq
D_{\bar{a}} f = 0
\label{holom}
\eeq
then is easy to see that such field is of the form $f = Vg$, where $g$ is a holomorphic function in the usual sense, i.e., $\p_{\bar{a}} g = 0$. In geometrical terms, the gauge field $A$ can be understood as a vector bundle $E$ over $\cam_{2n}$. Now, a vector bundle satisfying (\ref{HYM1}) is a holomorphic vector bundle, in the sense that their transition functions can be chosen to be holomorphic \cite{GSW2}. The converse is also true. Hence, considering Yang-Mills fields satisfying (\ref{HYM1}) amounts to studying holomorphic gauge bundles.

On the other hand, eq. (\ref{HYM2}) is related to the notion of $\mu$-stability by the theorems of Donaldson and Uhlenbeck-Yau. Indeed, let us consider an irreducible complex vector bundle $E$ over a compact K\"ahler manifold $\cam_{2n}$, and whose field strength is given by $F=dA$. The degree of $E$ is defined by
\beq
deg(E) = \int_{\cam_{2n}} c_1(E) \wedge \om^{n-1} = \int_{\cam_{2n}} \tr F \wedge \om^{n-1} 
\label{degree}
\eeq
where $c_1(E)$ is the first Chern class of $E$. The relevant quantity is the {\it slope} of $E$ which is defined as $\mu(E) = deg(E)/rank(E)$.\footnote{See the main text for a less abstract definition of slope in terms of magnetized compactifications.} Notice that $\mu$ depends on the K\"ahler moduli of $\cam_{2n}$, being independent of the complex structure moduli. A bundle $E$ is said to be {\it stable} if for any subbundle $S \subset E$ we have $\mu(S) \leq \mu(E)$. What the theorems above show is that any bundle $E$ satisfying (\ref{HYM1}) (that is, a holomorphic bundle) and being $\mu$-stable satisfies (\ref{HYM2}), and the other way round. 

\subsection{D-brane interpretation}

In general, we will consider a magnetic flux which is of the form
\beq
F = 
\left(
\begin{array}{ccc}
F_a \\
& F_b \\
& & \ddots
\end{array}
\right)
\label{genflux}
\eeq
Each box $F_\a$, $\a = a, b, \dots$  represents an irreducible component of the vector bundle (that is, a subbundle). Turning on the magnetic flux $F$ will break the initial gauge group $G$ to a direct product of smaller gauge groups $\prod_i G_i \subset G$. Generically, each of these smaller gauge groups $G_i$ will be an Abelian $U(1)_\a$ factor associated to the component $F_\a$. If we consider a compactification where $F_{\a^i} = F_\a$, $i = 1, \dots, N_\a$ then we may have the gauge enhancement $\prod_{\a^i} U(1)_{\a^i} \raw U(N_\a)$.\footnote{Such gauge unhiggsing occurs if we have in addition $A_{\a^i} = A_\a$, $i = 1, \dots, N_\a$, i.e., if all the Wilson lines are equal.}

This picture reminds of the gauge theory dynamics associated to D-brane physics and, in fact, it turns out that D-branes provide a nice physical realization of the mathematical results previously stated. \footnote{Recall, however, that our previous discussion is general and describes as well heterotic compactifications with non-trivial gauge fields $A$.} In order to see this, consider type IIB string theory compactified on $\cam_{6}$ with $N$ D9-branes filling up the non-compact and the compact dimensions of our ten-dimensional theory. In principle, this will yield a $D=4$ $U(N)$ gauge theory upon dimensional reduction. Now, we are again allowed to introduce magnetic fluxes of the form (\ref{genflux}) in the internal dimensions of the D9's worldvolume, again breaking the $U(N)$ gauge theory. To each factor $F_\a$ we associate a {\em dynamical} D9-brane $\a$ with gauge group $U(1)_\a$. Notice that in general a dynamical D9-brane will be composed of $k_\a = {\rm rank\ } F_\a \geq 1$ type IIB D9-branes. In fact, this is how we achieve gauge group rank reduction in this setup.

After turning on $F$, the spectrum of the dimensionally reduced theory can be arranged in several sectors:
\begin{itemize}

\item{Closed strings} \\
This will provide the gravity sector of the $D=4$ theory. 

\item{Open $D9_\a D9_\a$ strings} \\
Strings beginning and ending on the same dynamical brane will yield a $U(1)_\a$ gauge theory. A stack of $N_\a$ dynamical D9-branes with the same flux $F_{\a^i}$ will yield a gauge group $U(N_\a)$ if there is no relative Wilson lines between them. 

\item{Open $D9_\a D9_\b$ strings} \\
Strings stretching between two stacks of $N_\a$, $N_\b$ dynamical branes may yield $D=4$ chiral fermions, as well as scalars, transforming in the bifundamental representation $({N}_\a, \bar{N}_\b)$ of the corresponding $U(N_\a) \times U(N_\b)$ gauge group.

\end{itemize}

Let us now analyze the supersymmetry conditions above in terms of this physical picture. The closed string sector of the theory is mainly sensitive to the geometry of the compact manifold $\cam_6$. Hence, the condition of $\cam_6$ being a Calabi-Yau three-fold translates into a (tree-level) $D=4$ supersymmetric gravity sector.

Now let us introduce a magnetic flux satisfying eq.(\ref{HYM1}), that is, let us consider $F$ being a $(1,1)$-from. Without loss of generality, we can consider $F$ to be of the block-diagonal form (\ref{genflux}) with a total of $K$ blocks. Each block $\a$ corresponds to a stack of $N_\a$ dynamical branes, that is to a flux of the form $F_\a \otimes {\bf 1}_{N_\a}$, hence a $U(N_\a)$ gauge theory. The total flux $F$ will provide us with a low energy field theory with gauge group $\prod_{\a = 1}^K U(N_\a)$. Moreover, if each block $\a$ satisfies, by itself, the condition (\ref{HYM2}) we will actually have a low energy gauge theory which is a product of $U(N_\a)$ SYM theories.

Now, even if that is the case, the $D9_\a D9_\b$ spectrum may not be supersymmetric. Indeed, each block $\a$ will have a slope $c_\a = \mu (F_\a)$. If $c_\a = c_\b$, then the $D9_\a D9_\b$ sector will be supersymmetric.  To each chiral fermion in a $D9_\a D9_\b$ sector there will correspond a massless complex scalar arising from the same sector, completing a $D=4$ $\cn = 1$ chiral multiplet in the $(N_\a, N_\b)$ representation. On the other hand, if $c_\a \neq c_\b$, then the $D9_\a D9_\b$ sector will break supersymmetry, and the formerly supersymmetric partners of the chiral fermions will be either massive or tachyonic.

We then see that the conditions for supersymmetry, namely a covariantly constant spinor and the two Hermite Yang-Mills equations, can be matched with the supersymmetry conditions for the three different sectors of the theory at hand. They can be understood from demanding supersymmetry in the gravity, gauge and chiral sectors.

\subsection{Mirror symmetry and string corrections}

Actually, it turns out to be quite instructive to understand these supersymmetry conditions in terms of the T-dual or mirror symmetric picture. Indeed, T-duality relates type IIB superstring theory on the compact manifold $\cam_{6}$ with type IIA theory on its mirror $\cw_{6}$. Since we are supposing $H = d \phi = 0$ in both sides, the only way to achieve a supersymmetric gravity sector is to demand that both manifolds admit a covariantly constant spinor, that is, that they satisfy the ${\bf CY_3}$ condition. 

%
\EPSFIGURE{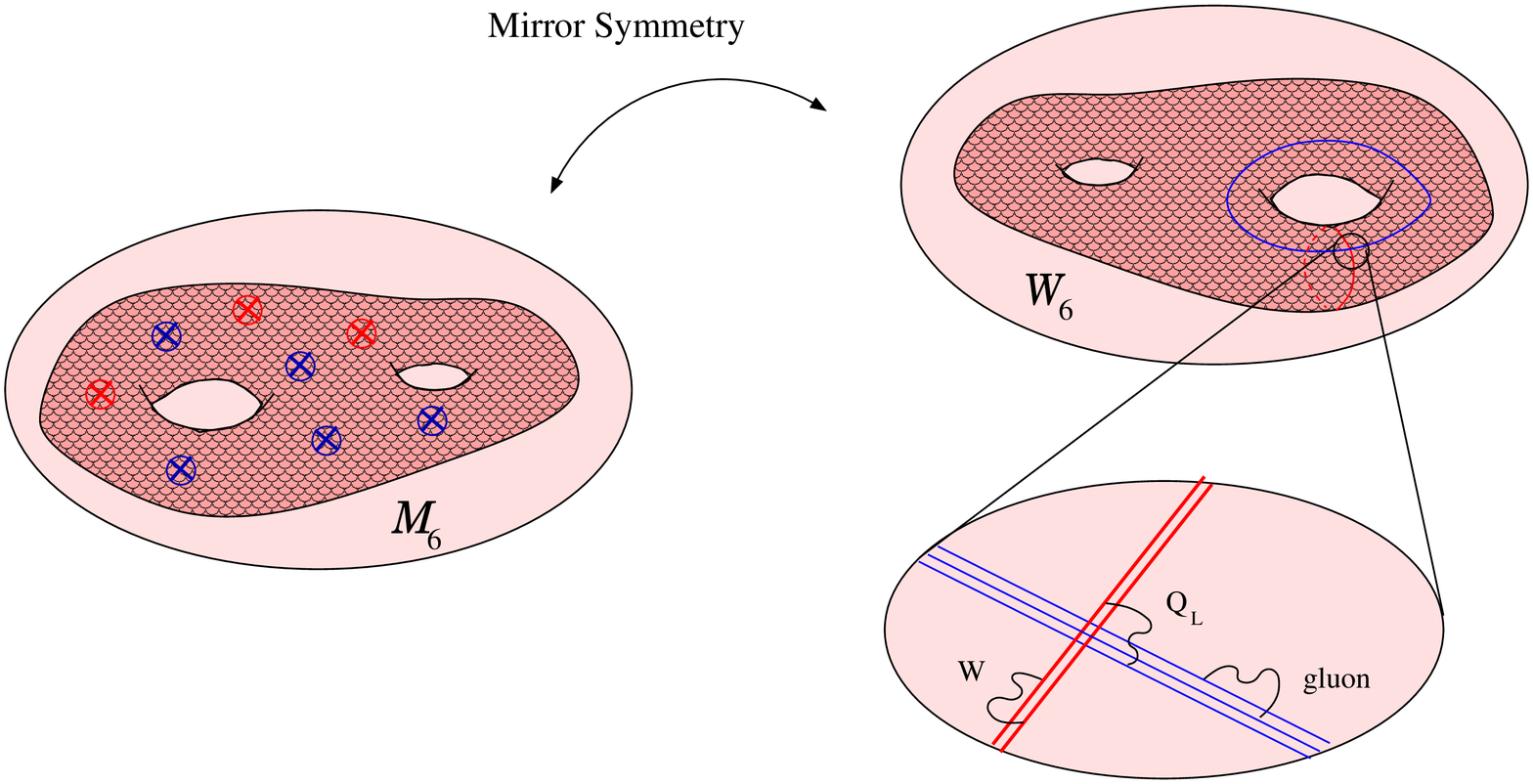, width=6.5in}
{\label{mirror}
Artist's view of open string mirror symmetry, which relates type IIB magnetized compactifications with type IIA intersecting D6-brane worlds. In general, $\cam_{6}$ and $\cw_{6}$ will have different topology.}
%

We now consider the open string sector of the theory. The 'mirror' D-branes of a  magnetized stack of D9-branes will be given by stacks of $N_\a$ D6-branes wrapping 3-cycles $\Pi_\a$ on $\cw_6$, and expanding the four non-compact dimensions. The first hermitian Yang-Mills condition, eq.(\ref{HYM1}), corresponds in this picture to imposing $\Pi_\a$ to be a Lagrangian cycle, that is, to satisfy $\om|_{\Pi_\a} \equiv 0$, where $\om$ is the K\"ahler form in $\cw_3$. On the other hand, the second hermitian Yang-Mills condition (\ref{HYM2}) will translate to $\Pi_\a$ being a Special Lagrangian (SL) submanifold, which in addition to $\Pi_\a$ being Lagrangian imposes the condition $\pim(e^{i\th_\a} \Om)|_{\Pi_a} \equiv 0$. Here $\Om$ stands for the holomorphic $(3,0)$-form of $\cw_6$ and $\th_\a$ is a constant phase. Finally, the chiral matter in the bifundamental representation $(N_\a,N_\b)$ arises again in the $D6_\a D6_\b$ sector of the theory. More precisely it is localized in the intersection points of the submanifolds $\Pi_\a$ and $\Pi_\b$ \cite{bdl}. See figure \ref{mirror}.

Again, we find that supersymmetry may be broken or not in different sectors of the theory. If we consider $\cw_6$ being a ${\bf CY_3}$ and $\Pi_\a$ being a Special Lagrangian submanifold for every stack $\a$ of $N_\a$ D6-branes, then at low energies we will recover a $D=4$ $\cn = 2$ supergravity sector and a gauge sector of $\prod_\a U(N_\a)$ $\cn=1$ SYM theories. However, if we consider the matter content at the intersection of, say, SL's $\Pi_\a$ and $\Pi_\b$ this may yield a supersymmetric spectrum or not, depending on the respective phases $\th_\a$ and $\th_\b$. If $\th_\a = \th_\b$ the $D6_\a D6_\b$ chiral matter will be arranged in $\cn=1$ supermultiplets, whereas if $\th_\a \neq \th_\b$ supersymmetry will be broken in this sector. From the point of view of the effective field theory, this can be understood as the appearance of a non-vanishing FI-term \cite{kg,doug9,cim1}.

In the rest of the paper we will be dealing with one of the most simplest cases of open string mirror symmetry. Namely, both $\cam_{6}$ and $\cw_{6}$ will be T-dual $6$-dimensional tori. Moreover, in the type IIB picture we will be considering D9-branes with constant magnetic fluxes, which in the type IIA picture correspond to flat D6-branes intersecting at angles. This class of configurations have been analyzed in great detail. In particular, in \cite{flatness} the two supersymmetry conditions for D6-brane wrapping a 3-cycle $\Pi$ in the intersecting picture were understood in terms of F- and D-flatness conditions on the worldvolume gauge theory of such D6-brane. This statement should also hold in the $D=4$ reduced theory and, by mirror symmetry, it should be matched
with the hermitian Yang-Mills equations (\ref{HYM1}), (\ref{HYM2}). We summarize the interpretation of the supersymmetry conditions in Table \ref{SUSY}.

\TABLE{\renewcommand{\arraystretch}{1.5}
\begin{tabular}{|c|c|c|} 
\hline 
 MBW & IBW & Field Theory \\ 
\hline 
\hline 
$\cam_6 = {\bf CY_3}$ & $\cw_6 = {\bf CY_3}$ & SUGRA \\
\hline
$F_{\bar{a}\bar{b}} = 0$ & $\om|_{\Pi_\a} \equiv 0$ & F-flatness \\ 
\hline 
$g^{a\bar{b}}F_{a\bar{b}} =  c \cdot {\bf 1}$ & $\pim(e^{i\th_\a} \Om)|_{\Pi_a} \equiv 0$ & D-flatness \\
\hline 
\end{tabular} 
\label{SUSY}
\caption{\small Supersymmetry conditions for both Magnetized Brane Worlds (MBW) and Intersecting Brane Worlds (IBW) in terms of the $D=4$ effective field theory. Notice that, in the MBW picture, the F-flatness condition imposes a constraint on the complex structure of $\cam_6$, whereas D-flatness concerns the K\"ahler structure. An opposite statements holds for the IBW mirror picture.}}

It is important to notice that, although the hermitian Yang-Mills equations and the calibration conditions match qualitatively in both sides of the mirror map, there will be a quantitative mismatch away from the limit of large volumes and diluted fluxes (or, equivalently, small angles). Indeed, the notion of $\mu$-stability and the hermitian Yang-Mills equations have been derived in the supergravity approximation of string theory. When leaving such regime, we would expect stringy $\a'$ corrections to the hermitian Yang-Mills conditions. Indeed, by considering the conditions for unbroken symmetry from the Dirac-Born-Infeld action instead of the SYM Lagrangian, the condition (\ref{HYM2b}) is replaced by the MMMS equation \cite{MMMS}
\beq
\pim e^{i\th} \Tr (\om + i \a' F)^n
\label{MMMS}
\eeq
which involves higher powers of the field strength $F$. These and other considerations led to a modification to the concept of $\mu$-stability, namely the $\Pi$-stability proposed in \cite{Fiol}. Roughly speaking, this proposal amounts to considering the stability condition in the intersecting D-brane picture and translate it back to the holomorphic bundle picture by means of the mirror symmetry map. As explicitly seen in \cite{aki,torons} in toroidal compactifications this implies substituting the slope $\mu$, which is related to the tangent of the angles between two D-branes, by the angles themselves. Due to this fact, our results concerning three-point functions should be only considered as accurate from the string theory point of view only in the limit of large compactification volume and diluted fluxes (which corresponds to small angles), where the ten-dimensional effective field theory captures all the physics of the underlying string theory.

\end{document}